\documentclass[printer]{aa}

\usepackage[dvips]{color}
\usepackage{graphicx,subfigure}
\usepackage[final]{pdfpages}
\usepackage{amsmath}
\usepackage[utf8]{inputenc}
\usepackage{epsfig,amsmath,amssymb}
\usepackage[varg]{txfonts}

\def\ga{\,\hbox{\hbox{$ > $}\kern -0.8em \lower 1.0ex\hbox{$\sim$}}\,}
\def\la{\,\hbox{\hbox{$ < $}\kern -0.8em \lower 1.0ex\hbox{$\sim$}}\,}

\def\beq{\begin{equation}}
\def\eeq{\end{equation}}

\titlerunning{Ambipolar diffusion effects on MHD filaments}
\authorrunning{Ntormousi, Hennebelle et al.}

\begin{document}

\title{The effect of ambipolar diffusion on low-density molecular ISM filaments}

\author{Evangelia Ntormousi\inst{1}, Patrick Hennebelle\inst{1,2}, Philippe Andr\'{e}\inst{1} \& Jacques Masson\inst{3}}

\institute{
Laboratoire AIM, 
Paris-Saclay, CEA/IRFU/SAp - CNRS - Universit\'e Paris Diderot, 91191, 
Gif-sur-Yvette Cedex, France \\
\and
LERMA (UMR CNRS 8112), Ecole Normale Sup\'erieure, 75231 Paris Cedex, France \\
\and
School of Physics and Astronomy, University of Exeter, Stocker Road, Exeter EX4 4QL, UK}


\abstract
{The filamentary structure of the molecular interstellar medium and the potential link of this morphology to star formation
have been brought into focus recently by high resolution observational surveys.
An especially puzzling matter is that local interstellar filaments appear to have the same thickness, 
independent of their column density.  This requires a theoretical understanding of their formation process and the physics
that governs their evolution.}
{In this work we explore a scenario in which filaments are dissipative structures of the large-scale interstellar turbulence cascade
and ion-neutral friction (also called ambipolar diffusion) is affecting their sizes by preventing small-scale compressions.}
{We employ high-resolution (512$^3$ and 1024$^3$), 3D MHD simulations, performed with the grid code RAMSES,  
to investigate non-ideal MHD turbulence as a filament formation mechanism.  We focus the
analysis on the mass and thickness distributions of the resulting filamentary structures.}
{Simulations of both driven and decaying MHD turbulence show that the morphologies of the density and the magnetic field 
are different when ambipolar diffusion is included in the models.
In particular, the densest structures are broader and more massive as an effect of ion-neutral friction and 
the power spectra of both the velocity and the density steepen at a smaller wavenumber.}
{The comparison between ideal and non-ideal MHD simulations 
shows that ambipolar diffusion causes a shift of the filament thickness distribution
towards higher values. However, none of the distributions exhibit the pronounced peak found in the 
observed local filaments.  Limitations in dynamical range and the absence of self-gravity in these numerical experiments
do not allow us to conclude at this time whether this is due to the different filament selection or due to 
the physics inherent of the filament formation. }


\maketitle


\section{Introduction}

The idea that the interstellar medium (ISM) is turbulent and hierarchically organised in sheets and filaments is not a new one at all.
Decades of observational data have shown that molecular clouds host supersonic internal motions 
\citep{Larson_1981,Falgarone_Phillips_1990, Williams_2000} and that dense gas structures tend to be elongated 
\citep{Barnard_1927,Schneider_Elmegreen_1979,Feitzinger_1986,Myers_2009} 

Nevertheless, recent high resolution submm mapping observations with the Herschel space observatory 
have revealed the filamentary nature of the dense interstellar 
gas in unprecedented detail \citep{Andre_2010, Molinari_2010}
and suggested that interstellar filaments likely play a key role in the star formation process 
(see \citealp{Andre_2014} for a review).  
For example, \citet{Konyves_2015}, find that ~75\% of the Herschel-identified prestellar cores in the Aquila cloud complex lie in supercritical filaments. 
While filaments are observed on every length scale throughout the Galactic plane \citep{Schisano_2014}, 
Herschel imaging of nearby molecular clouds in the Gould Belt led to the discovery that the central parts of interstellar filaments 
as traced by submm dust continuum have a constant inner thickness of about $0.1$ pc over a very large range of column densities \citep{Arzoumanian_2011}. 
In fact, the sample of nearby filaments from the Gould Belt survey to which this result applies includes both diffuse, unbound structures and self-gravitating, star-forming filaments (Arzoumanian 2012, PhD thesis, \cite{Andre_2014}, which is puzzling. 
In principle we would expect structures to become thinner and thinner as turbulent flows or gravity compress them and, as a result, their thickness to have a strong density dependence.  

At the same time though, such a puzzle provides a very good test for distinguishing between the numerous theories for interstellar filament formation.  Indeed, filamentary structure seems to appear naturally in all sorts of numerical experiments: colliding atomic flows \cite{Heitsch_2008,Hennebelle_2008,Gomez_2014}, supersonic turbulence \citep{Padoan_2004, Moeckel_2015}, expanding supershells \citep{Ntormousi_2011}, and gravitational collapse of either isolated structures \citep{Burkert_Hartmann_2004,Heitsch_2009} or portions of a galaxy \citep{Kim_Ostriker_2003,Smith_2014}.  None of these simulations, however, has yet predicted or explained the apparently constant thickness of the observed filaments.

Theoretical attempts to explain the constant thickness so far have focused on the 
self-gravitating star-forming filaments, which indeed show evidence of accretion in observations \citep{Palmeirim_2013}.
\citet{Heitsch_2013a} explored the evolution of accreting cylinders
through analytical models in which accretion drives turbulence in the interior of the filament.
These models explore a range of credible parameters and eventually produce a 
thickness distribution fairly independent of column density, which 
is attributed to the equilibrium between turbulence and gravitational fragmentation.
The produced filaments, however, do not cover the entire regime of column densities 
studied by \cite{Arzoumanian_2011}.  The challenge of explaining the constant filament thickness lies precisely
at its persistence over orders of magnitude in column density.

\citet{Smith_2014} studied the filament properties in hydrodynamic simulations
of galactic dynamics, extracting the filaments with a technique very similar to that used in 
observations.  However, they found a range of filament widths rather than a constant value.
This result is not surprising, given the self-similar behavior of both turbulence and gravity.

\citet{Hennebelle_2013} proposed that MHD turbulence describes the filamentary appearance of the ISM much better than hydrodynamic turbulence.  The shear provided by the turbulent flow tends to elongate small-scale structures and the magnetic tension keeps these filaments coherent.  At the same time, \citet{Hennebelle_Andre_2013} suggested through simple analytical calculations that, in the case of self-gravitating molecular clouds, ion-neutral friction (also called ambipolar diffusion) could be the process behind the constant filament thickness.  
In short, the argument in \citet{Hennebelle_Andre_2013} is that mass accretion onto a filament drives turbulence in its interior,
which is constantly dissipated through ambipolar diffusion. 
In contrast to \citet{Heitsch_2013a}, in this theory it is the balance between accretion-driven turbulence and dissipation that keeps the thickness of the filament almost independent of density as it gains mass. 

In this work we explore low-surface density, non self-gravitating filament formation in MHD turbulence, 
and observe the effects of ion-neutral friction on the filament properties.  
To this end, we perform a series of high-resolution numerical simulations of both decaying and driven MHD turbulence and solve the MHD equations with and without ambipolar diffusion.  
We focus the analysis on the thickness distribution of the densest elongated structures in the simulation (which here is 
defined differently than in observations), but 
we also explore the general behavior of a magnetized turbulent fluid with ambipolar diffusion.

Ion-neutral friction (or ambipolar diffusion), has been invoked many times as an important damping mechanism for interstellar turbulence
\citep{Zweibel_Josafarsson_1983,Brandenburg_Zweibel_1994,Tassis_Mouschovias_2004,Li_Houde_2008}.
In a low ionization plasma such as a molecular cloud, the neutrals make up most of the material and
the few ions are attached to the magnetic field.  
Collisions between the two species remove energy from both their motions, a process which can be expressed as a friction force.

This friction acts at a scale called the ambipolar diffusion length ($\lambda_d$), where the motions of the two species decouple. 
Below this length the propagation of MHD waves is prevented and for turbulence, this means a cutoff in the energy cascade.
This length was derived by \citet{Kulsrud_Pearce_1969} in the context of cosmic ray interactions with the interstellar medium.
However, it been used in star formation theory for estimating relevant timescales for prestellar cores \citep{Mouschovias_1991} and
was also the basis of the theory presented in \citet{Hennebelle_Andre_2013}:

\begin{equation}
\lambda_d = \frac{\pi v_A}{\gamma_{AD}\rho_i} \label{ldiss}
\end{equation}
where $v_A$ the Alfv\'{e}n speed and $\gamma_{AD}=9.2\times 10^{13}$ cm$^3$ s$^{-1}$ g$^{-1}$ is the coupling constant between ions and neutrals \citep{Draine_1983}.  
One can rewrite  Eq. (\ref{ldiss}) in terms of diffuse molecular gas parameters: $\rho_0\simeq 500~m_H$ gr cm$^{-3}$, $\rho_i\simeq C\sqrt\rho_0$ with $C=3\times10^{-16}$ cm$^{-3/2}$ g$^{1/2}$ {\bf{\citep{Elmegreen_1979}}} and $B\simeq10\mu$G:

\begin{equation}
\lambda_d = 0.13\times \left( \frac{B}{10\mathrm{\mu G}} \right) \left( \frac{\rho_0}{500~\mathrm{m_H~gr~cm^{-3}}} \right)^{-1}~\mathrm{pc}
\label{lambda2}
\end{equation}
assuming that $v_A$ in Eq. (\ref{ldiss}) is calculated with the total mass density.
A theory for the damping of linear MHD waves by ambipolar diffusion was proposed by \citet{Balsara_1996}, based on a one-dimensional dispersion analysis.
He suggested that at large scales, two-fluid MHD turbulence should behave like single-fluid turbulence, while at scales comparable to or smaller than $\lambda_d$, 
the damping of MHD waves should depend on the Alfv\'enic Mach number of the flow.
More specifically, his theory predicts that the Alfv\'en and slow magnetosonic waves will be damped when the Alfven speed is smaller than the sound speed,
while if the  Alfven speed is larger than the sound speed, the Alfv\'en and fast magnetosonic waves will be damped.

\citet{Burkhart_2015} confirmed these findings by high-resolution, two-fluid driven MHD turbulence simulations.  Using mode decomposition to
analyse the different MHD wave modes, they found that sub-Alfv\'enic turbulence is damped at the ambipolar diffusion scale, while in super-Alvf\'enic situations the cascade can continue to smaller scales.  Our studies include both forced and decaying turbulence with a mean Alfv\'en speed of about 1.1~km~sec$^{-1}$.  The decaying turbulence runs start super-Alfv\'{e}nic and finish sub-Alfv\'{e}nic, while
the driven runs are always sub-Alfv\'{e}nic.  The theory outlined above will be important for the interpretation of our results. 

There are many examples of turbulence studies that include ambipolar diffusion.
\citet{Li_McKee_2008} study the statistical behaviour of ideal MHD turbulence and MHD turbulence with ambipolar diffusion, while in
a follow-up paper, \citet{McKee_2010} study the behaviour of the dense clumps in the same simulations.
Their methods, both in the implementation of the ambipolar diffusion and in the analysis of the simulations are different 
from ours, so the results also differ.  We include a detailed comparison with our findings at the end of this paper. 

\citet{Oishi_MacLow_2006} simulated supersonic turbulence with ambipolar diffusion and reported the non-appearence 
of a characteristic scale.  
However, their study had limited resolution and may not have captured enough of the inertial range.  
Also, they did not include any explicit calculation of the filament thickness.  
On the other hand, \citet{Downes_OSullivan_2011} did observe a significant change in the morphology of their turbulence simulations when they included ambipolar diffusion.  \citet{Momferratos_2014} studied the properties of diffusive structures 
created by ambipolar diffusion with a spectral code and found them to be much broader than those created, for example, by Ohmic dissipation.  Clearly, there is a dependence on the simulation parameters which needs to be explored further.

The motivation for the present study is to examine whether ion-neutral friction can set a characteristic scale 
of the filaments created by MHD turbulence.  Its aim is not to reproduce the 0.1 pc filament width found in observations, but rather to 
help understand the physical mechanisms setting this characteristic width.
To this end, we perform simulations of MHD turbulence with and without ambipolar diffusion and with careful control of numerical dissipation
and compare the sizes and masses of the dense filaments.
The numerical method we use for our simulations is explained in Section \ref{numerics} and the main results are described in Section \ref{sims}.  
The calculations concerning the filament properties are presented in a dedicated section of the paper, Section \ref{thick}.  
In Section \ref{sec:resolution} we discuss resolution issues.
A comparison of our findings with previous work, as well as a discussion of their implications is presented in Section \ref{discuss}, 
while Section \ref{concl} concludes the paper.

\begin{figure*}[!ht]

 \includegraphics[width=0.49\linewidth]{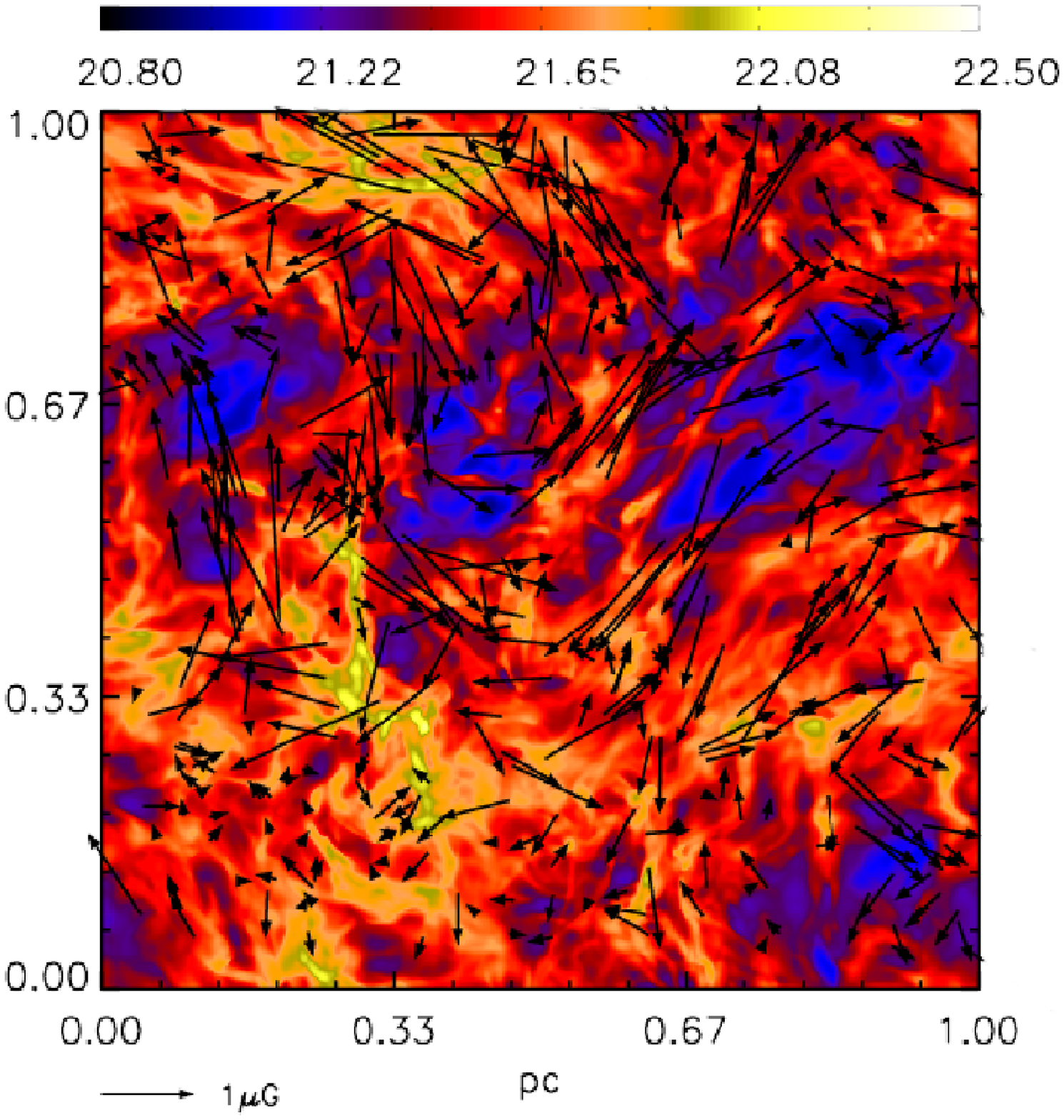}
   \includegraphics[width=0.49\linewidth]{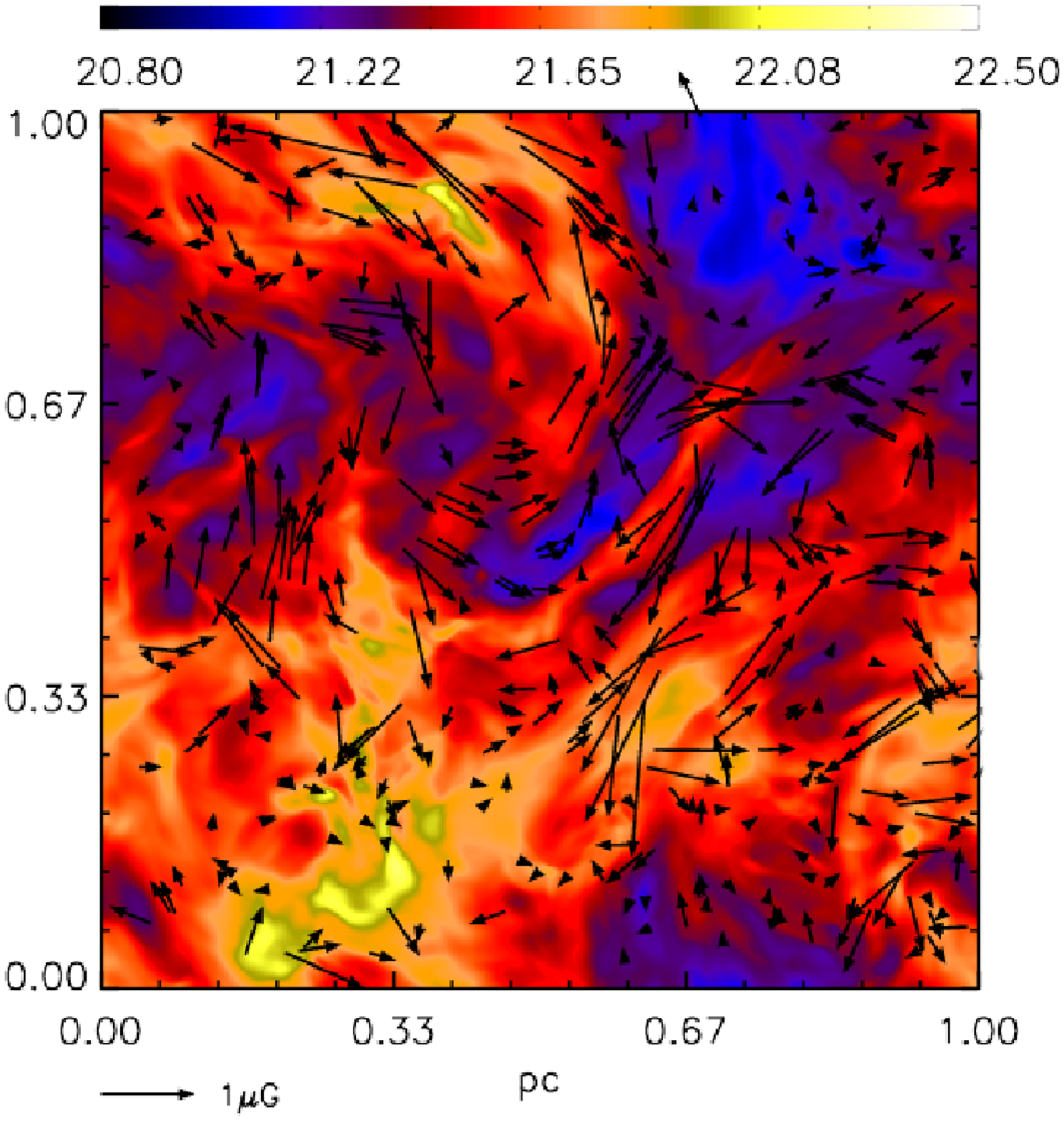}
   \includegraphics[width=0.49\linewidth]{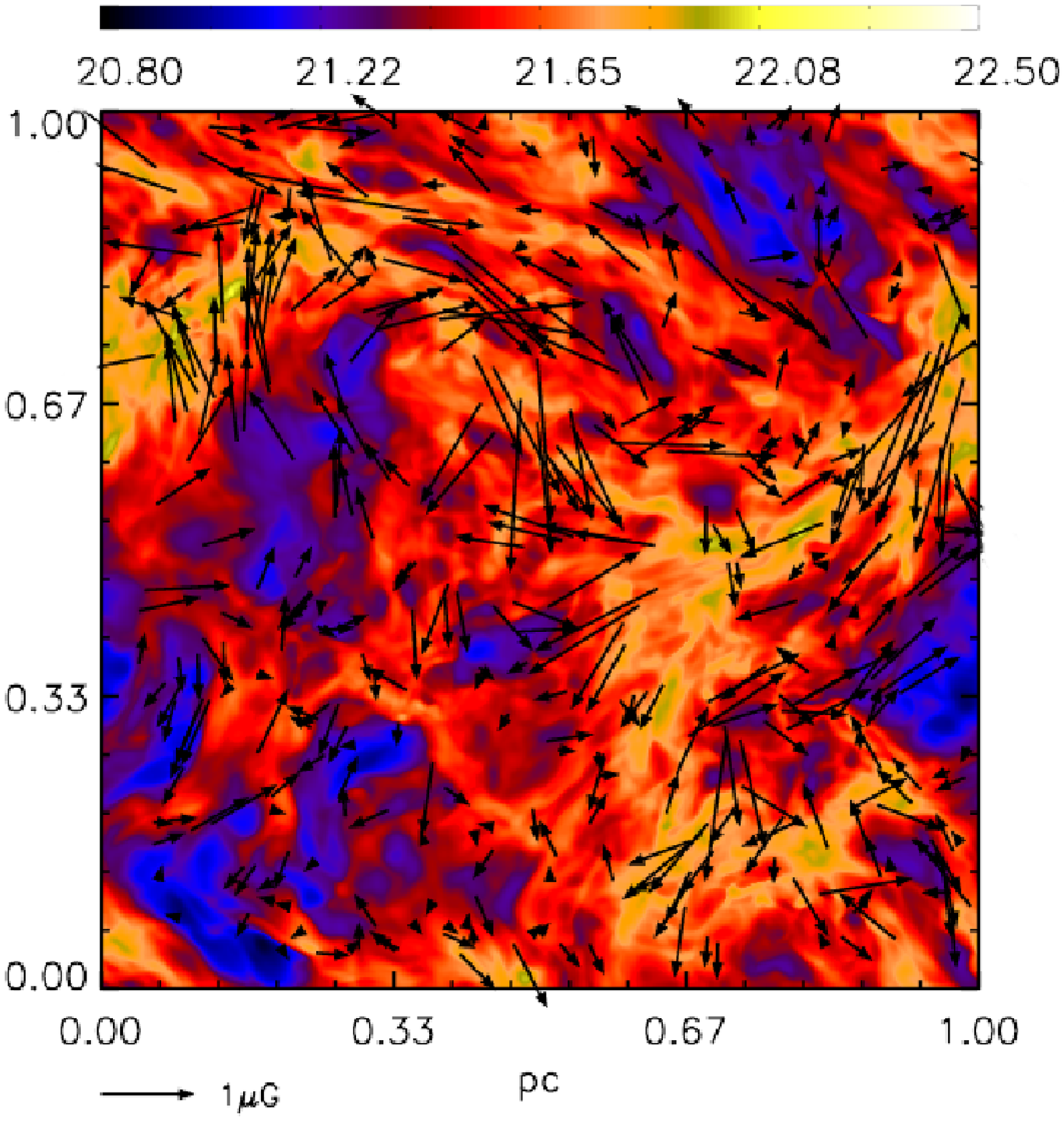}
   \includegraphics[width=0.49\linewidth]{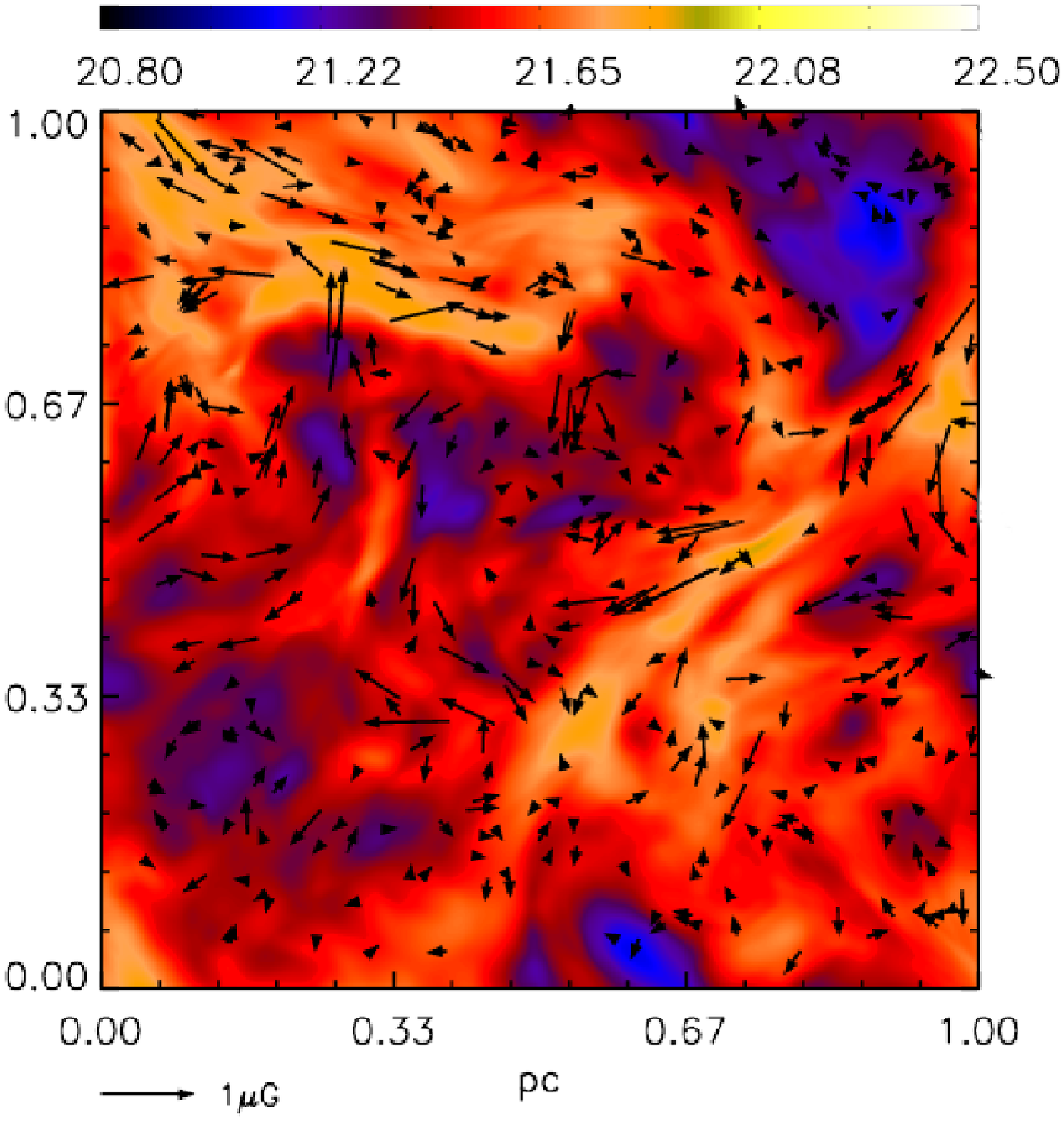}
   \caption{Logarithm of the total gas column density on the yz plane, in cm$^{-2}$, for the decaying turbulence runs A0 (ideal, left panel) and B0 (ambipolar diffusion, right panel) at t=$5\cdot10^{-3}$ (top)
  and t=$10^{-2}$ (bottom). The black arrows show the projected magnetic field on the same plane.}
   \label{turb_a0_b0}
\end{figure*}

\begin{figure}[!h]
   \includegraphics[width=\linewidth]{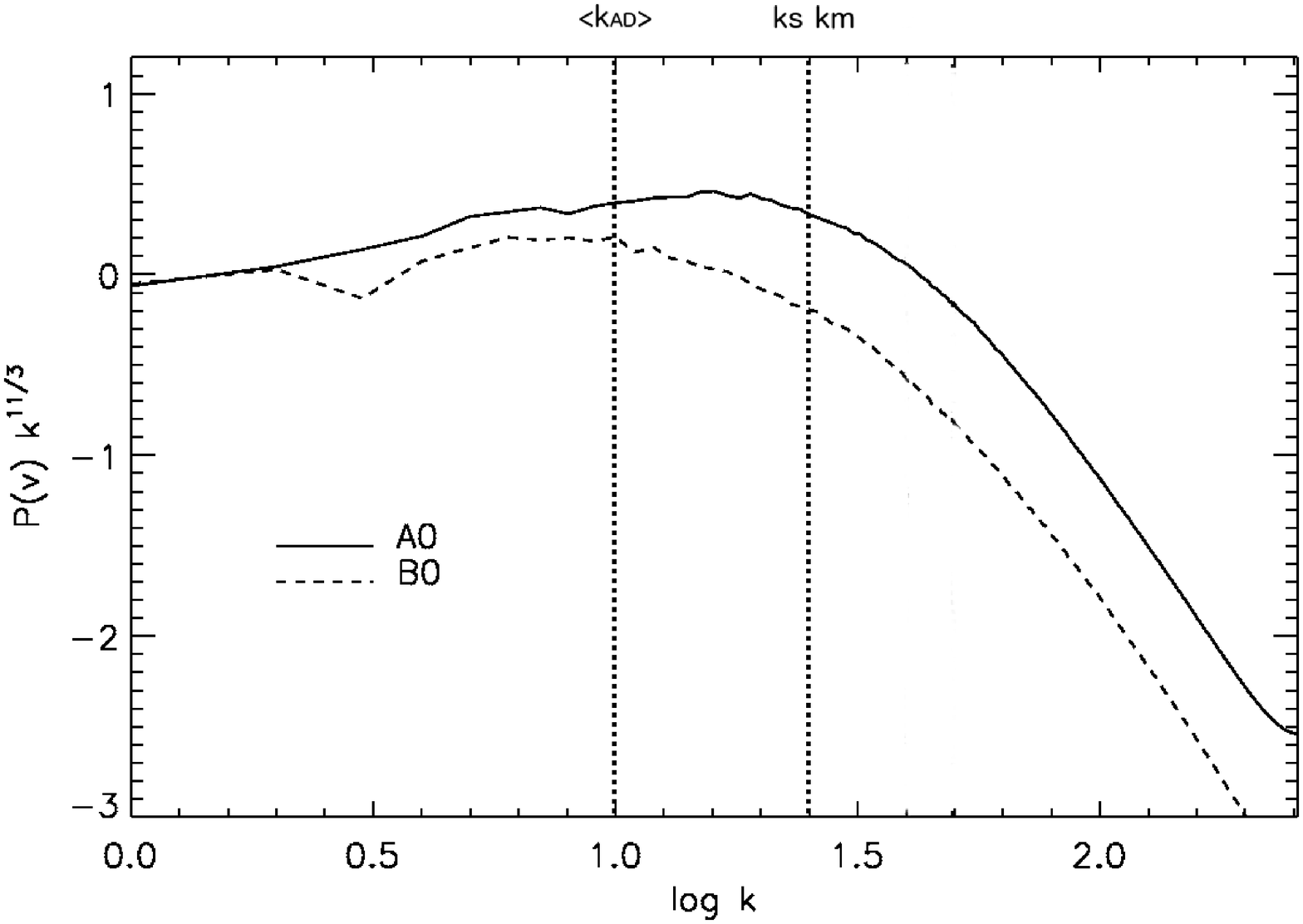}
   \includegraphics[width=\linewidth]{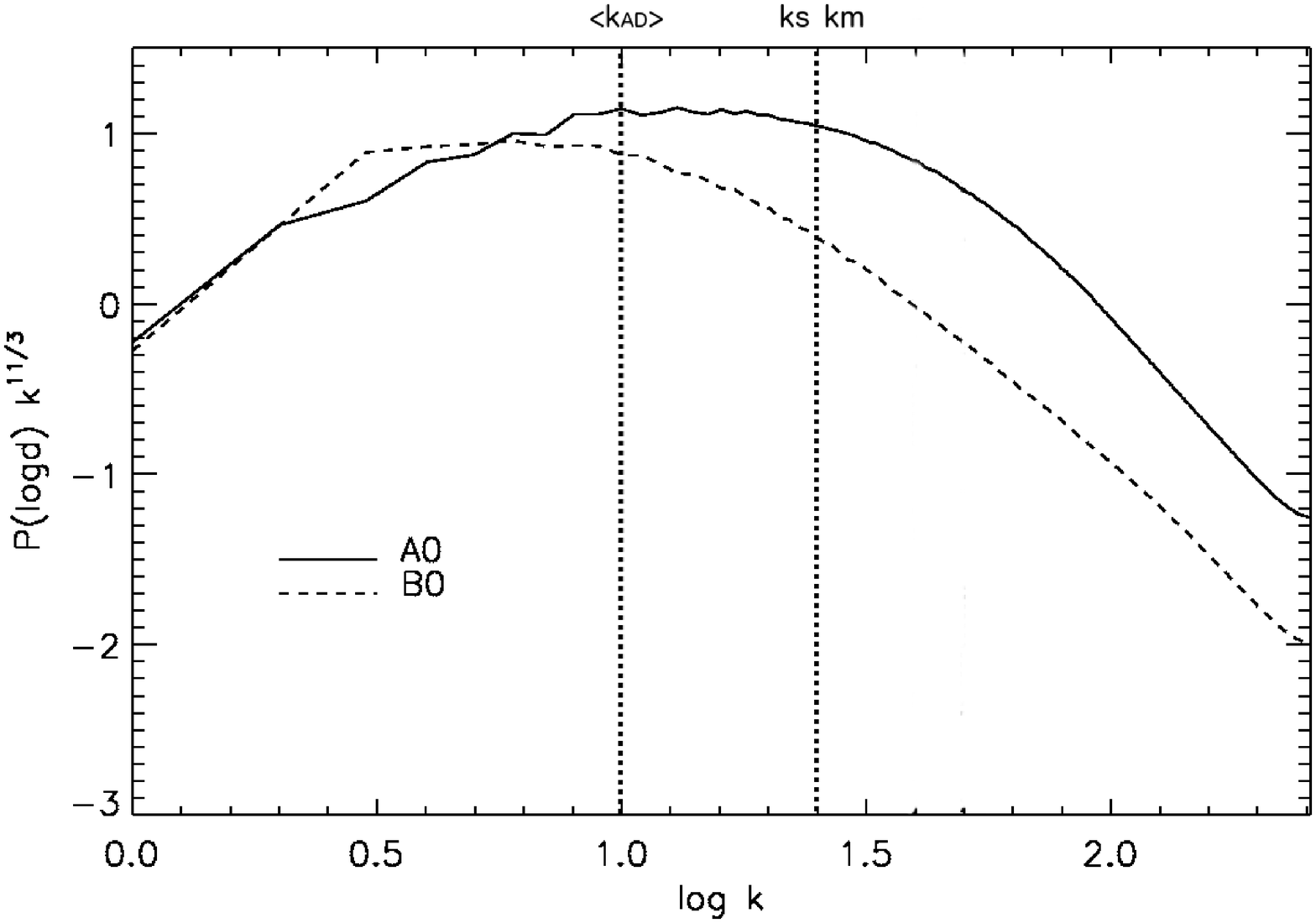}
   \caption{Compensated velocity (top) and log density (bottom) power spectra for runs A0 and B0.  
Solid lines correspond to run A0 and dashed lines to run B0. The vertical dotted lines show the ambipolar diffusion scale,
k$_{AD}$, calculated for the mean density in the simulation, n$_0$=500 cm$^{-3}$, the resolution scale, k$_m$, and the sonic scale, k$_s$.
The spectra are shown for a simulation time t=$10^{-2}$, which is 0.62 Myrs in physical units.
The turbulence crossing time is roughly 0.4 Myrs.}
   \label{pow_spec_a0_b0}
\end{figure}

\section{Code}
\label{numerics}

We use the publicly available MHD code RAMSES \citep{Teyssier_2002, Fromang_2006} to perform 
numerical simulations of driven or decaying MHD turbulence.  The RAMSES code solves the MHD equations on a Cartesian grid and has Adaptive Mesh Refinement (AMR) capabilities.

The standard version of the code does not include non-ideal MHD terms, so 
for the purposes of this study we have employed the magnetic diffusion module developed by \citet{Masson_2012}.
The details of the implementation are included in their paper, but we will outline the method here for completeness.


\subsection{Ambipolar diffusion implementation}

In order to simulate the effects of ambipolar diffusion, 
the motions of the ions and the neutrals need to be coupled through a friction force ${\bf{F_{fr}}}$ in the momentum
equations: 
\begin{equation}
\bf{F_{fr}} = \gamma_{AD}\rho_i\rho_n (\bf{v_i}-\bf{v_n})
\label{fric}
\end{equation}
where $\rho_i$ and $\rho_n$ are the mass densities of the ions and the neutrals, respectively, while $\bf{v_i}$ and $\bf{v_n}$ 
are the ion and neutral velocities. 

\begin{figure*}[!ht]
   \includegraphics[width=0.49\linewidth]{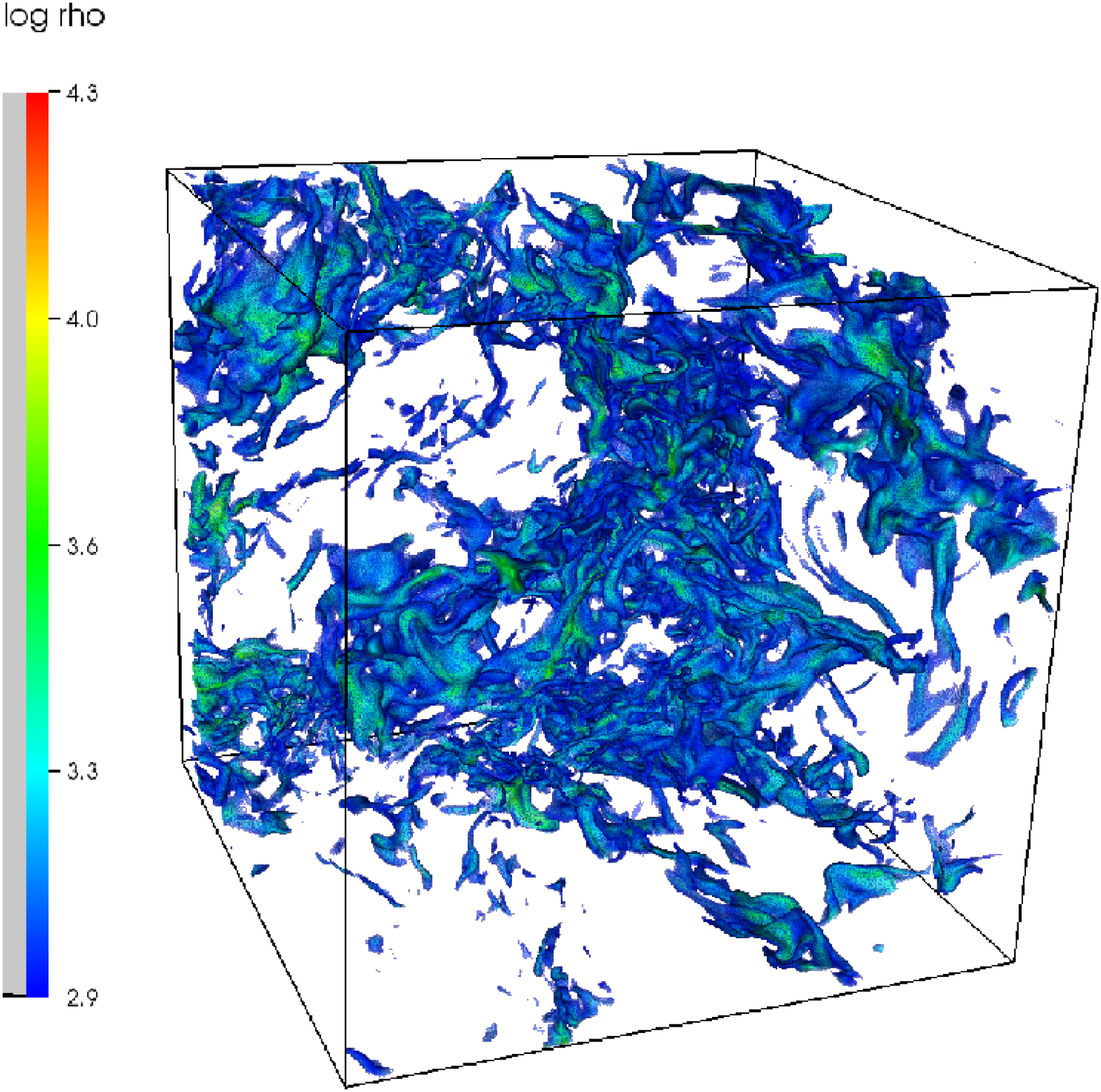}
   \includegraphics[width=0.49\linewidth]{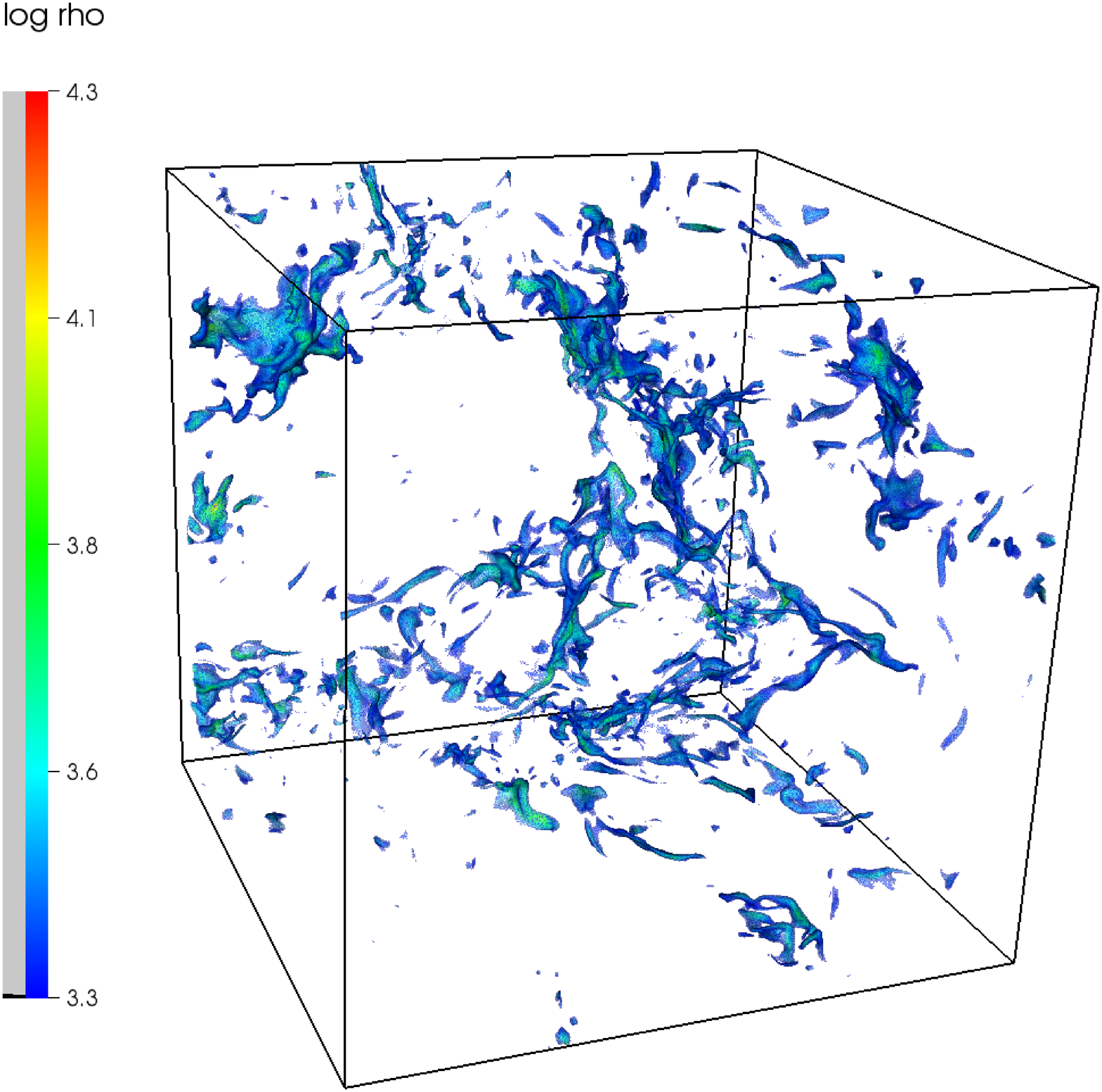}
   \includegraphics[width=0.49\linewidth]{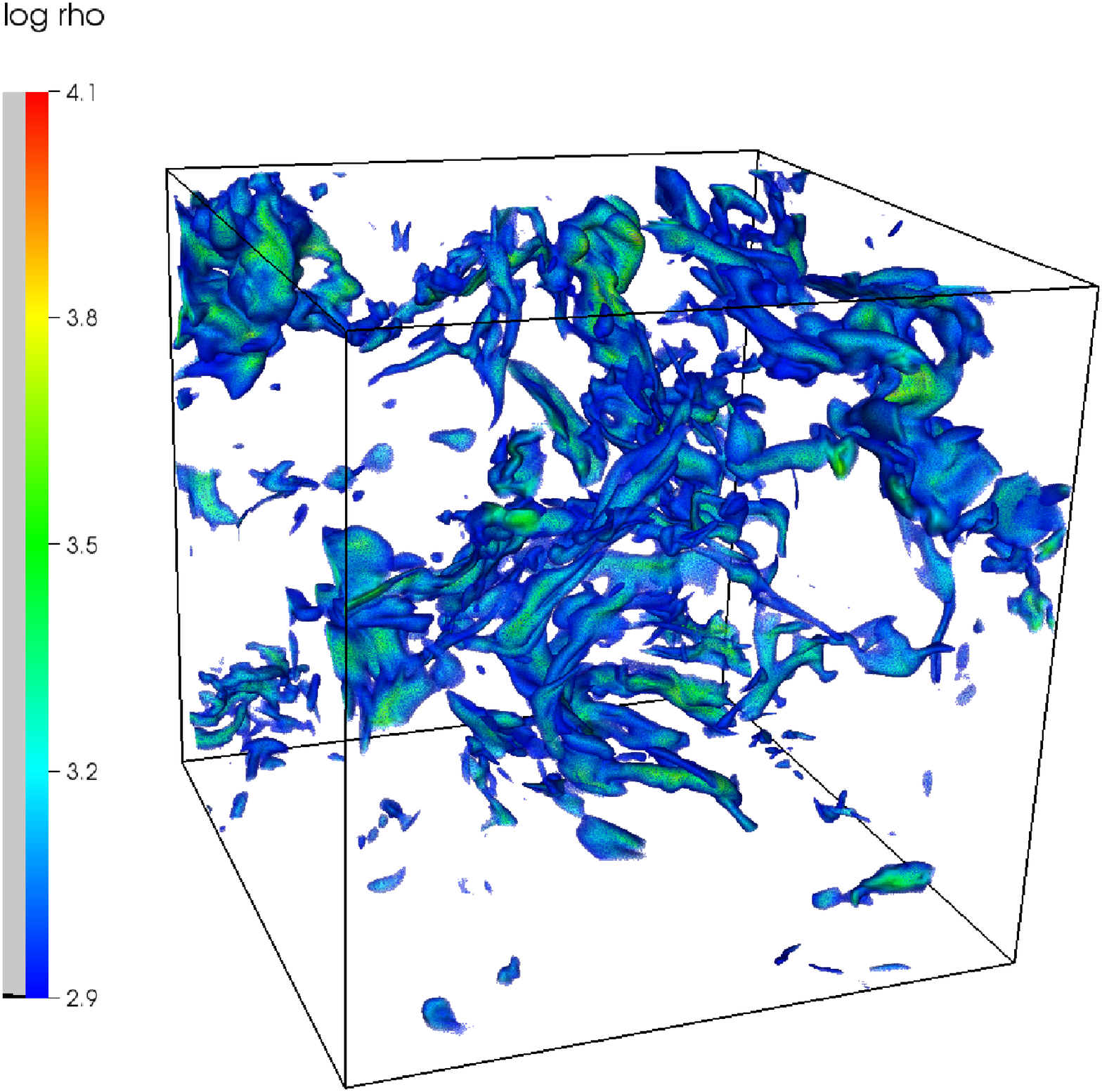}
   \includegraphics[width=0.49\linewidth]{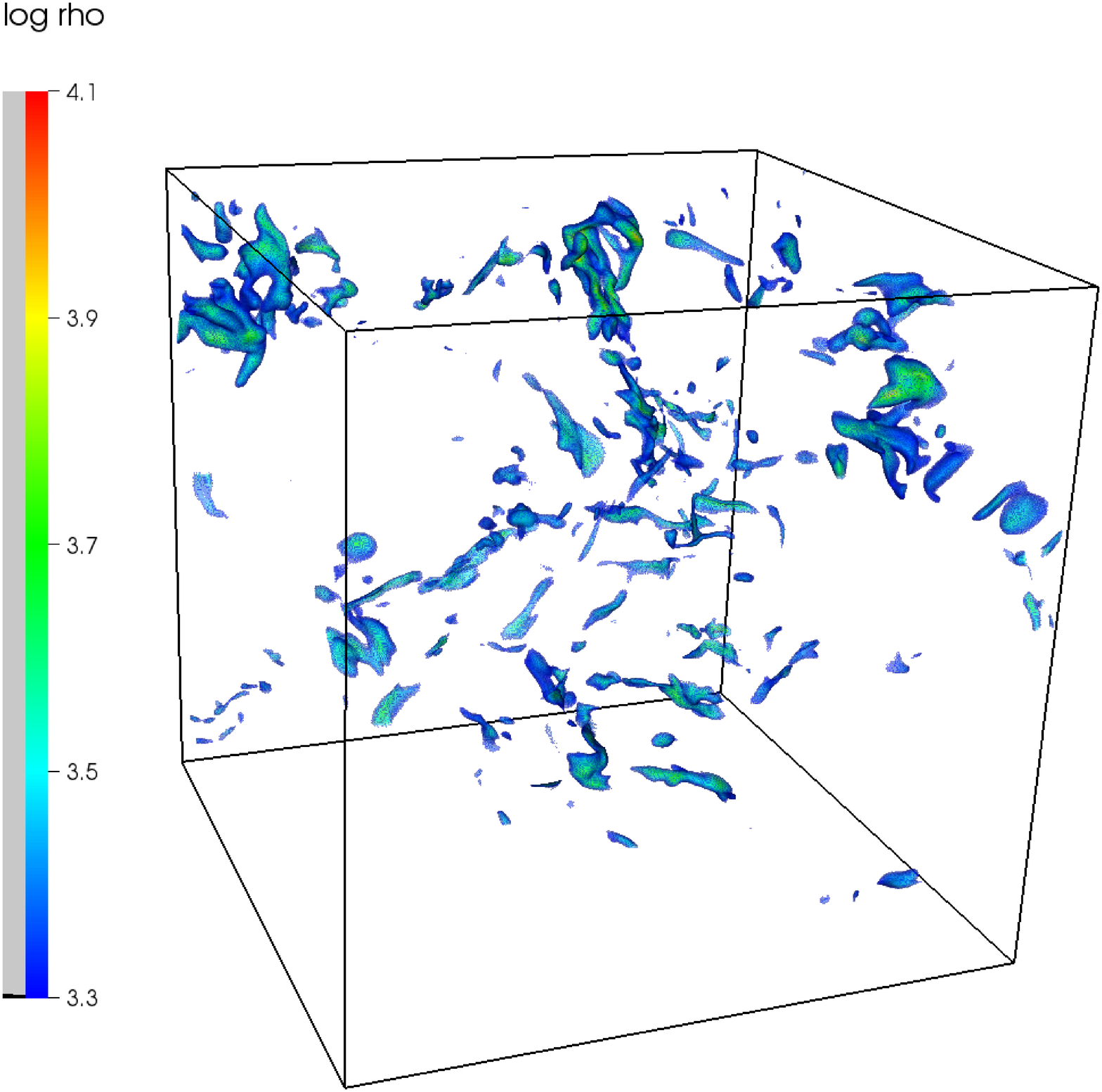}
    \caption{For the last snapshots of runs A0 (top) and B0 (bottom), at code time $t=5\cdot 10^{-3}$, 
volume plots of the logarithm of the mass density above a threshold of 2000~cm$^{-3}$ (left column) and 5000~cm$^{-3}$ (right column).}
 \label{volume_thresholds}
\end{figure*}

\begin{figure*}[!ht]
   \includegraphics[width=0.49\linewidth]{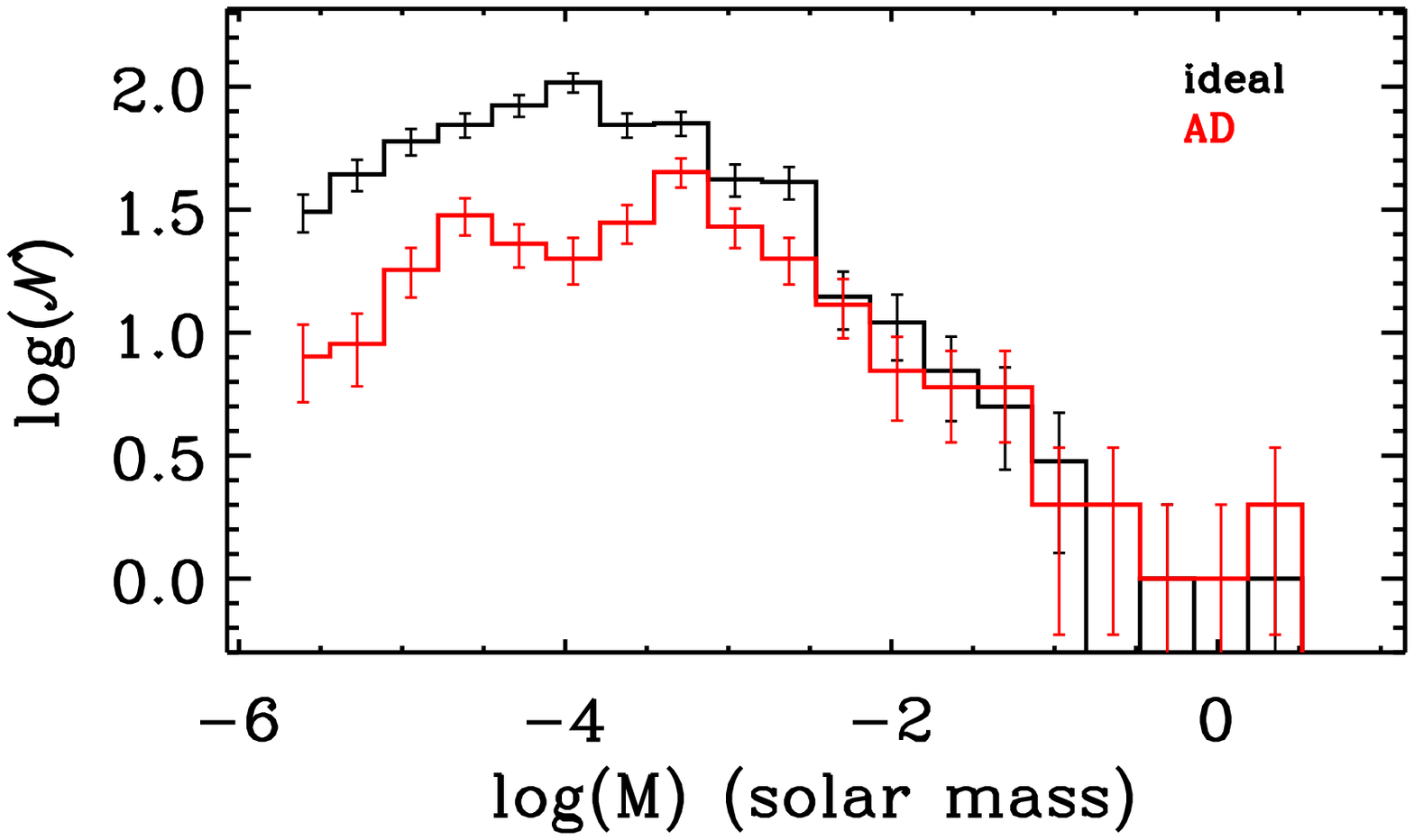}
   \includegraphics[width=0.49\linewidth]{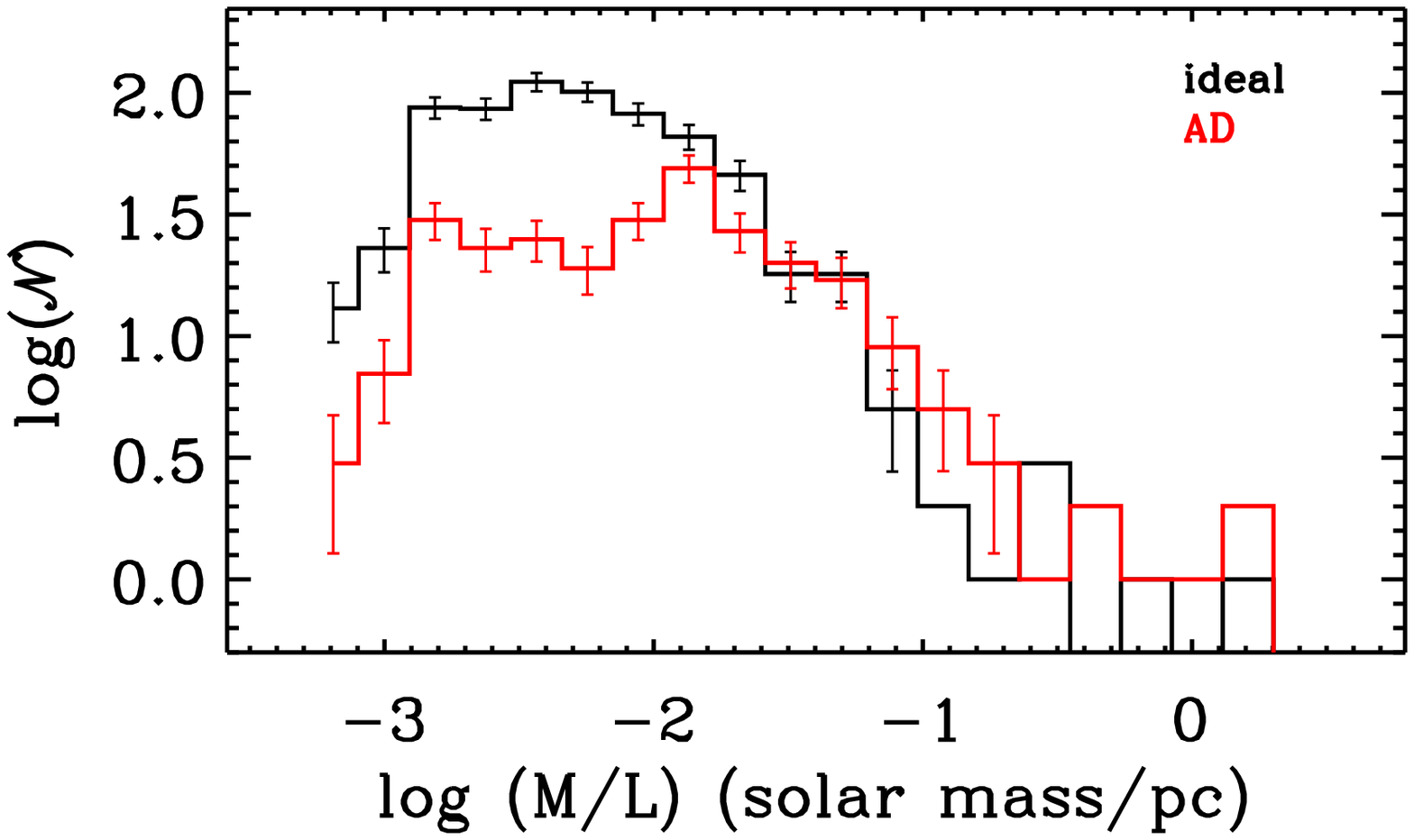}
   \includegraphics[width=0.49\linewidth]{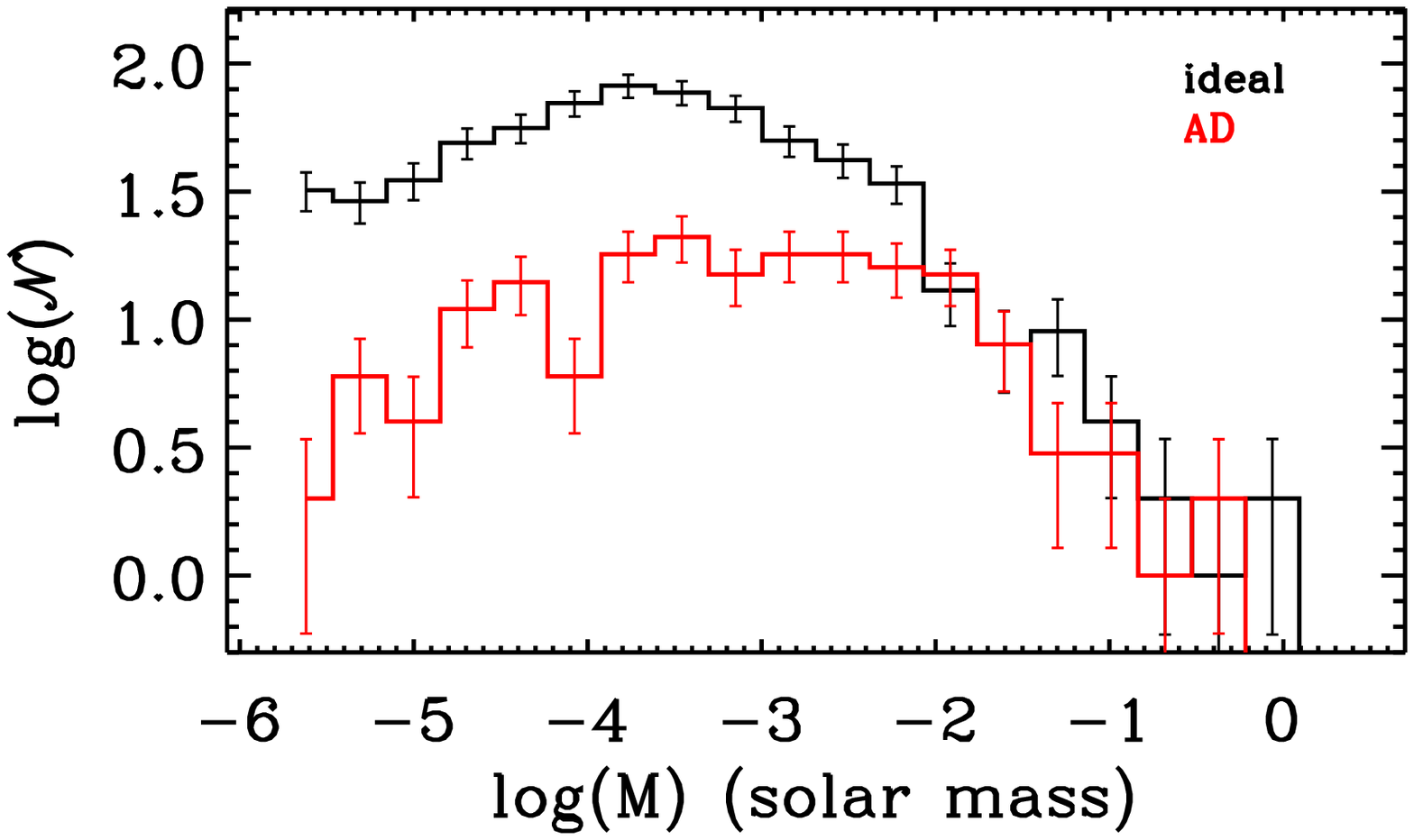}
   \includegraphics[width=0.49\linewidth]{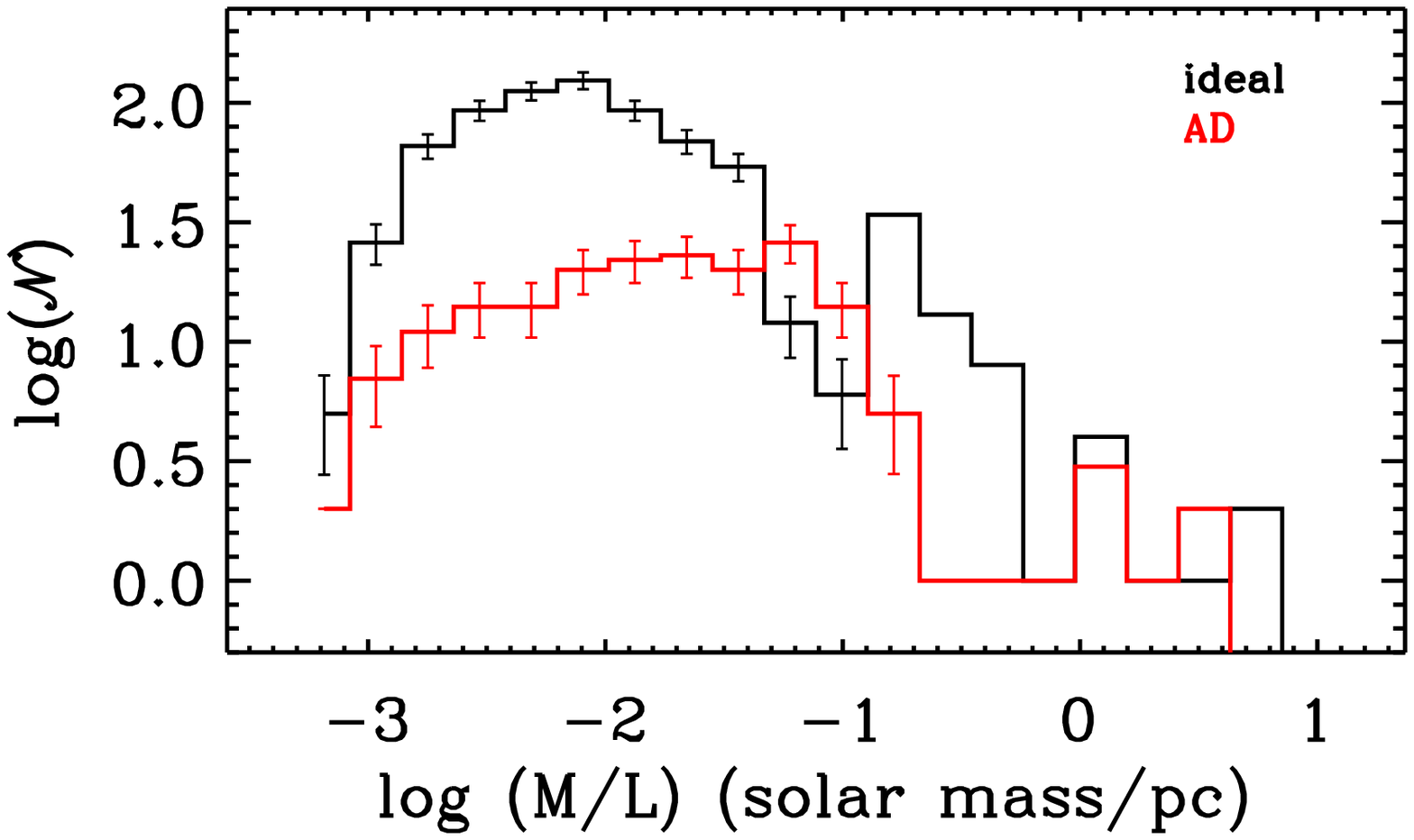}
    \caption{Filament mass (left) 
and mass per unit length (right) distributions. Black lines show run A0 and red lines run B0.  
  The top panel corresponds to $t=5\times 10^{-3}$ and the bottom to $t=10^{-2}$. 
  The density threshold for defining a filament is 2000~cm$^{-3}$}.
 \label{mass_histograms_1_2000}
\end{figure*}

There is more than one possible implementation of this basic principle.
One approach is to solve the two-fluid MHD equations and prevent
very small timesteps by assuming a low ionization fraction.  This implies
changing the coupling constant $\gamma_{AD}$ in Eq.~(\ref{fric}) accordingly, so this has been called 
the "heavy ion" approximation \citep{Li_McKee_klein_2006, Oishi_MacLow_2006,Li_McKee_2008}.
Another approach, used by \cite{Tilley_Balsara_2008} and \cite{Tilley_Balsara_2011},  is to implement
a semi-implicit scheme.  
This allows solving the two-fluid equations without having to modify the source terms through Eq.~(\ref{fric}).
 
In this work we are using the so-called "strong coupling" approximation:
In a weakly ionized plasma where the collision timescale between ions and neutrals is short compared to the timescales of the problem, 
the ion pressure and momentum are negligible compared to those of the neutrals.  
Then the friction force ${\bf{F_{fr}}}$ can be taken roughly equal to the Lorentz force and replaced  
in the equation of motion of the neutrals, so that one eventually solves one set of MHD equations. 
Since it requires low ionization conditions, this method is suitable 
for ambipolar diffusion studies in molecular clouds \citep{Shu_1987, Toth_1994, MacLow_1995, Choi_2009}.  

The equations in the strong coupling limit are:
\begin{eqnarray}
\frac{\partial\rho}{\partial t} +\nabla(\rho\bf v) = 0 \label{continuity} \\
\frac{\partial\bf{v}}{\partial t} + (\bf{v\cdot\nabla})\cdot\bf{v} + \frac{1}{\rho}\nabla P -\bf{F_{fr}} = 0 \label{momentum} \\
\frac{\partial E_{tot}}{\partial t} +\nabla(~(E_{tot}+P_{tot})\bf{v} -(\bf{v}\cdot\bf{B})\cdot\bf{B}-\bf{E_{AD}\times\bf{B}} = 0 \label{energy} \\
\frac{\partial\bf{B}}{\partial t} -\nabla\times(\bf{v}\times\bf{B}) - \nabla\times\bf{E_{AD}} = 0 \label{induction} \\
\nabla\cdot\bf{B} = 0 
\end{eqnarray}
where the additional friction force $\bf{F_{fr}}$ has been added to the momentum equation (Eq.(\ref{momentum})) 
and the electromotive force $\bf{E_{AD}}$ due to ambipolar diffusion is equal to
\begin{equation}
\bf{E_{AD}} = (\bf{v_i}-\bf{v_n})\times\bf{B}=\frac{1}{\gamma_{AD}\rho_i\rho}\bf{F_L}\times\bf{B}
\end{equation}
where $\bf{F_L}$ is the Lorenz force.

Like for the calculation of $\lambda_D$ in Eq. (\ref{lambda2}) above,
the density of the ions $\rho_i$ needed in Eq.~(\ref{fric}) to calculate the friction force between species
is calculated in the code as $\rho_i=C\sqrt{\rho_n}$, with $C=3\times10^{-16}$ cm$^{-3/2}$ g$^{1/2}$.
This approximation is only valid for densities higher than roughly $10^3$ cm$^{-3}$, when ionization
happens due to cosmic rays.  For lower densities the ionization fraction is approximately constant.  

However, in a turbulent environment, underdensities are created in several locations and there might be regions in the simulation domain where the ionization fraction is much higher and this leads to time steps which are pretty small making the code extremely slow. To tackle this issue the timestep has been limited to a minimum fraction of the Alfvén time by artificially modifying the ionisation degree in the regions where the timesteps is smaller than this threshold. A convergence test is presented in the Appendix.

We performed a convergence study for this timestep limit and concluded that
below a minimum dt=0.05t$_a$ (where t$_a$ the Alfv\'{e}n crossing time of a cell) the most sensitive results of 
the simulations (the power spectra) remained unchanged.
Although the mass distributions of the filaments and other statistical calculations presented in this work are not 
greatly affected by the choice of this limit, the simulations we use to draw our final conclusions were integrated with the 
converged timestep.
More details of the ambipolar diffusion module in RAMSES are contained in \cite{Masson_2012}.

\section{MHD simulations}
\label{sims}

In order to study the effects of ambipolar diffusion on the small-scale structure of turbulence we
compare between MHD turbulence simulations with and without the diffusive terms.
We simulate both driven and decaying turbulence.  
None of the simulations includes gravity.

All simulations are performed in three dimensions, 
in a periodic cubic box of 1 pc size.  The average density is 500 cm$^{-3}$ and the average temperature 10 K,
with an isothermal equation of state.  The magnetic field is initiated along the z direction with a magnitude of about $15~\mu$G,
which gives an initial plasma beta $\beta=0.1$.  Given these parameters, we ensure that the mean
$\lambda_d$ is always well resolved (see Eq.~(\ref{ldiss}) in the Introduction and the discussion later in this section).  

Table \ref{sim_table} summarizes the models included in this work. (AD stands for ambipolar diffusion.)

\begin{table}[!ht]
\caption{{\bf{Simulation parameters}} 
The last column shows in parentheses the 3D sonic rms Mach number at the beginning and at the end of the simulation.}
  \begin{tabular*}{\linewidth}{@{\extracolsep{\fill}}llllll}
    \hline
    \textbf{Name}  &  \textbf{MHD}  & \textbf{Turbulence} & \textbf{Resolution} & \textbf{Mach number} \\ 
    \hline
    \hline
     A0  & ideal   & decaying   & 512 & (10, 3)   \\
     B0  & AD      & decaying   &  512 & (10, 3) \\
     A1  & ideal   & driven   &  512 & (4, 4) \\ 
     B1  & AD      & driven   &  512 & (4 ,4) \\ 
     A0$_{l10} $ & ideal & decaying & 1024 & (10, 5) \\
     B0$_{l10}$  & AD & decaying & 1024 & (10, 5) \\
    \hline
  \end{tabular*}
\label{sim_table}
\end{table}

\begin{table*}[!ht]
\caption{{\bf{KS comparison.} }
The first column lists the compared distribution pairs, indicating the measured quantity (mass M or thickness r$_f$), 
the names of the simulations and the density threshold in cm$^{-3}$.  The second column lists the timestep 
in code units (see text) and the third lists the median values of the two distributions, ideal ($\mu_i$) and non-ideal ($\mu_{AD}$).  The last two columns contain the KS probability:
the fourth gives the result for the full samples and the last only for the high-value end ($10^{-1.7}$ pc for the thickness
and $10^{-3}$ M$_\odot$ for the mass). The same columns in parentheses include the probability in sigma levels.
We mention $<10^{-8}$ when the probability is essentially zero.}
  \begin{tabular*}{\linewidth}{@{\extracolsep{\fill}}llllll}
    \hline
    \textbf{Tested distributions} & \textbf{Time} & 
    \textbf{Median values} ($\mu_{i}$,$\mu_{AD}$) & \textbf{KS probability} & \textbf{KS probability, high values} \\
    \hline
    \hline 
     log(M), A0 vs B0, 2000   &  $5\times 10^{-3}$      &  (-3.44,-3.03)    & 1.37$\times 10^{-6}$ (4.8 $\sigma$) & 0.002 (2.76 $\sigma$)  \\
     log(M), A0 vs B0, 2000   &  $10^{-2}$                    &  (-3.43,-2.77)    & 9.2$\times10^{-6}$ (4.46 $\sigma$)                 & 0.005 (2.1 $\sigma$) \\
     log(M), A0 vs B0, 5000   & $5\times 10^{-3}$           & (-3.03,-2.68)     & 4.48$\times 10^{-5}$ (4.1 $\sigma$)    & 0.11 (1.5 $\sigma$) \\
     log(M), A0 vs B0, 5000   &  $10^{-2}$                       &  (-3.21,-2.67)    & 0.12 (1.2 $\sigma$)                             & 0.04 (2.1 $\sigma$) \\
     log(r$_f$), A0 vs B0, 2000   &  $5\times 10^{-3}$    & (0.01,0.014)      & $<10^{-8}$  (5.7 $\sigma$)          & 0.1 (1.7 $\sigma$) \\
     log(r$_f$) A0 vs B0, 2000   &  $10^{-2}$                 &  (0.014,0.019)   & $<10^{-8}$ (5.9 $\sigma$)  & 0.0004 (3.4 $\sigma$) \\
     log(r$_f$), A0 vs B0, 5000   & $5\times 10^{-3}$    & (0.01,0.013)        & 2.92$\times 10^{-7}$ (5.12 $\sigma$)   & 0.14 (1.45 $\sigma$) \\
     log(r$_f$), A0 vs B0, 5000   &  $10^{-2}$                &  (0.0112,0.0116) & 0.11  (1.56 $\sigma$)                         & 0.9 (0.05 $\sigma$)  \\
     log(mass), A0$_{l10}$ vs B0$_{l10}$, 2000   &  $5\times 10^{-3}$   & (-4.17, -3.73) & $<10^{-8}$($>7 \sigma$)   &  $<10^{-8}$ (2.9 $\sigma$)\\
     log(mass), A0$_{l10}$ vs B0$_{l10}$, 5000   & $5\times 10^{-3}$   &   (-3.85,-3.5) & $<10^{-8}$ ($>7 \sigma$)    & 8$\times 10{-7}$ (2.6 $\sigma$) \\
     log(r$_f$), A0$_{l10}$ vs B0$_{l10}$, 2000   &  $5\times 10^{-3}$   & (0.006,0.008) & $<10^{-8}$ ($>7 \sigma$) & 0.4 (0.87 $\sigma$) \\
     log(r$_f$), A0$_{l10}$ vs B0$_{l10}$, 5000   & $5\times 10^{-3}$   & (0.005,0.007) & $<10^{-8}$ ($>7 \sigma$) & 0.38 (2 $\sigma$) \\
     log(M), A1 vs B1, 2000   &  $5\times 10^{-3}$       &  (-3.29, -2.92)     & 0.001 (3.19 $\sigma$)  & 0.04 (2 $\sigma$)  \\
     log(M), A1 vs B1, 2000   &  $10^{-2}$                    & (-3.17, -2.86)      & 0.002 (3.1 $\sigma$)      & 0.33 (1 $\sigma$)  \\
     log(M), A1 vs B1, 5000   & $5\times 10^{-3}$         & (-3.29, -2.86)      & 0.0014 (3.1 $\sigma$)    & 0.9 (0.1 $\sigma$) \\ 
     log(M), A1 vs B1, 5000   &  $10^{-2}$                     & (-2.92, -2.72)       & 0.13  (1.5 $\sigma$)     & 0.12  (1.54 $\sigma$)  \\
     log(r$_f$), A1 vs B1, 2000   &  $5\times 10^{-3}$   &  (0.011,0.015)      & 0.004 (2.9 $\sigma$)       & 0.61 (0.5 $\sigma$)\\
     log(r$_f$), A1 vs B1, 2000   &  $10^{-2}$                &    (0.013,0.016)    & 0.003 (3 $\sigma$)     & 0.2 (1.3 $\sigma$) \\
     log(r$_f$), A1 vs B1, 5000   & $5\times 10^{-3}$   & (0.01,0.012)           & 0.01 (2.56 $\sigma$)     & 0.6 (0.6 $\sigma$)\\
     log(r$_f$), A1 vs B1, 5000   &  $10^{-2}$               & (0.011,0.012)         & 0.21 (1.24 $\sigma$)     &  0.01 (2.54 $\sigma$)\\
    \hline
  \end{tabular*}
\label{stat_table}
\end{table*}

In the main body of the paper we will present the decaying turbulence results, using runs A0 and B0 as our main reference.
The case for driven turbulence is presented in Appendix \ref{app:driven}.

The turbulent field is created by following the classic recipe from \cite{MacLow_1998}, 
where random phases are given to wave numbers $k=1-4$ in Fourier space and amplitudes are chosen to give the desired rms Mach number.  
We allow the code to iterate for a while until the turbulence has had time to isotropise in the initially uniform density field.
All runs are initiated with the same velocity field pattern, multiplied by the appropriate factor to set the desired Mach number or account for energy losses.

The reference runs have an initial Mach number of about 10, decaying to about 3 at the end of the simulation.  
The driven turbulence runs are kept 
at a Mach number of 4 by  by calculating the difference in
kinetic energy with respect to the initial conditions and then adding a static velocity field, multiplied by the appropriate coefficient to account for the energy losses.\

\subsection{Resolving the physical dissipation length}

The challenge we are faced with is to resolve the physical dissipation length
while still keeping the size of the problem manageable in terms of computational resources.

The first considerations are to minimize the numerical diffusion
and to resolve the inertial range adequately for statistical studies.
No matter how great the resolution, 
in all numerical simulations the smallest scales are affected by numerical viscosity, 
so one must focus the analysis only on well-resolved structures.  
The most conservative demand for accurate turbulence studies has been suggested by 
 \citet{Federrath_2011}, who argue that only scales containing more than 30 cells should be considered resolved. 
Ideally, we would want all scales down to the ambipolar diffusion length to
be resolved with at least 20 cells.  However, as we will show, this is not always possible. 
For reference we have indicated the resolution limit $l_m$ (or the associated wavenumber, k$_m$) as the length of 20 cells in the power spectra and the
thickness histograms throughout the paper.

Another relevant scale in molecular cloud studies is the sonic length, $l_s$, defined as the length scale
at which the rms sonic Mach number of turbulence equals unity \citep{Padoan_1995,McKee_2010}.  
Assuming that the one-dimensional velocity dispersion scales with length like $\sigma_v\propto l^\alpha$, then the
sonic Mach number can be written like:
\begin{equation}
\mathcal{M} = \frac{\sigma_0}{c_s} = \left(\frac{l_0}{l_s}\right)^{\alpha} 
\end{equation} 
where $\sigma_0$ the velocity dispersion at the integral scale, $l_0$.
We can assume $\sigma_0$ corresponds to the mean rms sonic Mach number of the turbulence.
Then
\begin{equation}
l_s = \frac{l_0}{\mathcal{M}^{1/ \alpha}}
\label{ls}
\end{equation}
For our setup, $l_0$=1 pc and, given the Kolmogorov-like behavior 
of the power spectra (see following sections), $\alpha\simeq 1/3$. 
Then $l_s\simeq$ 0.04 pc for a global rms Mach number of 3, at $l_s\simeq$ 0.015 pc for a global
rms Mach number of 4 and at $l_s\simeq$ 0.008 pc for a global rms Mach number of 5.

\begin{figure*}[!ht]
   \includegraphics[width=0.49\linewidth]{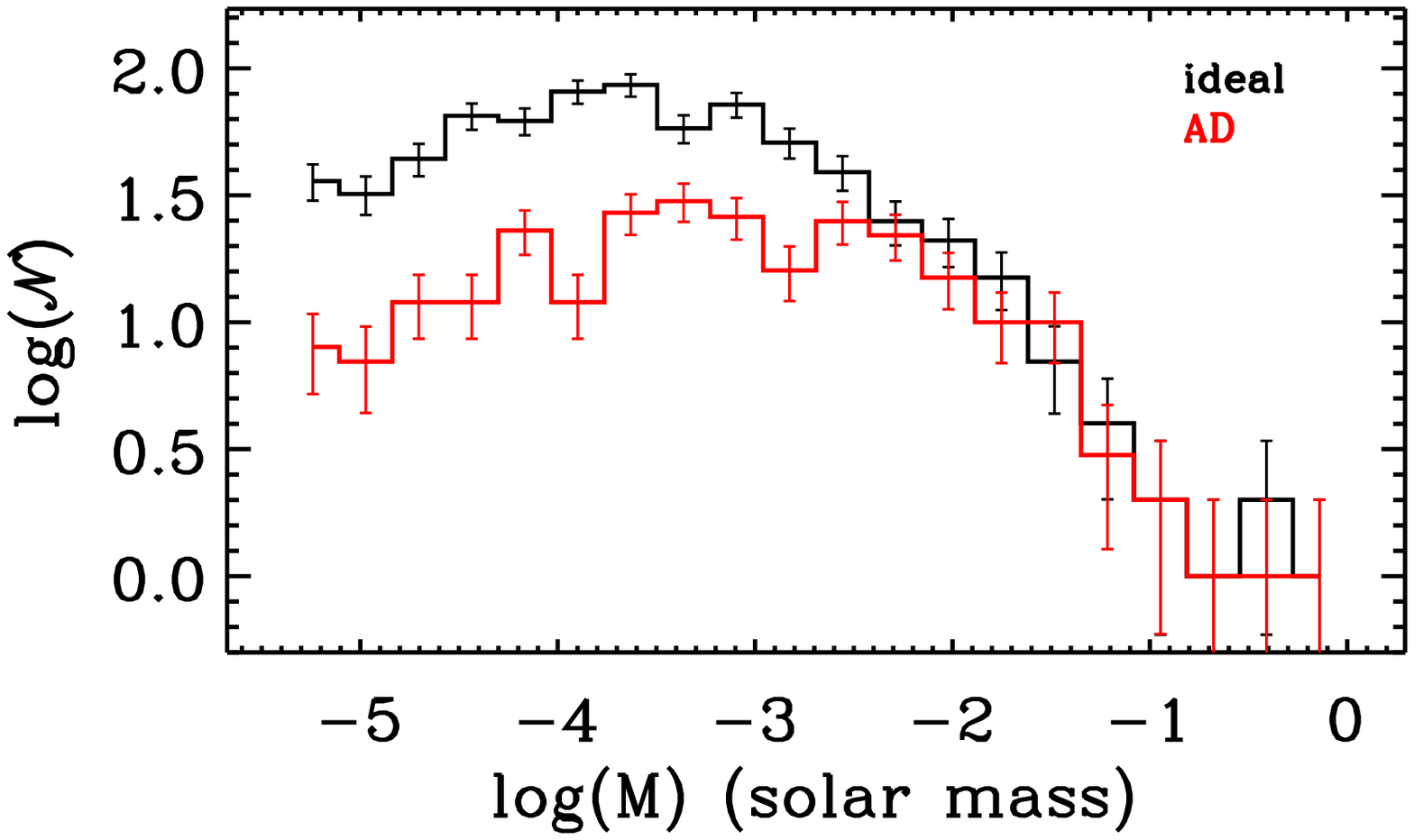}
   \includegraphics[width=0.49\linewidth]{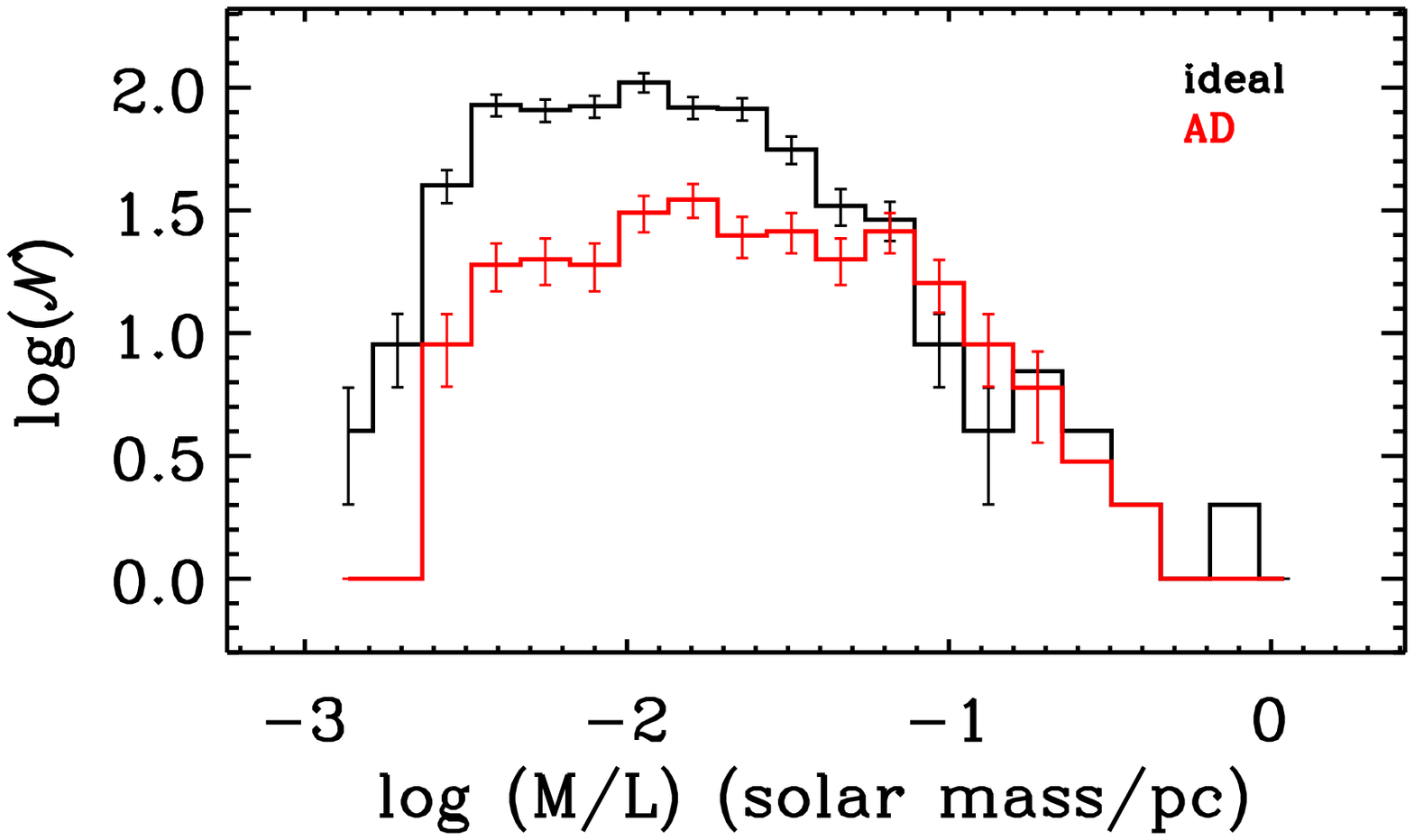}
   \includegraphics[width=0.49\linewidth]{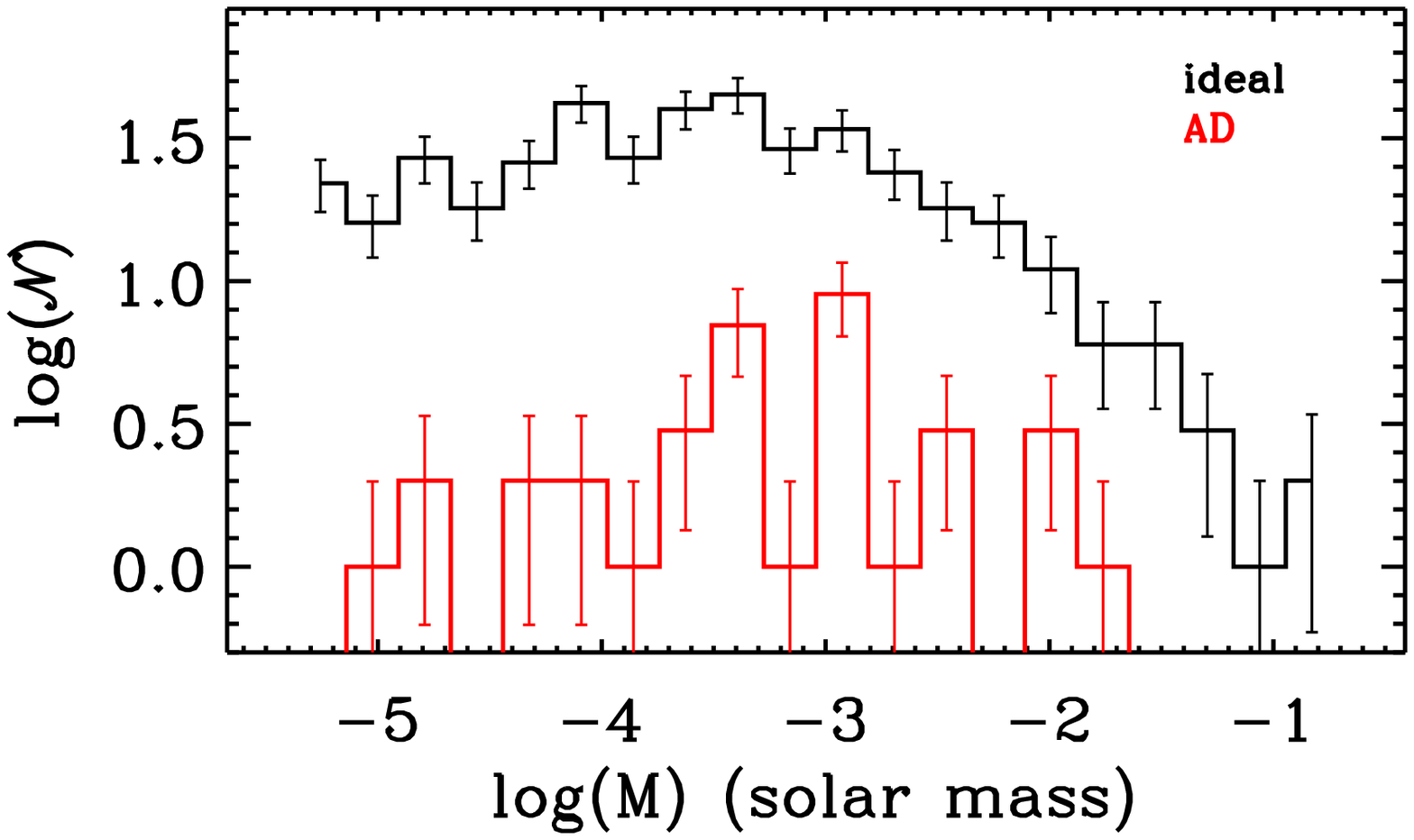}
   \includegraphics[width=0.49\linewidth]{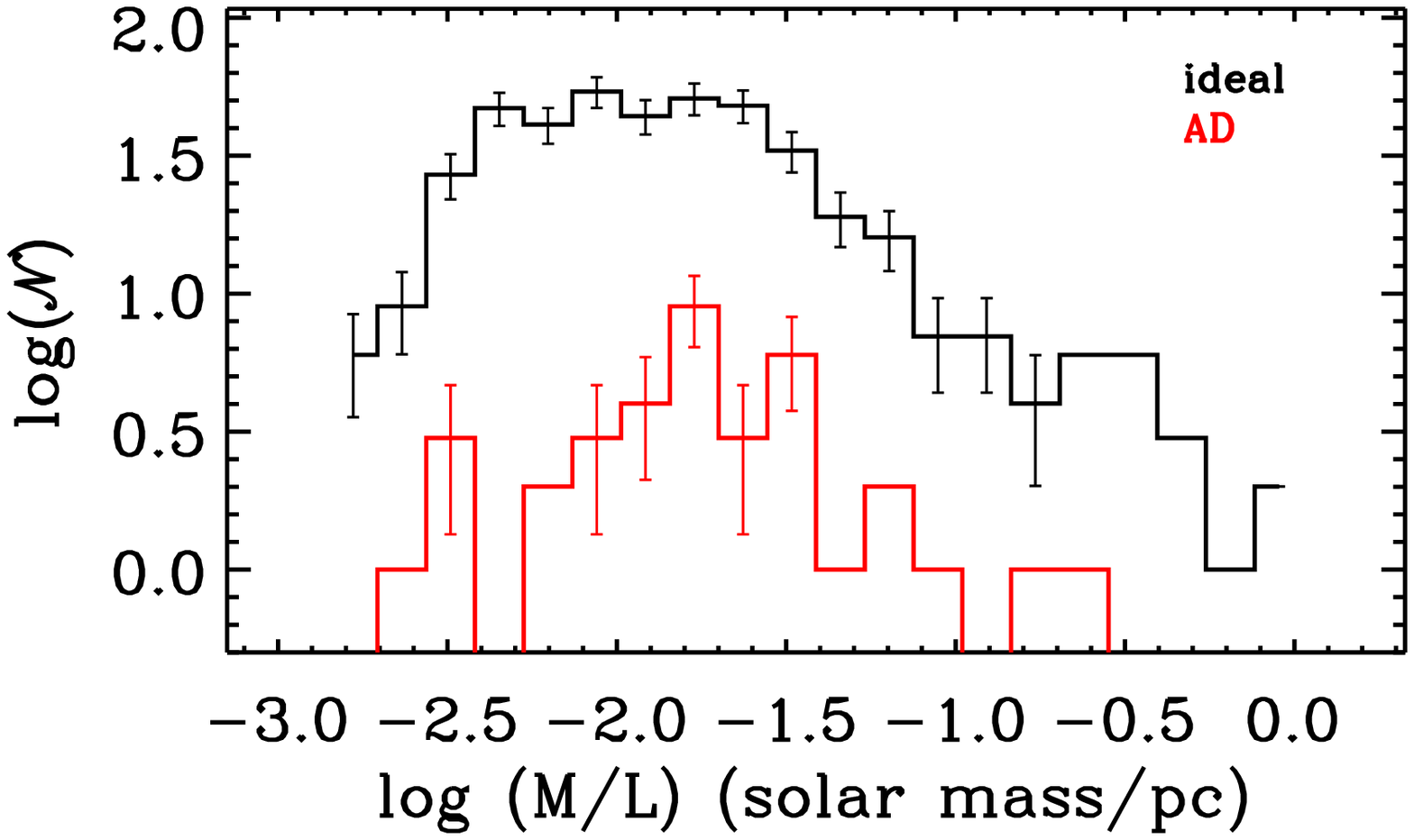}
  \caption{Same as Figure \ref{mass_histograms_1_2000}, for a density threshold of 5000~cm$^{-3}$.}
 \label{mass_histograms_1_5000}
\end{figure*}

An estimate of the mean dissipation length due to ambipolar diffusion in our setup comes from Eq.~(\ref{ldiss}).
Using the mean initial values of the experiments, $\rho_0=2\times10^{-21}$~gr~cm$^{-3}$,
 $B=15~\mu$G, and $\rho_i\simeq1.3\times10^{-26}$~gr~cm$^{-3}$, we calculate a mean dissipation length of about $\lambda_{d}\simeq0.1$~pc.

A resolution of 512$^3$ gives a physical grid size of $2\times10^{-3}$ pc.
Given the values we calculated above, the mean dissipation length is resolved with about 
50 cells. Nonetheless, according to Eq.~(\ref{ldiss}), the dissipation length is roughly inversely proportional to the density.  
This means that at small scales, where the gas is compressed, the dissipation length becomes even smaller.  
Even if we assume that the magnetic field strength increases in the high-density areas as the square root of the density \citep{Crutcher_2012}, which is not strictly true everywhere, a small dependence on the density is left in the expression for the dissipation length.

To study this effect we repeated runs A0 and B0 at a resolution of $1024^3$, although 
for a shorter time (Runs $A0_{l10}$ and $B0_{l10}$ in Table \ref{sim_table}).
The results of that comparison are presented in Section \ref{sec:resolution}.

\subsection{Evolution of the turbulent field}

We initialize the simulations with a sonic rms Mach number equal to ${\cal{M}}=10$ and an Alfvenic Mach number of
about ${\cal{M}_A}=2$. 

Figure~\ref{turb_a0_b0} shows column density contours of the simulation domain on the yz plane, with  
black arrows indicating the projected magnetic field. The arrow sizes are scaled to the maximum magnitude of the field in the plane.
The left panel corresponds to run A0 and the right panel to run B0.  
Two snapshots are represented in this figure: one at time $5\times 10^{-3}$ (in code units) at the top, 
and another at time $10^{-2}$ at the bottom.  
The code time unit corresponds to 62~Myrs in physical units, so these correspond to physical times of about 0.3 and 0.6~Myrs. 
The crossing time of the turbulence at Mach 10 is 0.4~Myrs, so the second snapshot is at just over one turbulence crossing time.

It is clear that ambipolar diffusion alters the morphology of the density field substantially:
Small-scale structure is smoothed out and dense features appear much broader with respect to the ideal MHD run.  
As the simulation evolves, this effect becomes clearer.  
The decay is much faster in the non-ideal MHD run and it eliminates small-scale features much more efficiently.

The projected magnetic field topology also changes from one run to another.  
While at early times A0 and B0 show similar magnetic field configurations,
at later times the magnetic field in B0 exhibits a very different behavior.  
Not only is its magnitude smaller, but at many locations the field
lines also have different orientations.  
This happens due to the loss of both kinetic and magnetic energy through diffusion.

In order to study the physical processes that cause the broadening of the small-scale structure we can look at the power spectra.
Power spectra of a given quantity $v$ are defined as the mean power of that quantity at each wavenumber k in Fourier space, 
and are calculated as $P_v(k) = \left|\tilde{v_k}\right|^2$.  
Here $\tilde{v_k}$ is the Fourier transform of the quantity $v$ at wave number k.
For incompressible MHD turbulence, we expect the power spectrum of the total velocity  
to have a $k^{-11/3}$ dependence on the wave number.

When a physical dissipation process such as ambipolar diffusion is modeled in a turbulence simulation, 
the energy cascade loses power below its scale of influence, which does not depend on resolution.
In simulations of ideal MHD turbulence this behavior is mimicked by numerical dissipation, 
which acts at the scale of a few grid cells, and therefore depends on resolution.  

Figure \ref{pow_spec_a0_b0} shows the power spectra of the velocity field and the logarithm of the density field, 
for runs A0 and B0, both divided by the expected $k^{-11/3}$ dependence. 
This Figure corresponds to the second snapshot (bottom row of Figure \ref{turb_a0_b0}), at simulation time $t=10^{-2}$.
The same spectra for this simulation at time  $t=5\cdot10^{-3}$ are included in Figure \ref{spectra_comparison_2} of Section \ref{sec:resolution}, as well
as in Appendix \ref{app:timestep}.
The solid lines correspond to run A0 and the dashed lines to run B0.  


\begin{figure}[hbt]
   \includegraphics[width=\linewidth]{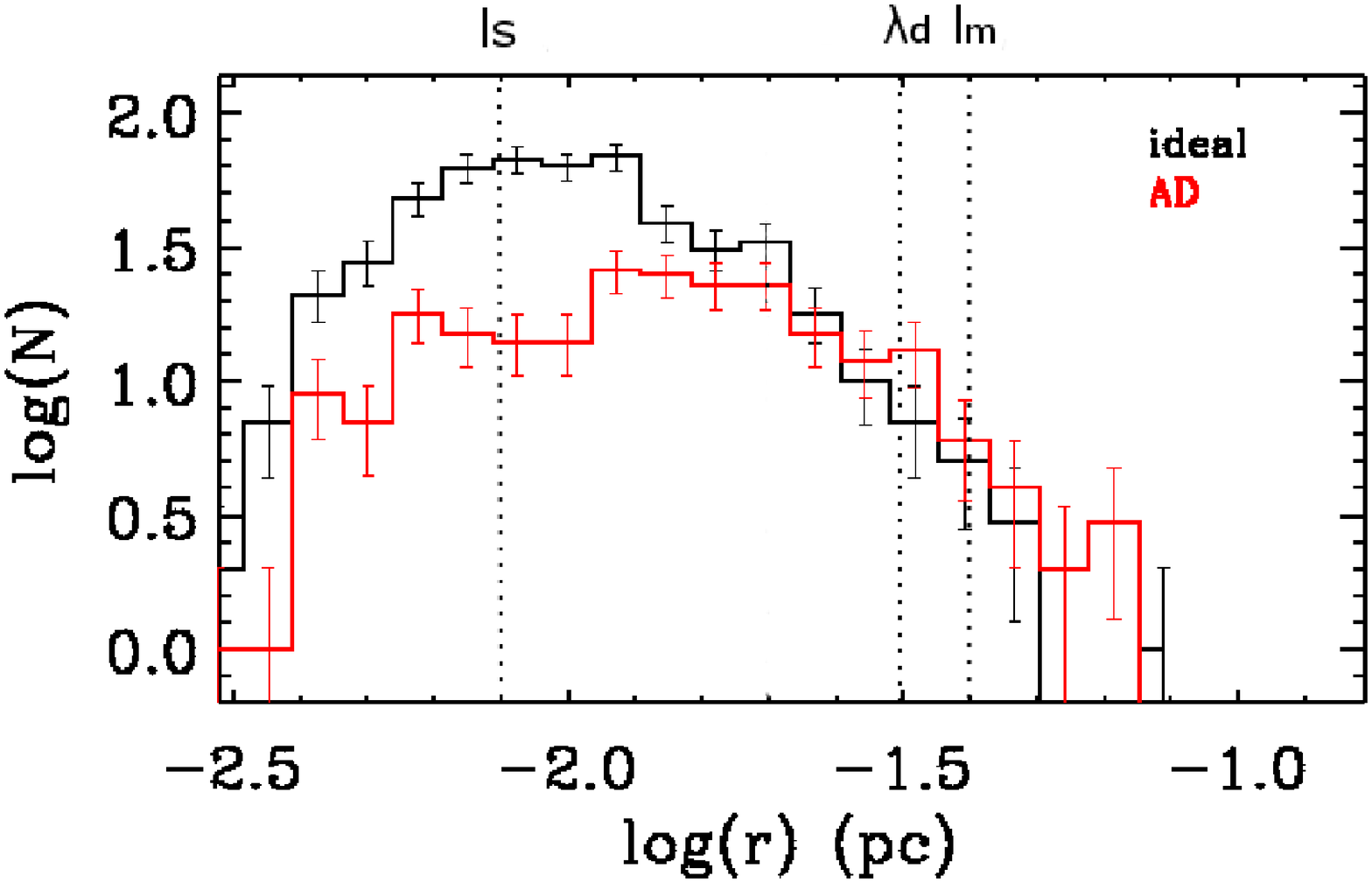}
   \includegraphics[width=\linewidth]{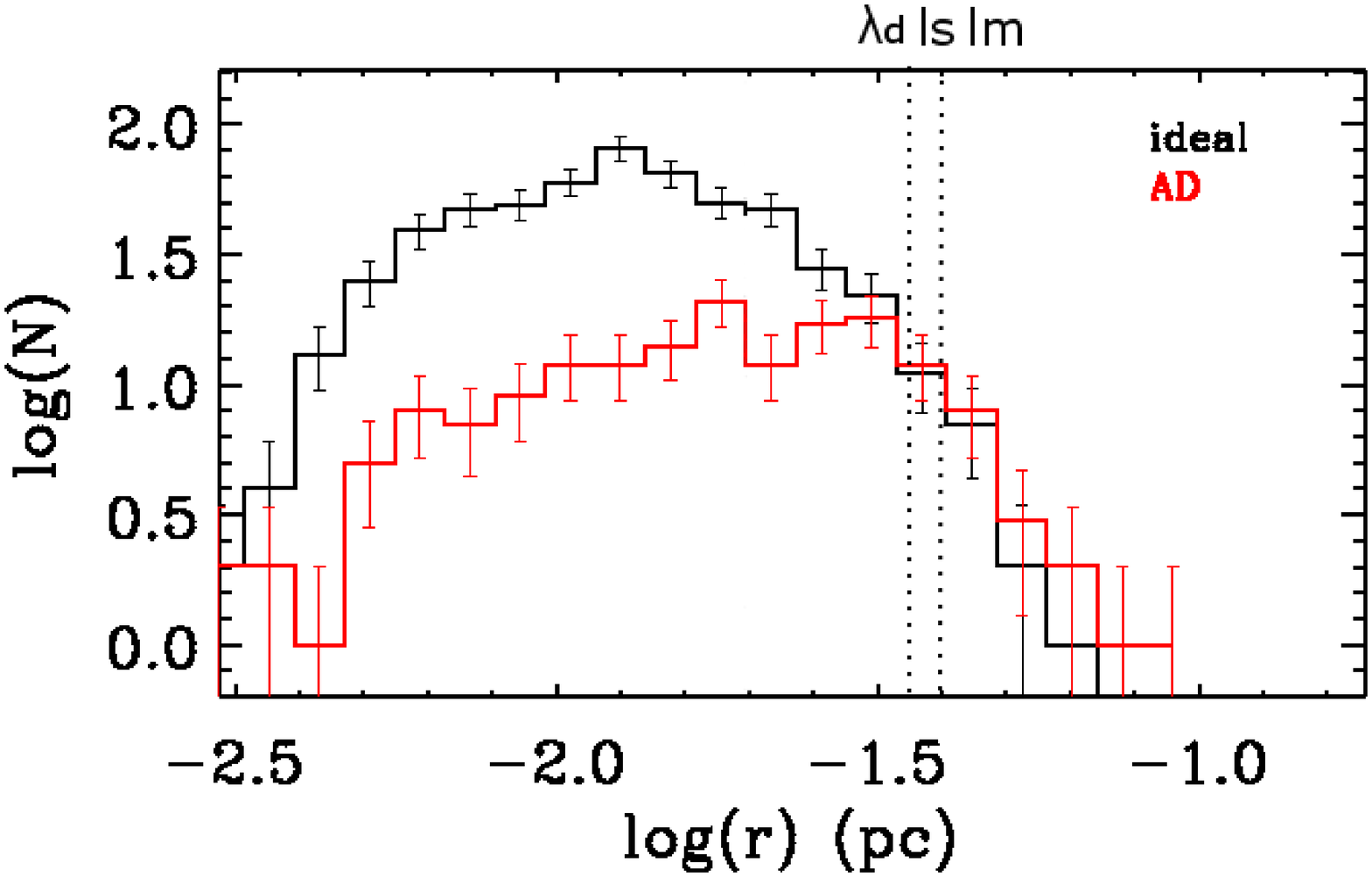}
  \caption{Filament thickness distributions in runs A0 (black lines) and B0 (red lines) at times $t=5\times10^{-3}$ (top) and  $t=10^{-2}$ (bottom).  The density threshold for defining a filament is 2000~cm$^{-3}$.  The dotted lines indicate the mean ambipolar diffusion dissipation length $\lambda_d$, calculated from Eq. (\ref{ldiss}) for this density threshold, the resolution length $l_m$, equal to the length of 20 cells, and the sonic length $l_s$, defined by Eq. (\ref{ls}), for an rms Mach number of 3 (at $t=10^{-2}$, bottom) and 5 (at $t=5\times10^{-3}$, top).}
   \label{logthick_histograms_1_2000}
\end{figure}
%
\begin{figure}[!h]
   \includegraphics[width=\linewidth]{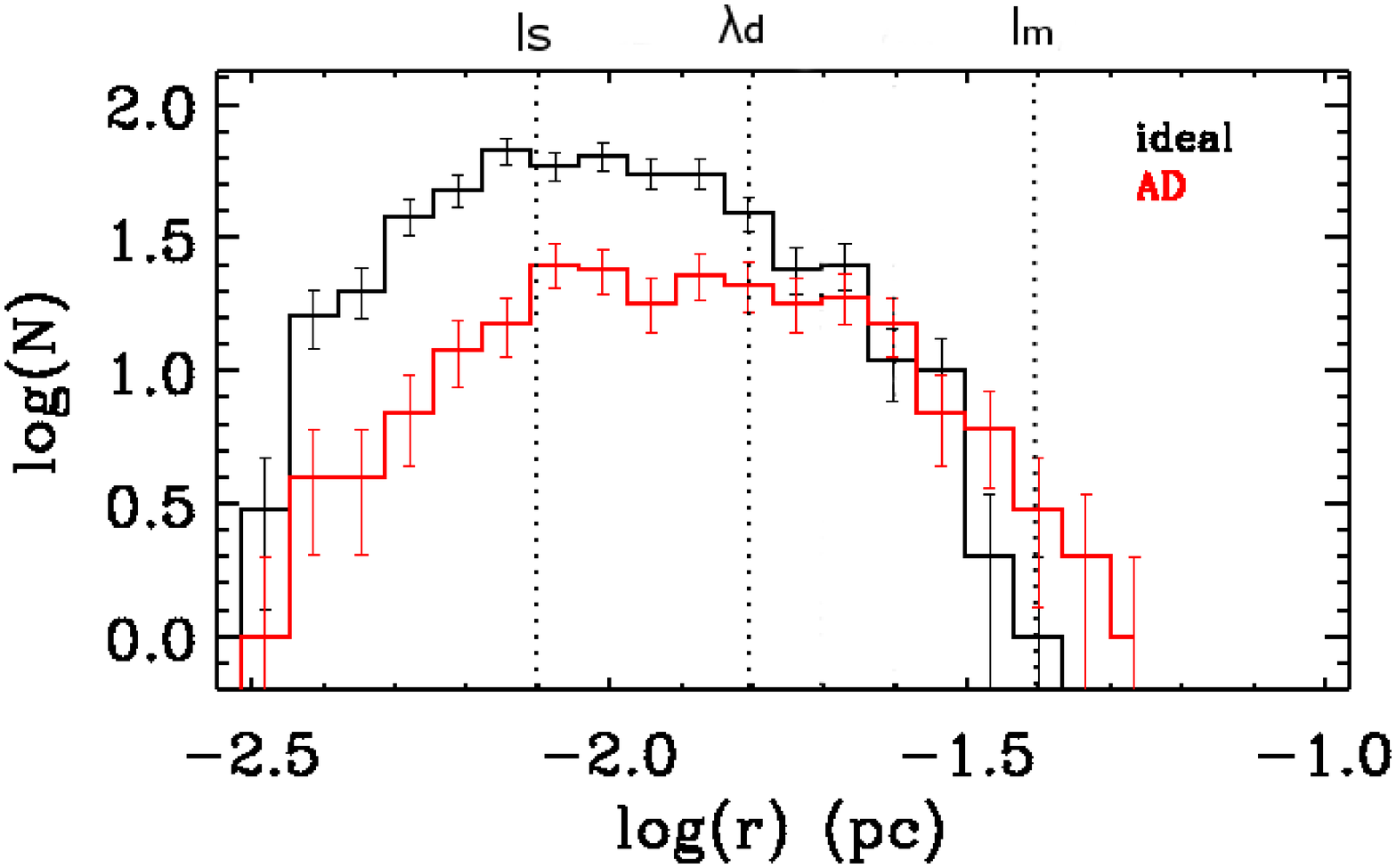}
   \includegraphics[width=\linewidth]{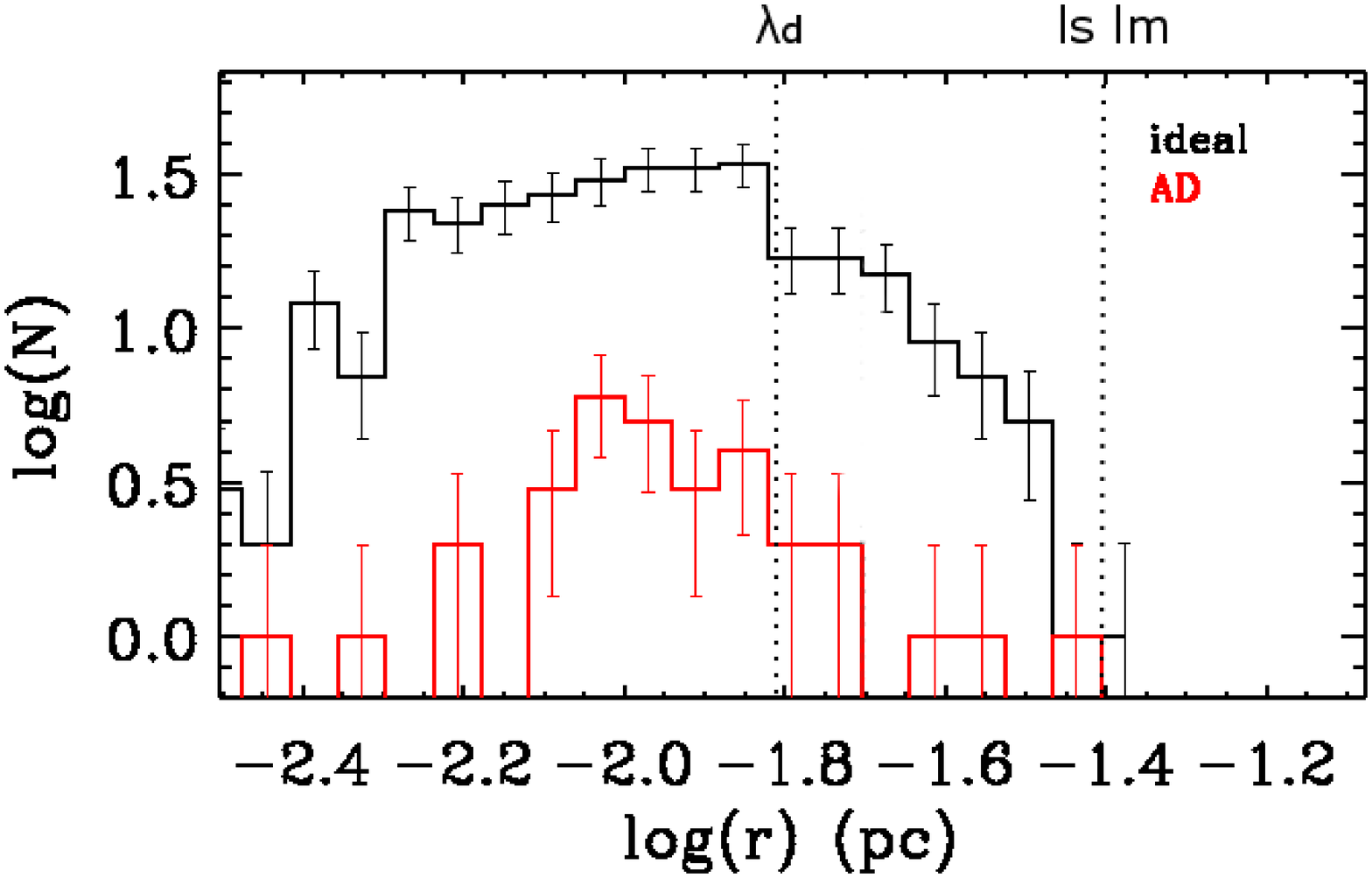}
  \caption{Same as Figure \ref{logthick_histograms_1_2000}, with a density threshold of 5000~cm$^{-3}$. The dotted lines indicate the mean ambipolar diffusion dissipation length $\lambda_d$, calculated from Eq. (\ref{ldiss}) for this density threshold, the resolution length $l_m$, equal to the length of 20 cells, and the sonic length $l_s$, defined by Eq. (\ref{ls}), for an rms Mach number of 3 (at $t=10^{-2}$, bottom) and 5 (at $t=5\times10^{-3}$, top).}
   \label{logthick_histograms_1_5000}
\end{figure}

\begin{figure}[!h]
   \includegraphics[width=0.9\linewidth]{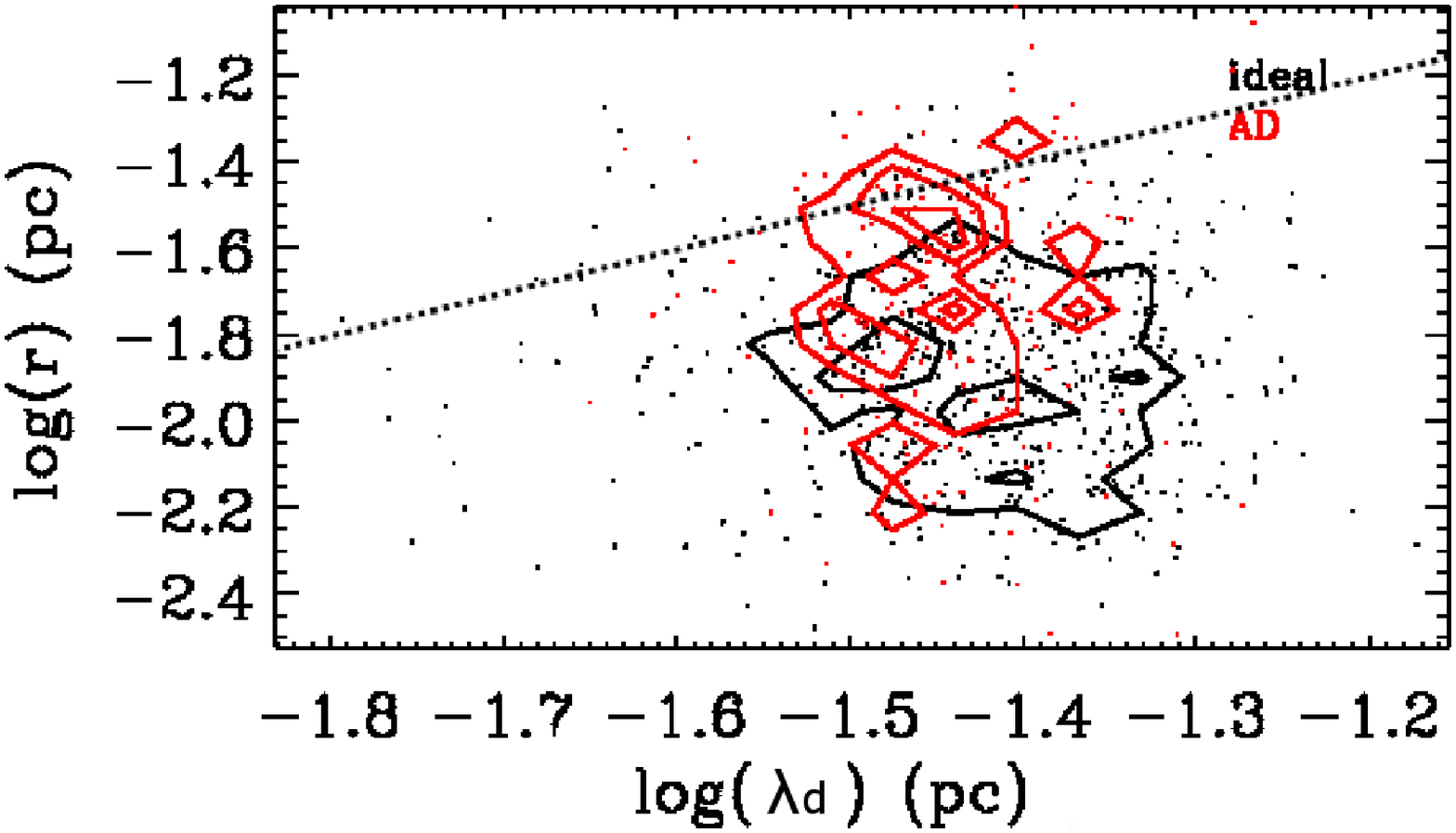}
   \includegraphics[width=0.9\linewidth]{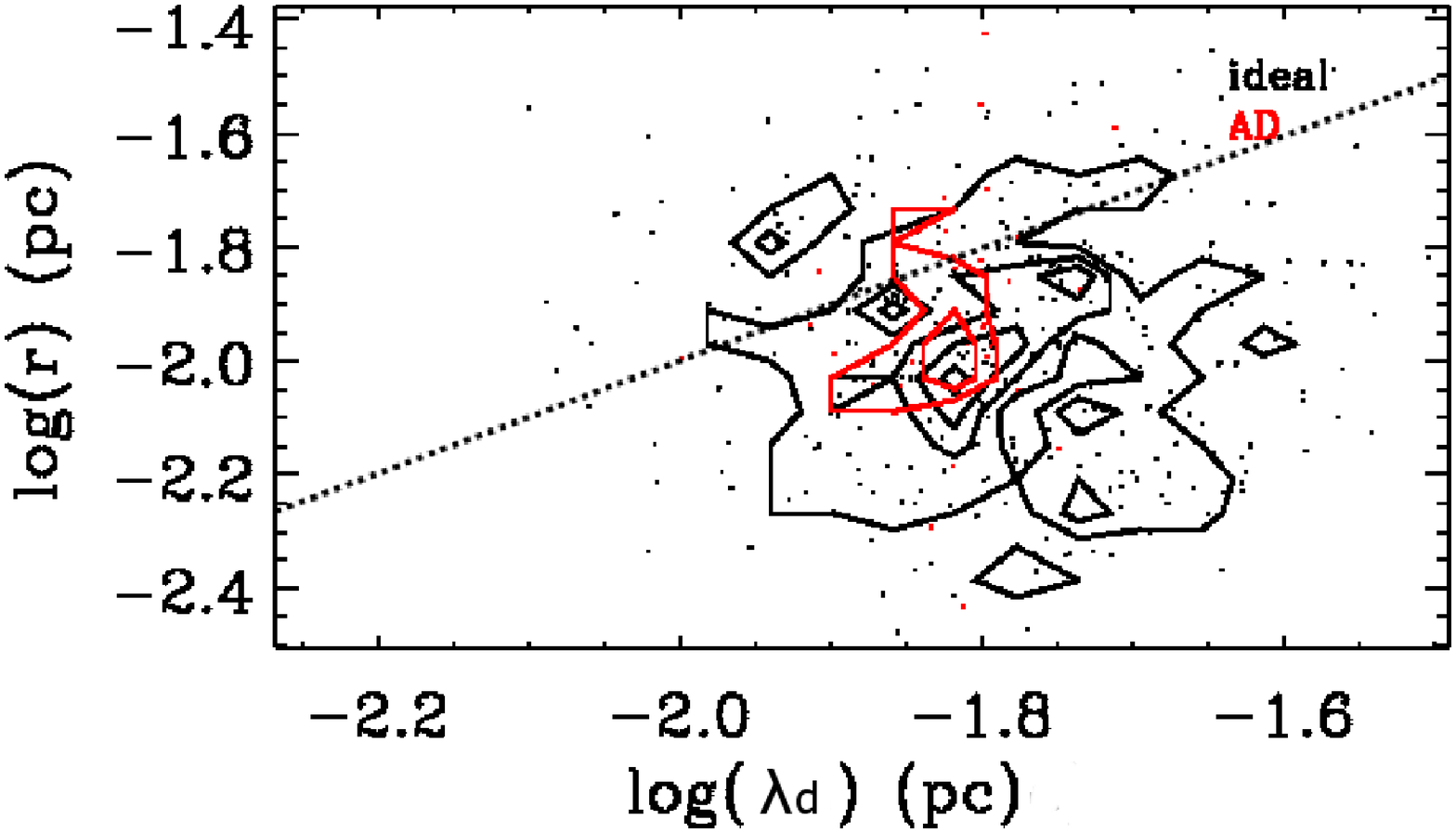}
   \caption{Scatter plots (dots) of the logarithm of the mean filament thickness versus the logarithm of $\lambda_d$ in runs A0 (black) and B0 (red). 
The simulation time is $t=10^{-2}$. Shown here are density thresholds 2000 cm$^{-3}$ (top) and 5000cm $^{-3}$ (bottom).  The contours
show the surface density distribution of the points and the dotted lines show the relation $\lambda_d$=r.}
 \label{thick_ldiss_decay}
\end{figure}

\begin{figure}[!h]
   \includegraphics[width=0.9\linewidth]{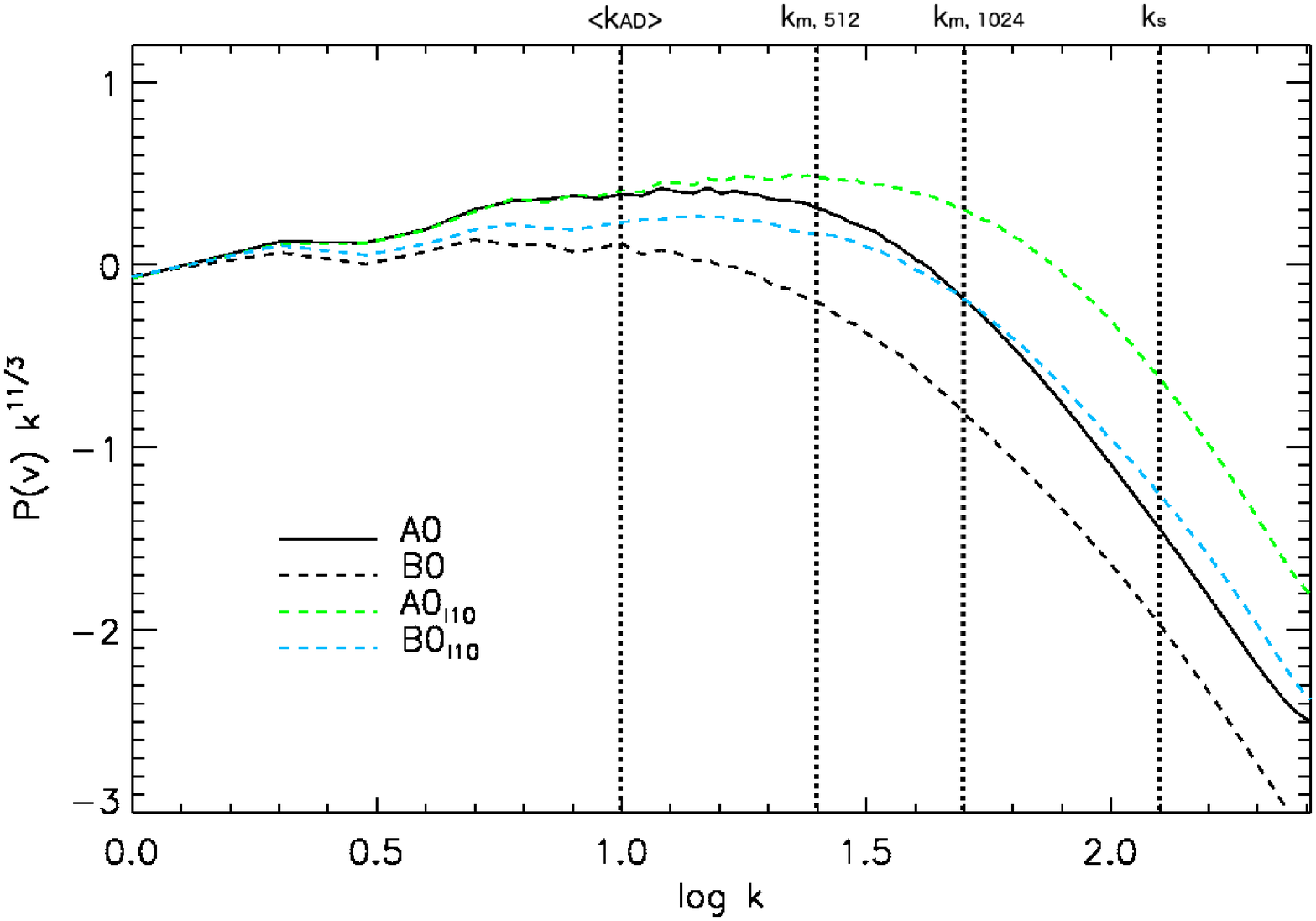}
   \includegraphics[width=0.9\linewidth]{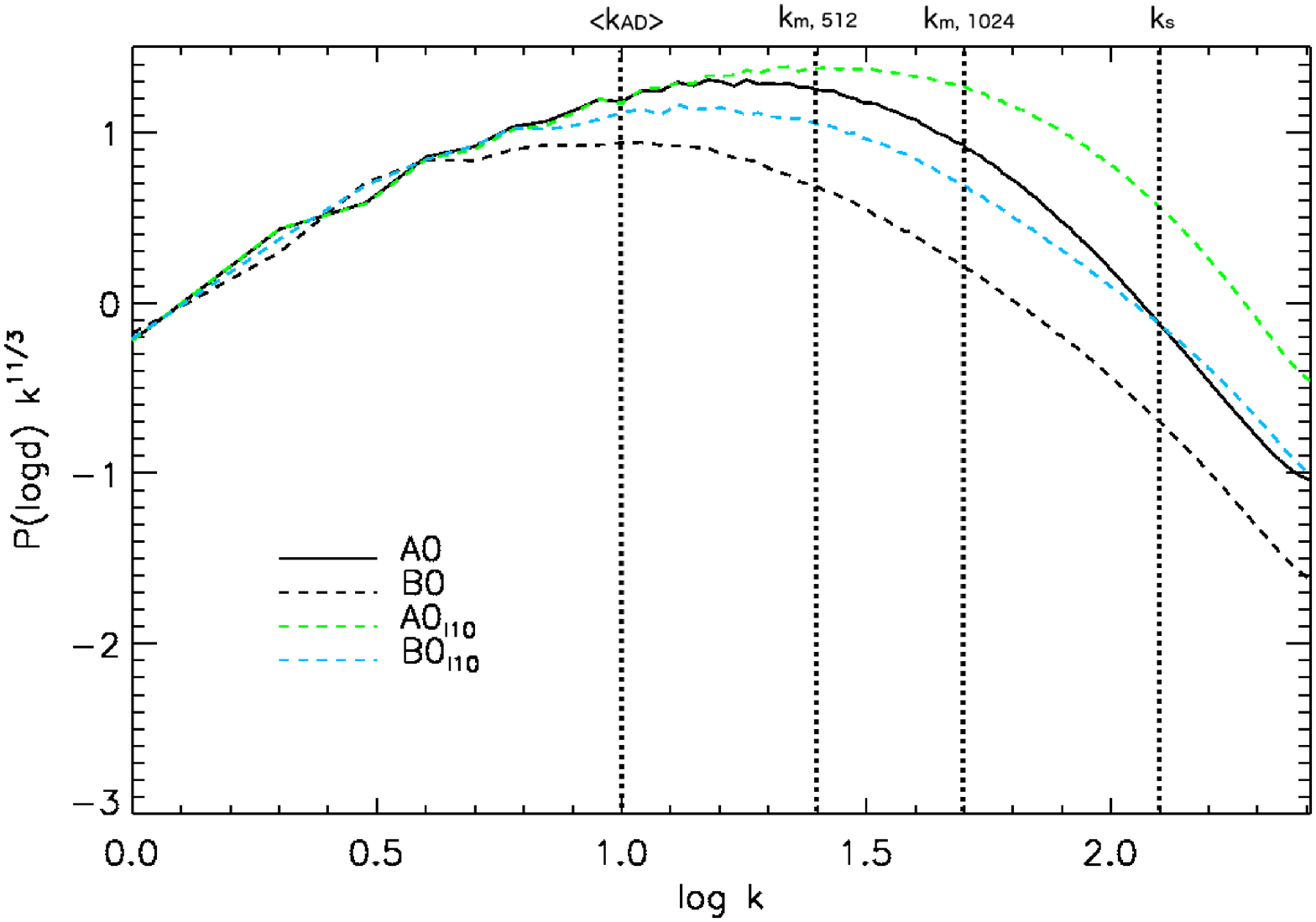}
   \caption{Comparison of the compensated power spectra in different resolutions. At the top, velocity power spectra, and at the bottom, power spectra of the logarithm of the density.  As indicated on the label, the solid black line refers to run A0 and the dashed lines refer to runs B0, A0$_{l10}$ and B0$_{l10}$. The vertical dotted lines show the mean ambipolar diffusion scale,
$<k_{AD}>$, the numerical dissipation scales, k$_{m,512}$ and k$_{m,1024}$ of the two resolutions and the sonic scale, k$_s$.
Runs A0$_{l10}$ and B0$_{l10}$ are run at double spatial resolution compared to A0 and B0.}
   \label{spectra_comparison_2}
\end{figure}

\begin{figure}[!h]
   \includegraphics[width=0.9\linewidth]{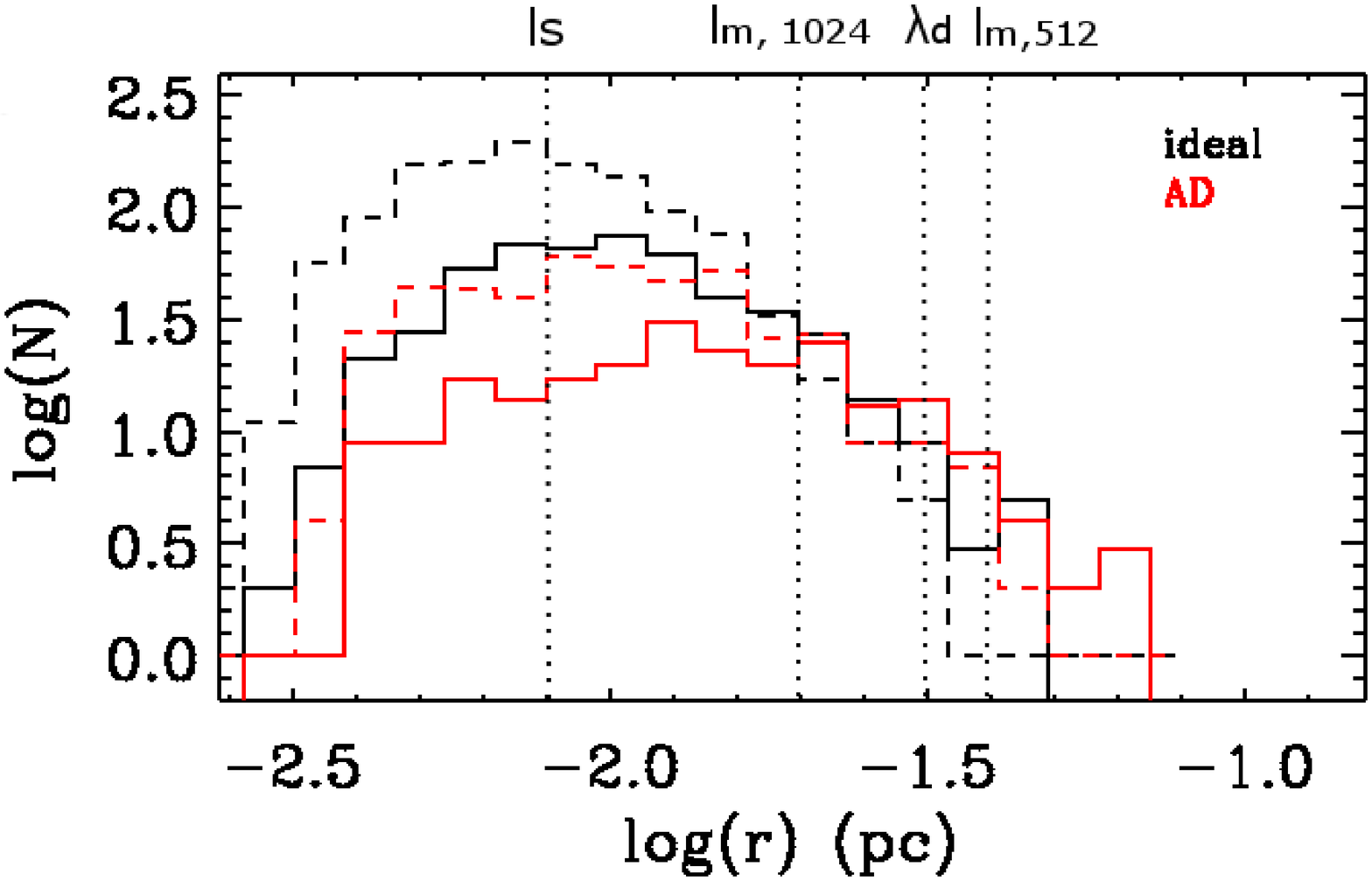}
   \includegraphics[width=0.9\linewidth]{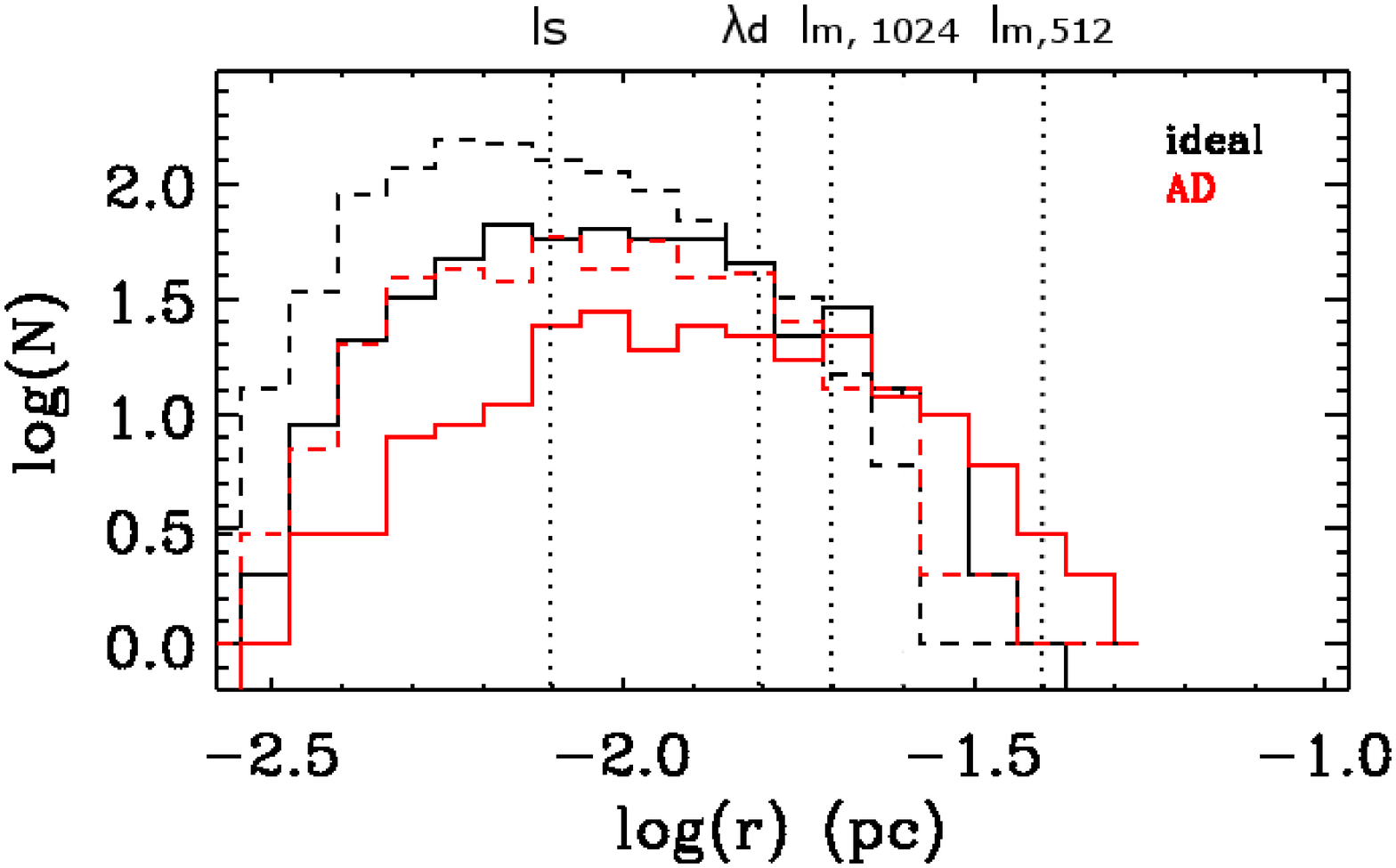}
   \caption{Thickness distributions of runs A0 and B0 at time $\text{t=}5\times 10^{-3}$.  Black lines correspond to the ideal MHD runs and red lines to the runs with ambipolar diffusion.  The corresponding dashed lines show the $1024^3$ cases.  Above, threshold 2000 cm$^{-3}$ for filament identification, below, threshold of 5000 cm$^{-3}$.  The dotted lines indicate the mean ambipolar diffusion dissipation length $\lambda_d$, calculated from Eq. (\ref{ldiss}) for each density threshold, the resolution lengths $l_{m,512}$ and $l_{m,1024}$, equal to the length of 20 cells, and the sonic length $l_s$, defined by Eq. (\ref{ls}), for an rms Mach number of 5.}
   \label{thickness_comparison_2}
\end{figure}
%

First, we observe that the inertial range is resolved in all simulations.
We also see that the non-ideal simulation exhibits loss of power at small scales, a trend that appears both in the velocity and the log density power spectra.
The spectra of the non-ideal MHD run descend steeply above a wave number of about 10.  This places the turbulence dissipation
scale at around 0.1~pc, in complete agreement with our prediction.  
In contrast, the ideal MHD run enters the dissipation regime at a smaller scale, roughly at a wave number of 25.  This corresponds to
about 20 cells, which marks the scales affected by numerical dissipation.
Clearly, ion-neutral friction introduces a characteristic dissipation scale, 
which is likely responsible for the disappearance of small and dense structures.

\section{Filament properties}
\label{thick}

In this section we compare the filaments formed in runs A0 and B0.  
The process of filament identification and the analysis is the same for all the runs in this paper.

In order to identify filaments in the simulations, we need to have a physical definition of an ISM filament in mind.  
However, the current definitions for filaments vary in different works.                 
In \citet{Andre_2014} a filament is defined as "any elongated ISM structure with an aspect 
ratio larger than 5 - 10 that is significantly overdense with respect to its surroundings".
\citet{Konyves_2015}, refine this definition to include structures of significant aspect ratios (greater than 3) and
column density  contrasts of at least 10\% with respect to the local background.
In the work of \citet{Arzoumanian_2011}, the Herschel filaments are located by their DisPerSE \citep{Sousbie_2011} skeletons in column density and
then their radial density profiles are extracted at different locations along their length to define a thickness.  
\cite{Panopoulou_2014}, working with $^{13}$~CO data (that involves a smaller dynamical range but 
includes the velocity information), employ a technique similar to that of \citet{Arzoumanian_2011} to 
identify filaments in Taurus.

In this work we define filaments by first setting a density threshold above which a location is considered to belong
to a structure of interest.  This definition of the structures of interest relates directly to the density-dependent length scale 
$\lambda_d$ from Eq.~(\ref{lambda2}), which facilitates a first interpretation of the results.
Direct comparison to the observations requires definitions of the filament and the thickness more similar to the observational ones.
Figure \ref{volume_thresholds} shows in logarithm of density the locations remaining once we
apply a threshold of 2000 or 5000 cm$^{-3}$.  We will be using these two thresholds throughout this work.
Since the mean density of the simulation box is 500 cm$^{-3}$, these thresholds correspond to 
density contrasts of 4 and 10, respectively, for all the results included in this paper.
The locations above a threshold are then connected into filaments by means of a friend-of-friends algorithm.

For each filament thus defined we find the eigenvalues of its inertial matrix.  
The eigenvector associated with the largest of these eigenvalues defines the longest axis of the structure. 
The filamentary morphology of the structures is clear in Figure \ref{volume_thresholds},
but we have also confirmed it by their aspect ratio distributions, which peak at values larger than 3 for all the simulations.

Each of the three-dimensional filaments is separated into segments along its longest axis and the local center of mass is found for each fragment.  
The local thickness is calculated by measuring the mass-weighted mean distance of the cells from the local center of mass.  
Finally, the thickness of the filament is found by taking twice the mean of these measurements 
over all the segments. This is the same technique used in \citet{Hennebelle_2013}.

In the following we will compare statistical distributions of the filament properties 
between ideal and non-ideal MHD runs.  In order to quantify the difference between each pair of distributions,
we performed a Kolmogorov-Smirnov (KS) two-sample test on each pair of distributions.  
This test gives the probability that two distributions were drawn from the same 
sample;  in other words, the closer this probability is to zero, the less likely it is that the two distributions were drawn from the same sample (see, for example, \cite{Press_1993}).
KS probability values above about 0.1 indicate that the two distributions are statistically indistinguishable.
Table \ref{stat_table} lists the KS probabilities, along with other properties of the distributions.

The mass and mass per unit length distributions of the filaments in runs A0 and B0 are shown in 
Figures \ref{mass_histograms_1_2000} and \ref{mass_histograms_1_5000}, for thresholds equal to 2000~cm$^{-3}$ and 5000~cm$^{-3}$, respectively.  
The mass per unit length is calculated as the total mass of a structure divided by the length of its longest axis.
The black lines in all these plots refer to ideal MHD and the red lines to runs with ambipolar diffusion.

Looking at the lower density threshold (2000~cm$^{-3}$) first, we notice that there are
clearly fewer low-mass structures when ambipolar diffusion is modeled.  
In run B0 the peak of the mass distribution shifts to higher values.  
The distributions evolve slightly with time, but the difference of a factor of 3-10 between the peak masses 
in the two runs persists.
The higher density threshold (5000 cm$^{-3}$) eliminates many structures in the non-ideal case, 
confirming the visual impression from Figure \ref{turb_a0_b0}.  
At late times, the statistics are too poor to make any strong statement about
the characteristic mass, as indicated by the error bars and by the KS test result 
in Table \ref{stat_table}, but the effect still persists. 

The distributions of mass per unit length (right panels of Figures \ref{mass_histograms_1_2000} and \ref{mass_histograms_1_5000})
greatly resemble the mass distributions.  This is not surprising for filamentary structures of similar density.
Although with poor statistics, the non-ideal MHD run seems to produce filaments of slightly higher mass per unit length.

The local thicknesses calculated for each center of mass are averaged per structure.
Figures \ref{logthick_histograms_1_2000} and \ref{logthick_histograms_1_5000} show the distributions of the mean thickness of the identified filaments
in runs A0 and B0 for the two different thresholds, 2000 cm$^{-3}$ and 5000 cm$^{-3}$, respectively, and for two different times.

From these plots and the corresponding ones for driven turbulence in Appendix \ref{app:driven}, 
we recognize a general trend for the ideal MHD simulations to produce a peak in the thickness distribution.  
This happens because at large scales the turbulence sets
a power law, while at scales comparable to the grid size the numerical dissipation gradually removes structure.  
With ambipolar diffusion this small-scale dependence is less noticeable. 

In Figure \ref{logthick_histograms_1_2000} we observe that the ion-neutral friction causes the peak of 
the thickness distribution to move to higher values.  The relative shift between the two histograms starts from 
a factor of 1.6 and at later times evolves to factor of 2.  

The statistics with a 5000 cm $^{-3}$ threshold are again not as good.  
Even so, we can notice in Figure \ref{logthick_histograms_1_5000} that a lot of low-thickness filaments
disappear with ambipolar diffusion and in general, the thickness distribution is flatter.  

In the Introduction we defined the dissipation length $\lambda_d$.  
For each dense location in the simulations we calculate $\lambda_d$ from Eq.~(\ref{ldiss}) 
and plot it against the mean thickness of each filament.  
Figure \ref{thick_ldiss_decay} (and Figure \ref{thick_ldiss_driven} in Appendix \ref{app:driven}) show the resulting distributions for the two thresholds, 2000 cm$^{-3}$ and 5000 cm$^{-3}$ at the last snapshot of each simulation.  
Each symbol in these plots corresponds to one filament and the contours show the probability distributions for the same data.  
Black symbols and lines indicate ideal MHD, red symbols and lines indicate ambipolar diffusion runs.
The dotted lines show the relation $\lambda_d$=r.
Although strictly speaking there is no definition of $\lambda_d$ in the ideal runs, we do plot the quantity expressed by Eq~(\ref{ldiss}) for comparison.

Although the maximum surface density of points in the non-ideal run falls closer to the $\lambda_d$=r line than the 
corresponding maximum in the ideal run, 
there is a tendency in all these diagrams for the measured thickness to be slightly lower than the ambipolar 
diffusion critical length.  This tendency is less pronounced for the higher threshold and for driven turbulence case.
We also observe that the distribution of $\lambda_d$ values is narrower for the runs with ambipolar diffusion and move to smaller values 
with higher threshold, in agreement with the theory.  
The same distributions for the ideal runs spread over a larger range of $\lambda_d$ values.  

\section{Resolution study and convergence}
\label{sec:resolution}

We emphasized before that resolving the physical dissipation regime is essential in this work. 
But while the power spectra in Figure \ref{pow_spec_a0_b0}
(and also Figures \ref{vel_spec_a1_b1} and \ref{den_spec_a1_b1} in the Appendix) show that the inertial range of turbulence is well resolved
with a $512^3$ grid, numerical diffusion could still be affecting the smallest scales. It is therefore 
important to study the behavior of the dense structures at higher resolution.

Three additional pairs of runs, ideal and non-ideal MHD, were used to study the effects of numerical diffusion,
with the same setup as runs A0 and B0.  
Two of them were done using AMR with two levels of refinement, two with a lower Courant condition,
and two were run entirely on a $1024^3$ grid, without AMR.
The clearest effect was shown in this last case, runs A0$_{l10}$ and B0$_{l10}$.

These high-resolution simulations are computationally very demanding, especially the cases with ambipolar diffusion.
This made long integrations very challenging.  
We were able to evolve the simulations to a time $t=5\times 10^{-3}$, the time of the first snapshots in the previous sections.

In Section \ref{numerics} we discussed the need for a timestep limitation in order to stabilize the 
numerical scheme with the implementation of ambipolar diffusion.  We found that the results were 
converged below a minimum timestep equal to 5 percent of the Alfv\'{e}n crossing time of the cell.
Unfortunately, we were only able to afford a 12.5 percent limit for run B0$_{l10}$, due to lack of computational resources.
Although the mass and thickness distributions are fairly insensitive to the choice of the timestep limit (see Appendix \ref{app:timestep}),
the power spectra could still change with a smaller timestep.

The velocity and log density power spectra of runs A0, B0, A0$_{l10}$ and B0$_{l10}$ are plotted together in Figure \ref{spectra_comparison_2}.
Despite the obvious expansion of the inertial range with the increase in resolution, 
the general trend between ideal and non-ideal MHD is unaltered.
With ambipolar diffusion the spectra steepen earlier than in the ideal MHD situation.  This happens 
because the physical dissipation length acts on scales larger than those affected by the grid. 

Although the difference between runs A0 and A0$_{l10}$ is marginally higher than that between runs B0 and B0$_{l10}$, 
it is evident that the runs have not fully converged in spatial resolution.
The power spectra of the non-ideal runs also change when we increase the resolution, extending the inertial range to higher wavenumbers.
This either means that the numerical dissipation is affecting also scales we were considering dominated by ambipolar diffusion,
or that, even with the ion-neutral friction included, there are modes still capable of transporting power to smaller scales 
(magnetosonic waves), or both.

The corresponding comparison of the filament thickness distributions between 
A0, B0, A0$_{l10}$ and B0$_{l10}$ is shown in Figure \ref{thickness_comparison_2} and it reflects the same picture.
The increase in resolution produces much more structure in both cases.
The difference between runs A0 and A0$_{l10}$ is very pronounced, with the peak of the distribution shifting
to smaller values by almost 0.5 dex.  The peak of the ambipolar diffusion simulations, B0 and B0$_{l10}$, also shifts, but
by a smaller factor, of about 0.3 dex.  The largest differences between runs of different resolution are observed
towards the smallest values of the filament thickness, in the regime affected by the grid.

Although the relative shift between ideal and non-ideal runs does not change by increasing the resolution, 
both distributions move to smaller values with respect to their lower resolution counterparts. 
This is not against our initial expectations.  It appears that numerical diffusion still affects the smallest scales. 
At the same time, the physical dissipation length $\lambda_d$ is inversely proportional to the density.  
Higher resolution allows higher densities to exist in the box, thus the local $\lambda_d$ also shrinks. 
This may indicate that some modes still propagate to smaller scales, an issue we discuss in Section \ref{discuss}.


\section{Summary and Discussion}
\label{discuss}

We have presented high-resolution, ideal and non-ideal MHD simulations of decaying and driven turbulence, 
with the aim to trace the physics behind the formation of low-density molecular filaments.

The main result of this work is that, in accordance with theoretical predictions, ambipolar diffusion alters the behavior 
of MHD turbulence.  Small-scale structure is smoothed out and 
the power spectra exhibit clear differences in the extent of the inertial range. 

In all these simulations we identified filaments by setting a density threshold and connecting the
locations above that threshold with a friends-of-friends algorithm.  Then we calculated their masses and thicknesses.
We found that both the mass and the thickness distributions of the filaments consistently  shift towards higher values with ambipolar diffusion.
This result was tested with a Kolmogorov-Smirnov test, which showed that the mass and thickness distributions
of the filaments in ideal and non-ideal simulations are genuinely different in most cases, with the exceptions appearing
only when the filament statistics are poor.

The parameters used for the turbulent gas in these simulations were chosen to fall close to observed molecular cloud properties, give reasonable range for statistics, 
but at the same time also allow a very good resolution of the dissipation length for ion-neutral friction, $\lambda_d$, as it is calculated from Eq. (\ref{ldiss}).  
Trying to compromise between all these different criteria, with a particular focus on the later, has set a few limitations.
The length of the box, for example, is set to 1~pc, which is clearly smaller than most molecular clouds.  This choice leads to very small filament masses and masses per unit length , as one can
observe in Figures \ref{mass_histograms_1_2000}, \ref{mass_histograms_1_5000}, \ref{mass_histograms_2_2000} and \ref{mass_histograms_2_5000}. 

In addition, the simulation volume includes several Jeans masses.  This makes no difference here because self-gravity is not included, but it would
very likely change the picture if it were.  Even with these models being in a periodic box with supersonic turbulence we would expect dense regions to collapse rapidly,
probably faster than the ambipolar diffusion timescale.  On the other hand, self-gravity would also modify the density profiles of some of the filaments, which would allow 
a more realistic comparison to the observed structures. 
These matters deserve further investigation, which is outside the scope of this paper.
What matters for our present conclusions is the comparison between the ideal and non-ideal MHD situations.

It is instructive to compare our results directly to theories about the dissipation of MHD waves through ambipolar diffusion.
\cite{Balsara_1996} predicted that, when the Alfven speed is larger than the sound speed, as it happens in all our simulations, 
ion-neutral friction damps the Alfv\'{e}n and fast magnetosonic waves, allowing only the propagation of slow magnetosonic waves.
In agreement with this prediction, \citet{Burkhart_2015} find that in 
sub-Alfv\'{e}nic turbulence the turbulent cascade is completely cut-off below the characteristic ambipolar diffusion scale, while super-Alfv\'{e}nic turbulence can persist below it.

In this work we have used sub-Alfv\'{e}nic turbulence for all the models, but the actual situation inside molecular clouds is not known: \citet{Padoan_1999} and \citet{Luntilla_2009}, for example, argue that molecular clouds could be super-Alfv\'{e}nic, but recent Planck results are more consistent  with models of sub-Alfv\'{e}nic turbulence \citep{Soler_2013,Planck2015}.  Although there is no particular reason for our choice other than our current inability to fully explore the parameter space, a different choice could affect the result.

Our models suggest neither a complete cut-off nor a continuous cascade: On the one hand, the power spectra clearly steepen close to the ambipolar
diffusion length $\lambda_d$.  (This result is consistent, for example, with the decaying non-ideal turbulence simulations of \citealt{Momferratos_2014}).
On the other, we do find structures with thicknesses smaller than the local ambipolar diffusion dissipation length 
(see, for example, Figures \ref{thick_ldiss_decay} and \ref{thick_ldiss_driven}).  
There are two effects that could contribute to this result.
One is our filament identification algorithm, which is based on a density threshold.  
Selecting filaments only above a certain density threshold (roughly equivalent in this study to a density contrast threshold) 
could be causing a bias towards the densest and thinner part of larger structures.
Another is, as \citet{Burkhart_2015} argue, that there is a potential dependence of the dissipation scale on the driving scale, 
which could be the cause of the differences between the freely decaying and the driven runs.  

\citet{McKee_2010} studied the mass spectra of clumps in non-ideal MHD simulations, 
{varying the so-called ambipolar diffusion Reynolds number, which they define as $L/\lambda_d$} (Here $L$
is the integral scale and $\lambda_d$ the ambipolar diffusion scale.).
They found that, as the ambipolar diffusion Reynolds number decreases, which means ambipolar diffusion becomes stronger,
the slope of the higher-mass portion of the clump mass spectrum increases.  
In our simulations we do not observe such behavior in the high-mass portion of the clump mass function.
However, the slope of the intermediate mass regime does seem to change  
(Figures \ref{mass_histograms_1_2000}, \ref{mass_histograms_1_5000}, \ref{mass_histograms_2_2000} and \ref{mass_histograms_2_5000}).
Again, it is possible that the source of this apparent discrepancy are the different algorithms for identifying clumps.
\citet{McKee_2010} use clumpfind, which separates cores in a hierarchical manner.  Our algorithm
counts everything above a certain density threshold as one structure.

The flattening we observe in the intermediate mass regime of the distributions can be explained if we refer to the density power spectra. 
\citet{Hennebelle_Chabrier_2008} predicted that the exponent of the power law turbulent mass distribution
for non-self-gravitating structures should be $x=2-n'/3$, where $n'$ is the power spectrum index of log d.
The decrease of the power-law slope in the mass spectra with AD 
is consistent with the change of power-law slope in the log d power spectrum.

Regarding the very important issue of resolution, we dedicated Section \ref{sec:resolution} 
to the study of convergence of the dissipation length.
Although our simulations in principle are resolving the mean characteristic length for ambipolar diffusion, they do not necessarily 
do so for the densest regions, due to its inverse dependence on the density.  
We used two simulations with a 1024$^3$ resolution, which 
showed that the runs with ambipolar diffusion are also affected by an increase in resolution. 
This implies that there are still modes that transport momentum to small scales, in agreement with both \cite{Balsara_1996}'s and \citet{Burkhart_2015}'s conclusions.

Finally, by increasing the resolution we found that the peak of both the ideal
and the non-ideal turbulence distributions shift to lower densities.  Whether 
this is because we have not reached convergence or because the non-ideal 
thickness distribution also inherently spreads to smaller values is something we could not fully test.  
In principle, even higher resolution would help resolve this issue, but this is extremely difficult given the current computational resources.


\section{Conclusions}
\label{concl}

The main conclusion of this work is that the introduction of ambipolar diffusion in simulations
of supersonic turbulence changes the appearence of the density field and the 
properties of the densest filaments.

More specifically, we find that the power spectra of both the density and the velocity field show 
a steeper cutoff, which happens at a smaller wave number in non-ideal MHD than in ideal MHD.  
This suggests different morphologies and apparent sizes of the dissipative structures with ambipolar diffusion.

The distributions of the filament masses tend to flatten in the high-mass end of the spectrum, which we attribute to
the different density scaling.  

A very important motivation for this work was the observational finding that the central regions of
local molecular filaments as observed in the sub-mm dust continuum with Herschel seem to have a roughly constant thickness of 0.1~pc.  
As a first step towards understanding the origin of this result, we have sought for any measurable effect of ambipolar diffusion on the distribution
of the total width of filaments formed by MHD turbulence.  Since our models are limited in dynamical range and do not include dust chemistry or self-gravity, 
in this attempt we have not used the same definition of the filament width or the same methods as those used
in deriving the observational result.

We have defined a filament as a structure of significant overdensity with respect to its surroundings and
its width as the mean distance of its points from the local barycenters along its main axis.
Using this definition, we find that ambipolar diffusion causes the thickness distributions of the filaments to shift
towards higher values.  A Kolmogorov-Smirnov test confirms that the differences between almost all
pairs of distributions in this paper are statistically highly significant.
Moreover, we notice that ambipolar diffusion does not cause a sharp peak around any particular value of the thickness.
Due to limitations in spatial dynamical range, we cannot directly compare our findings with, for instance, Herschel 
observed molecular clouds.
Future simulations of larger volumes will allow such comparisons, employing the same tools 
and adopting the same definitions as those used in analysing observational data.

Finally, we investigated the matter of resolution. 
Our results show that, although there is clearly a difference between ideal and non-ideal
simulations independently of resolution, even the 1024$^3$ simulations with ambipolar diffusion are not converged.
This is a very important warning for studying the sizes of 
small-scale structure in numerical simulations of turbulence, especially when no physical dissipation is employed.


\emph{Acknowledgments}
The authors would like to thank the referee, 
Mordecai-Mark Mac Low, for very useful discussions and a very constructive report that improved this work and helped put our results better into context.
This research has received funding from the European Research Council  (FP7/2007-2013 Grant Agreement "ORISTARS" --no. 291294--
and Grant Agreement no. 306483).

\bibliographystyle{aa}
\bibliography{ntormousi_nonidealturb}
\label{lastpage}

\appendix

\section{Timestep convergence}
\label{app:timestep}

\begin{figure}[!h]
   \includegraphics[width=\linewidth]{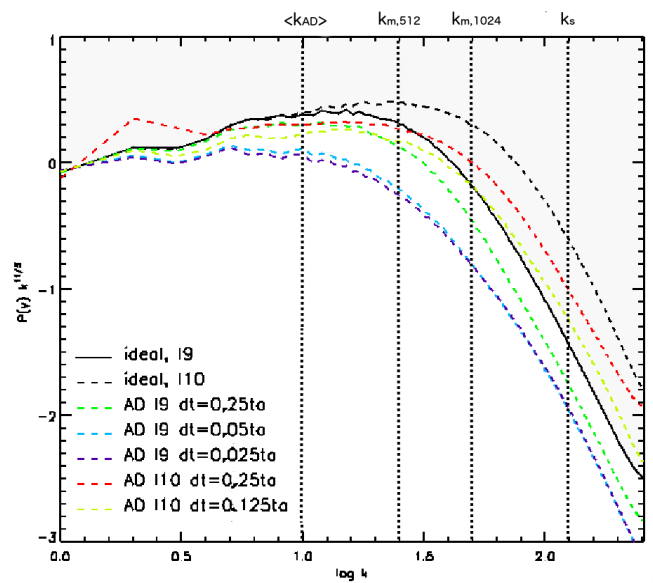}
   \caption{Compensated velocity power spectra for decaying turbulence simulations at time t=5$\cdot 10^{-3}$ using different timestep limits.  
The black solid line corresponds to the ideal 512$^3$ run and the black dashed line to the ideal 1024$^3$ run.
The vertical dotted lines show the mean ambipolar diffusion scale,
$<k_{AD}>$, the numerical dissipation scales, k$_{m,512}$ and k$_{m,1024}$ of the two resolutions and the sonic scale, k$_s$ (k$_s$ and k$_{m,1024}$ are practically the same).
Run B0, used as reference in the main body of the paper, uses a timestep limit dt=0.05t$_a$, where t$_a$ the Alfv\'{e}n crossing time of a cell.
}
\label{cts_powspec}
\end{figure}

\begin{figure}[!h]
   \includegraphics[width=\linewidth]{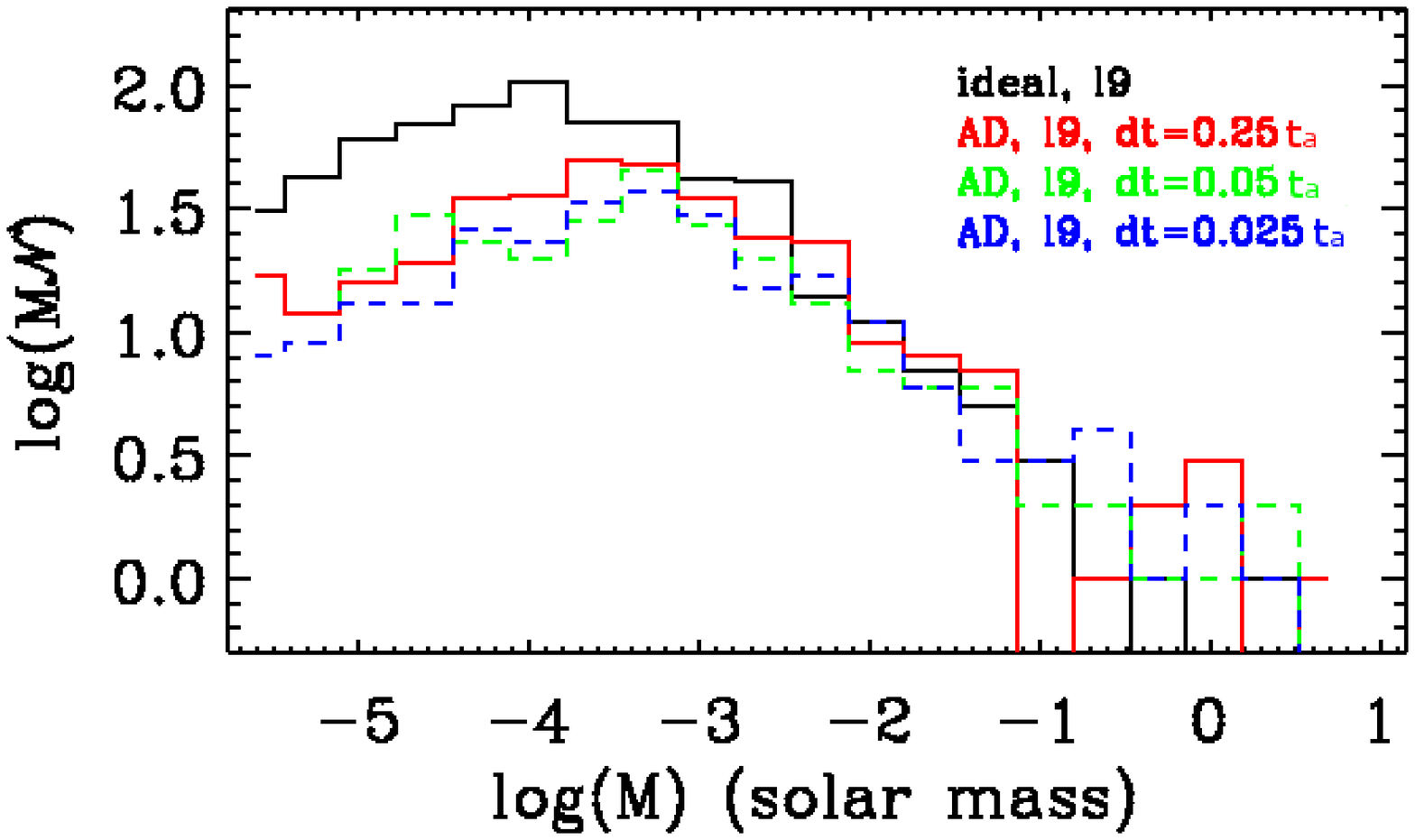}
   \includegraphics[width=\linewidth]{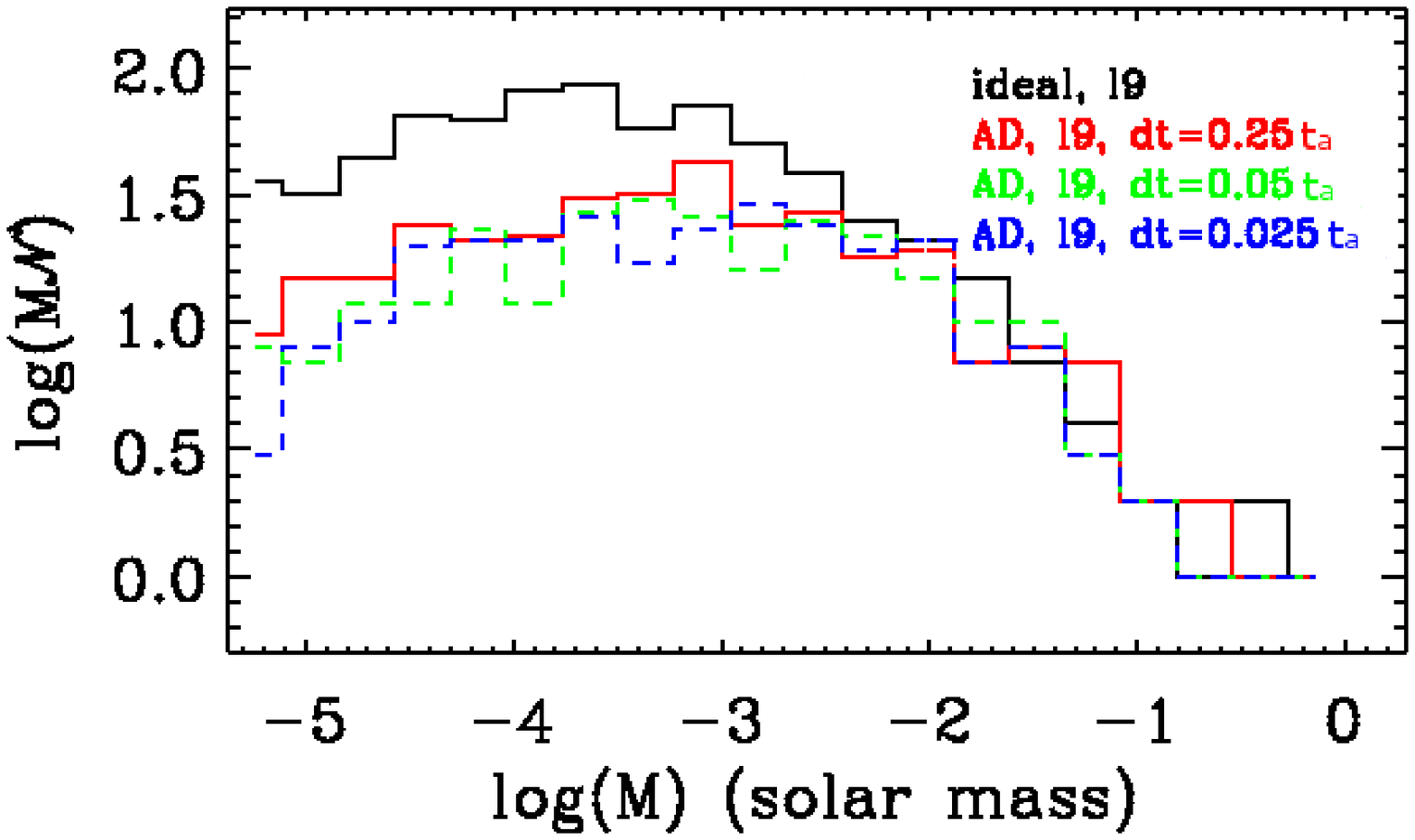}
   \caption{Mass distributions for decaying turbulence simulations at time t=5$\cdot 10^{-3}$ using different timestep limits.  
The black solid line corresponds to the ideal 512$^3$ run and the colored lines to runs with ambipolar diffusion and different timestep
limits.  t$_a$ is the Alfv\'{e}n crossing time of a cell.  Top: Density threshold of 2000 cm$^{-3}$, bottom, 5000 cm$^{-3}$.}
   \label{cts_mass}
\end{figure}

\begin{figure}[!h]
   \includegraphics[width=\linewidth]{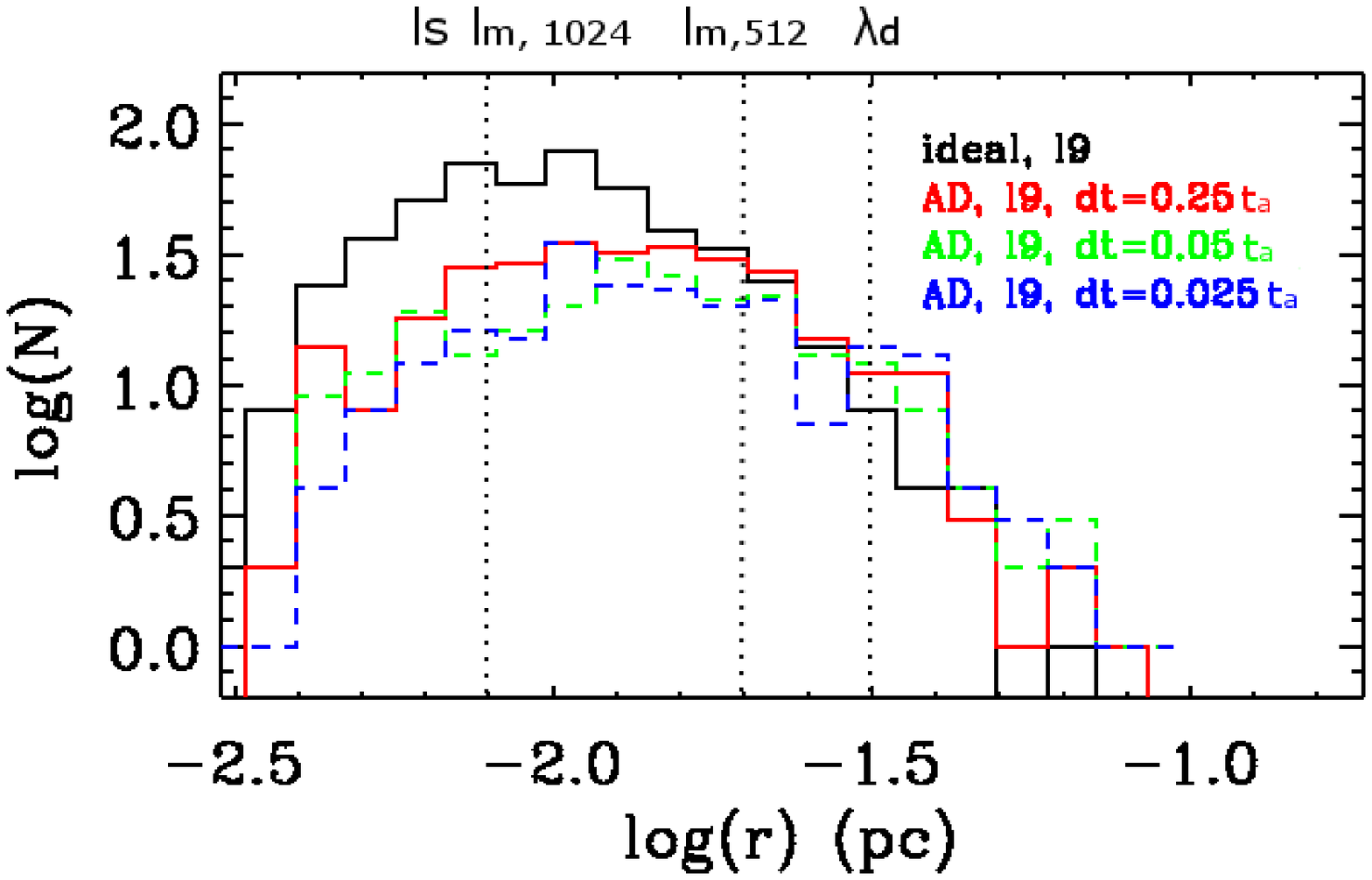}
   \includegraphics[width=\linewidth]{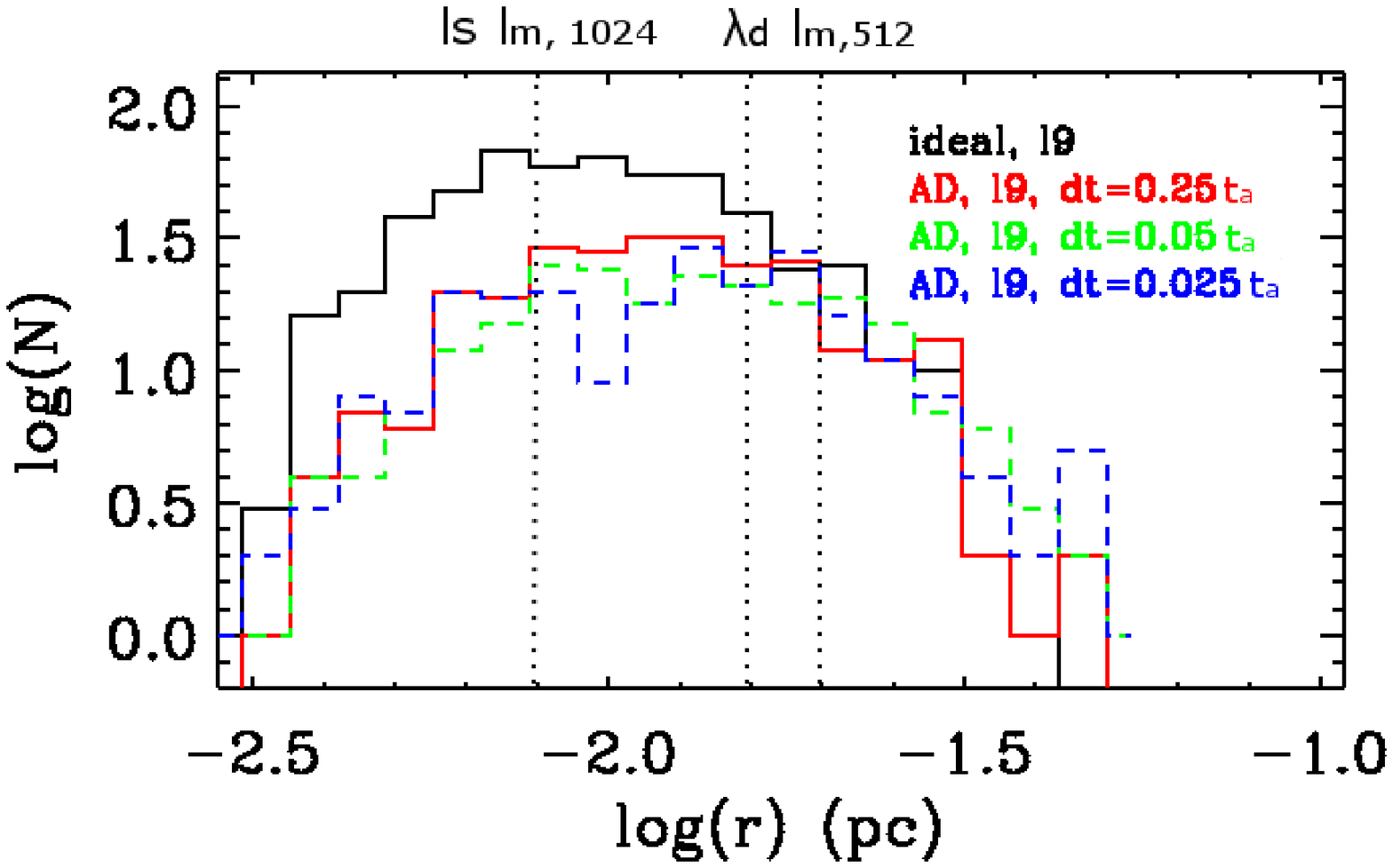}
   \caption{Filament thickness distributions for decaying turbulence simulations at time t=5$\cdot 10^{-3}$ using different timestep limits.  
The black solid line corresponds to the ideal 512$^3$ run and the colored lines to runs with ambipolar diffusion and different timestep
limits.  t$_a$ is the Alfv\'{e}n crossing time of a cell. Top: Density threshold of 2000 cm$^{-3}$, bottom, 5000 cm$^{-3}$. The dotted lines indicate the mean ambipolar diffusion dissipation length $\lambda_d$, calculated from Eq. (\ref{ldiss}) for each density threshold, the resolution lengths $l_{m,512}$ and $l_{m,1024}$, equal to the length of 20 cells, and the sonic length $l_s$, defined by Eq. (\ref{ls}), for an rms Mach number of 5.}
   \label{cts_thickn}
\end{figure}

In Section \ref{numerics} we mentioned the necessary timestep limitation as part of the non-ideal MHD implementation we used for this 
work.  Although this limitation stabilizes the numerical scheme, it can also affect the accuracy of the result.
Here we include some results of a convergence study on this timestep limit which led to the adopted dt=0.05t$_a$ 
(where t$_a$ the Alfv\'{e}n crossing time of a cell) in the reference run B0.

Figure \ref{cts_powspec} shows power spectra of decaying turbulence models with decreasing timestep limits for the
ambipolar diffusion scheme.  Figures \ref{cts_mass} and \ref{cts_thickn} show the filament mass and thickness distributions 
for the same models, like in Section \ref{thick} of this paper.

Not surprisingly, the velocity power spectra seem to be much more sensitive 
to the change of timestep limit than the mass and thickness distributions.  The velocities are
a direct output of the code, while the mass and thickness distributions are derived quantities.
It is clear that allowing timesteps smaller than dt=0.05t$_a$ does
not affect the power spectra, so we have adopted this limit for the reference run.

Since the mass and thickness distributions seem to be insensitive to the timestep limit,
we do include some results of the high resolution run B0$_{l10}$ in this paper, however keeping these small uncertainties in mind. 


\section{Driven turbulence}
\label{app:driven}

Turbulence in an ideal MHD simulation decays through numerical resistivity, existent in all numerical codes.  
When a physical diffusive process is explicitly modeled, there can be some doubts on whether or 
not it behaves as a scaled-up version of the numerical resistivity,  causing the same effect at a faster rate, 
or if it qualitatively alters the behavior of the simulation. 
If the former is true, one would expect the ideal MHD run to eventually reach a state of the AD run at a later time. 

In order to investigate this, we performed two numerical experiments of forced turbulence.  
In these simulations, turbulent kinetic energy is replenished at each coarse time step by the same amount that was dissipated in the last timestep.  

\begin{figure*}[!ht]
   \includegraphics[width=0.49\linewidth]{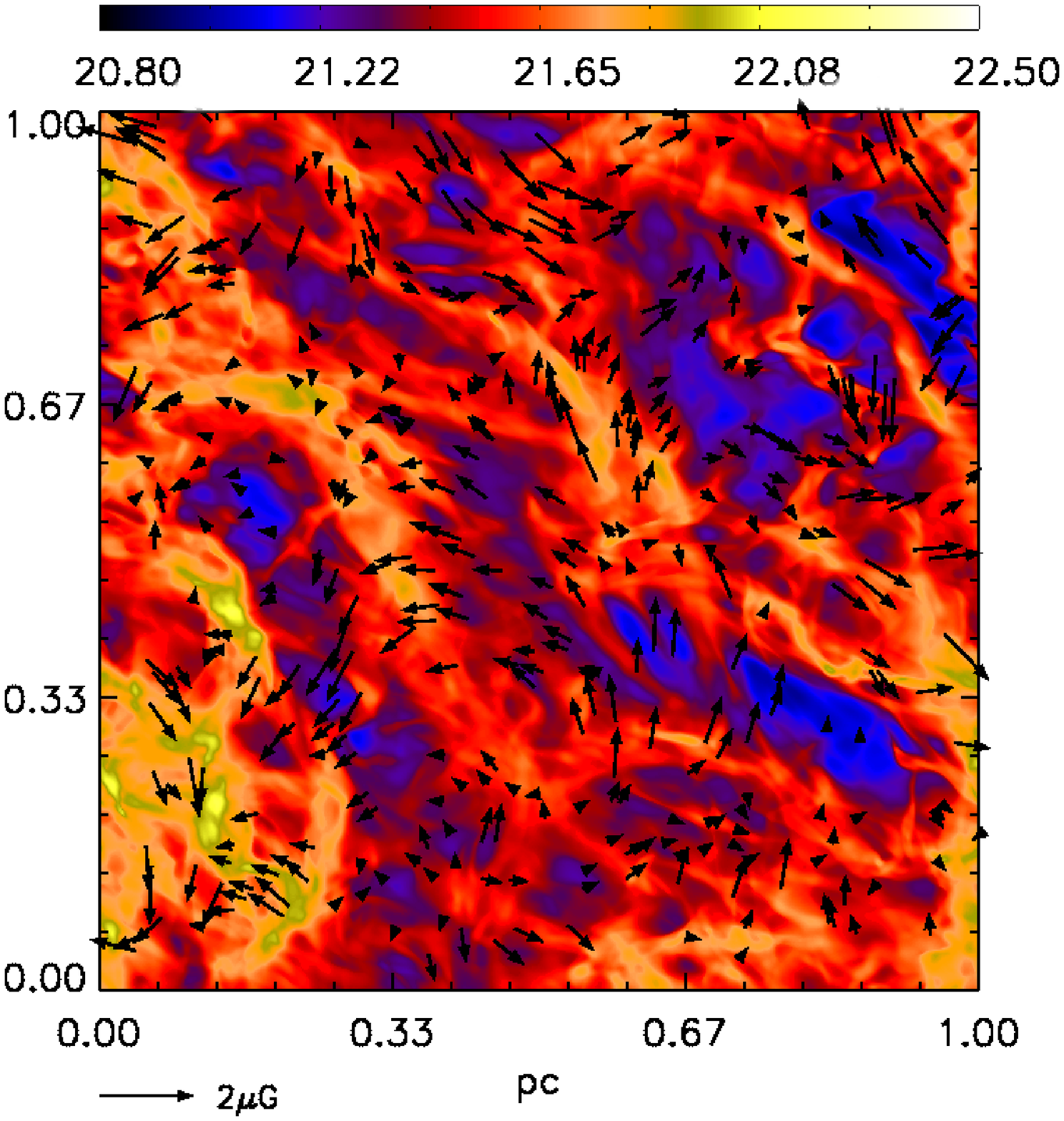}
   \includegraphics[width=0.49\linewidth]{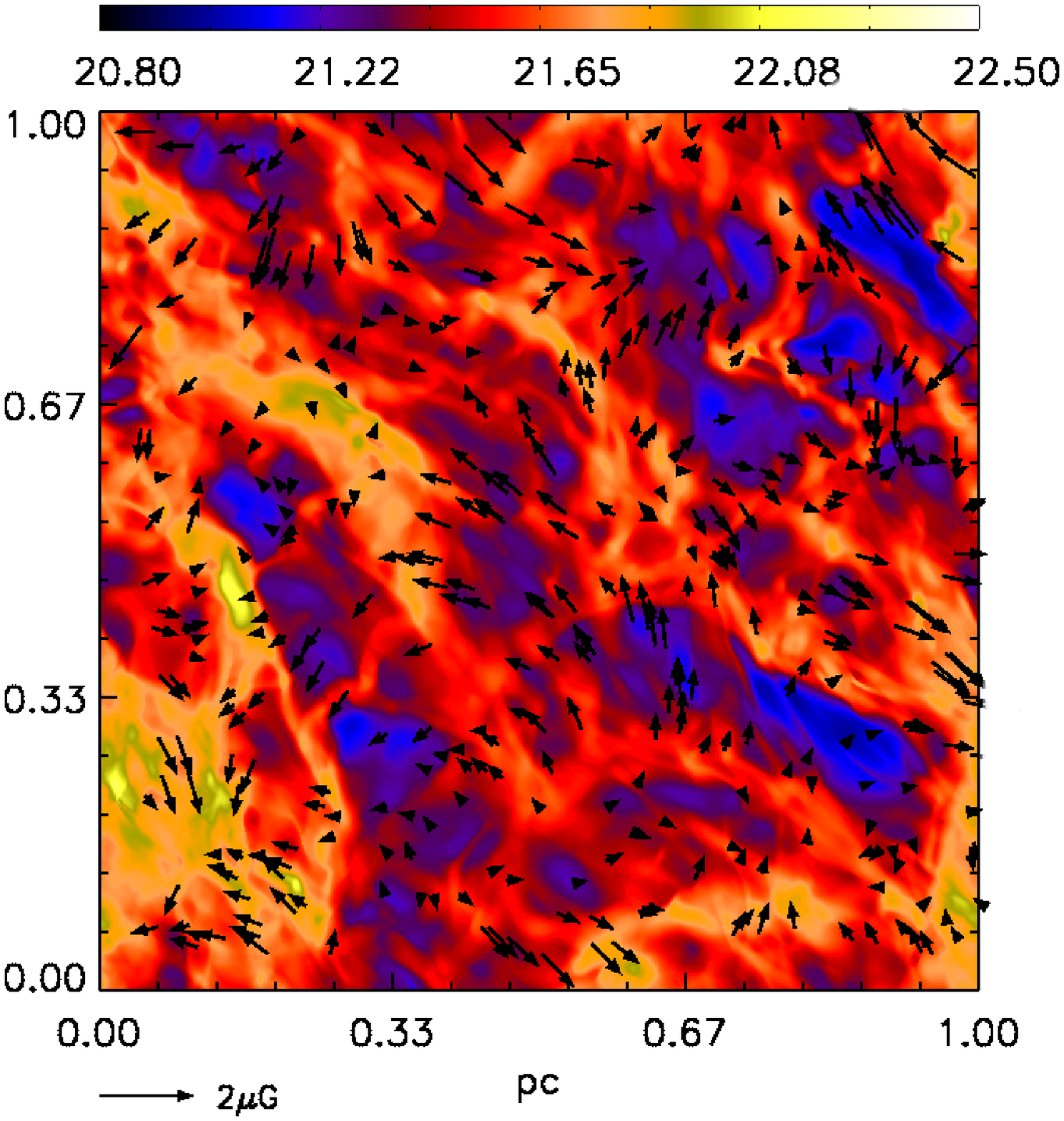}
   \includegraphics[width=0.49\linewidth]{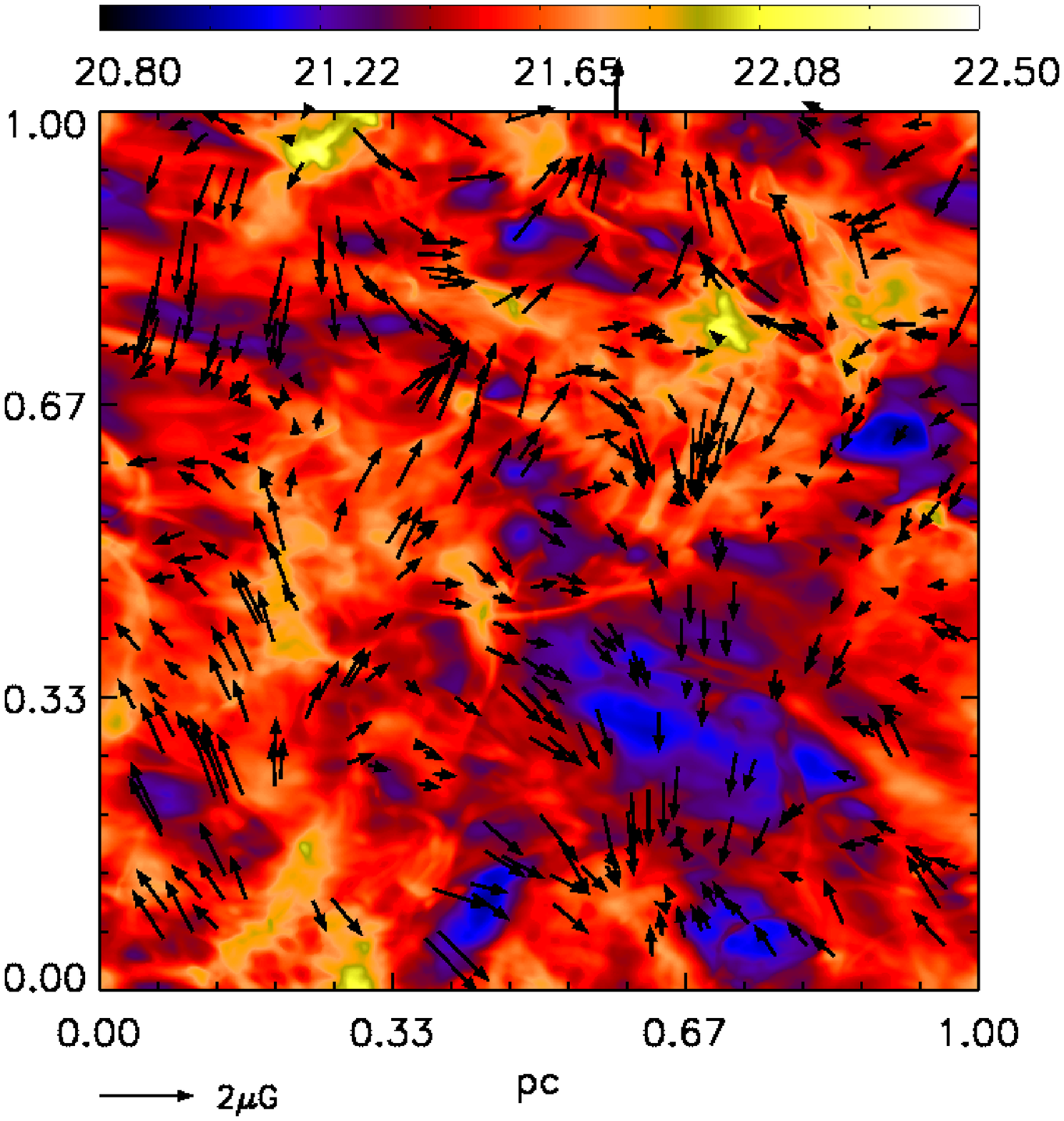}
   \includegraphics[width=0.49\linewidth]{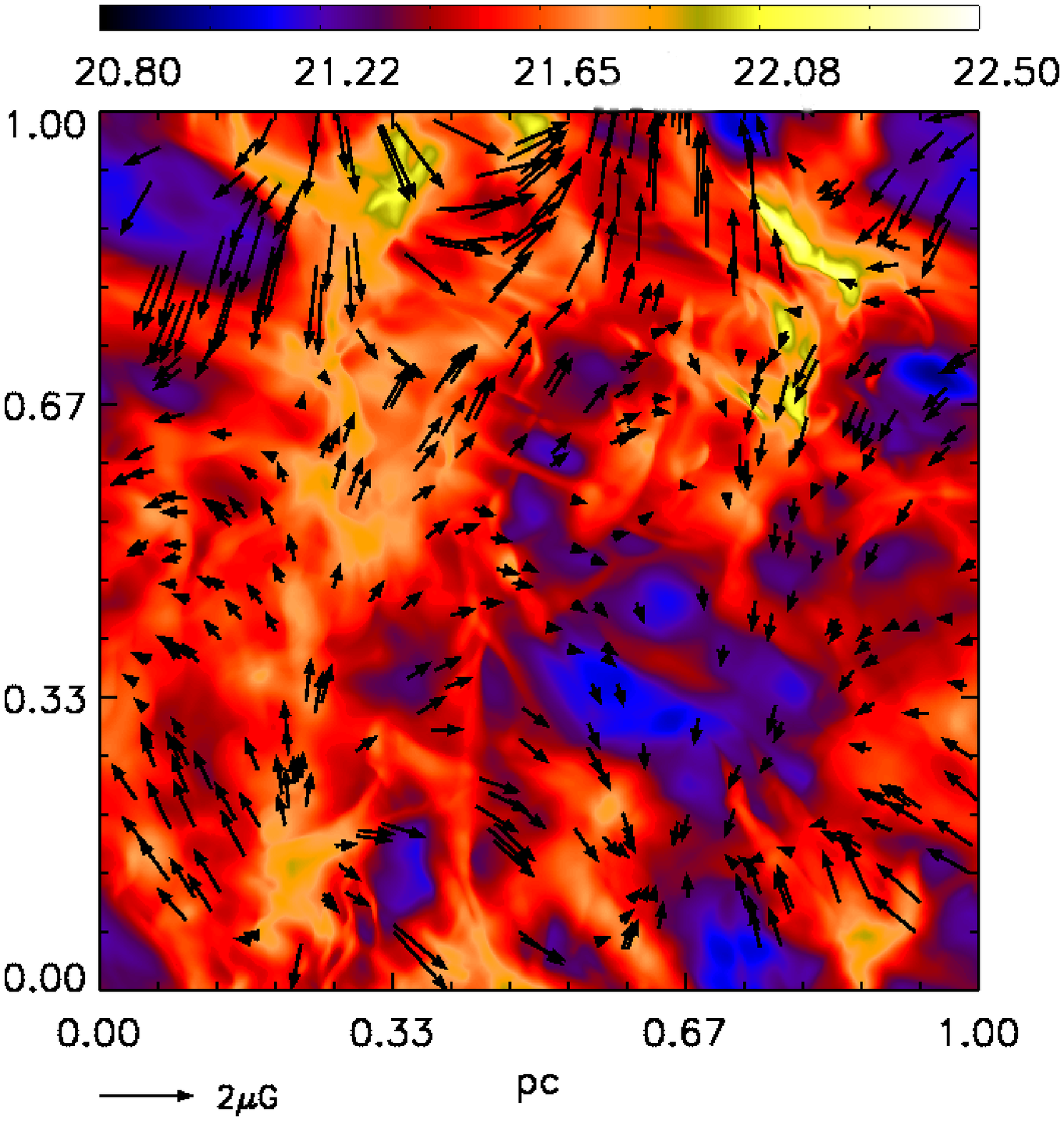}
   \caption{Logarithm of the total gas column density on the xy plane, in cm$^{-2}$, for driven turbulence runs A1 (ideal, left panel) and B1 (ambipolar diffusion, right panel) at t=$5\cdot10^{-3}$ (top)
  and t=$10^{-2}$ (bottom). The black arrows show the projected magnetic field on the same plane.}
   \label{turb_a1_b1}
\end{figure*}

Figure ~\ref{turb_a1_b1} shows the logarithm of the column density for runs A1 and B1, overplotted
with black arrows to illustrate the projected magnetic field.  The snapshots were taken at the same
simulation times as the ones in Figure ~\ref{turb_a0_b0} for the decaying run.  

Although less pronounced, the disappearance of small-scale features and the relative 
lack of dense regions that we observed between runs A0 and B0 are still evident in run B1 with respect to run A1.
However, contrary to the decaying turbulence situation, in this case the projected 
magnetic field maintains a very similar morphology between ideal and non-ideal runs.  
This probably happens because we keep adding velocities with the same
spatial configuration, so the magnetic field is also constantly re-created.

\begin{figure}[!ht]
   \includegraphics[width=\linewidth]{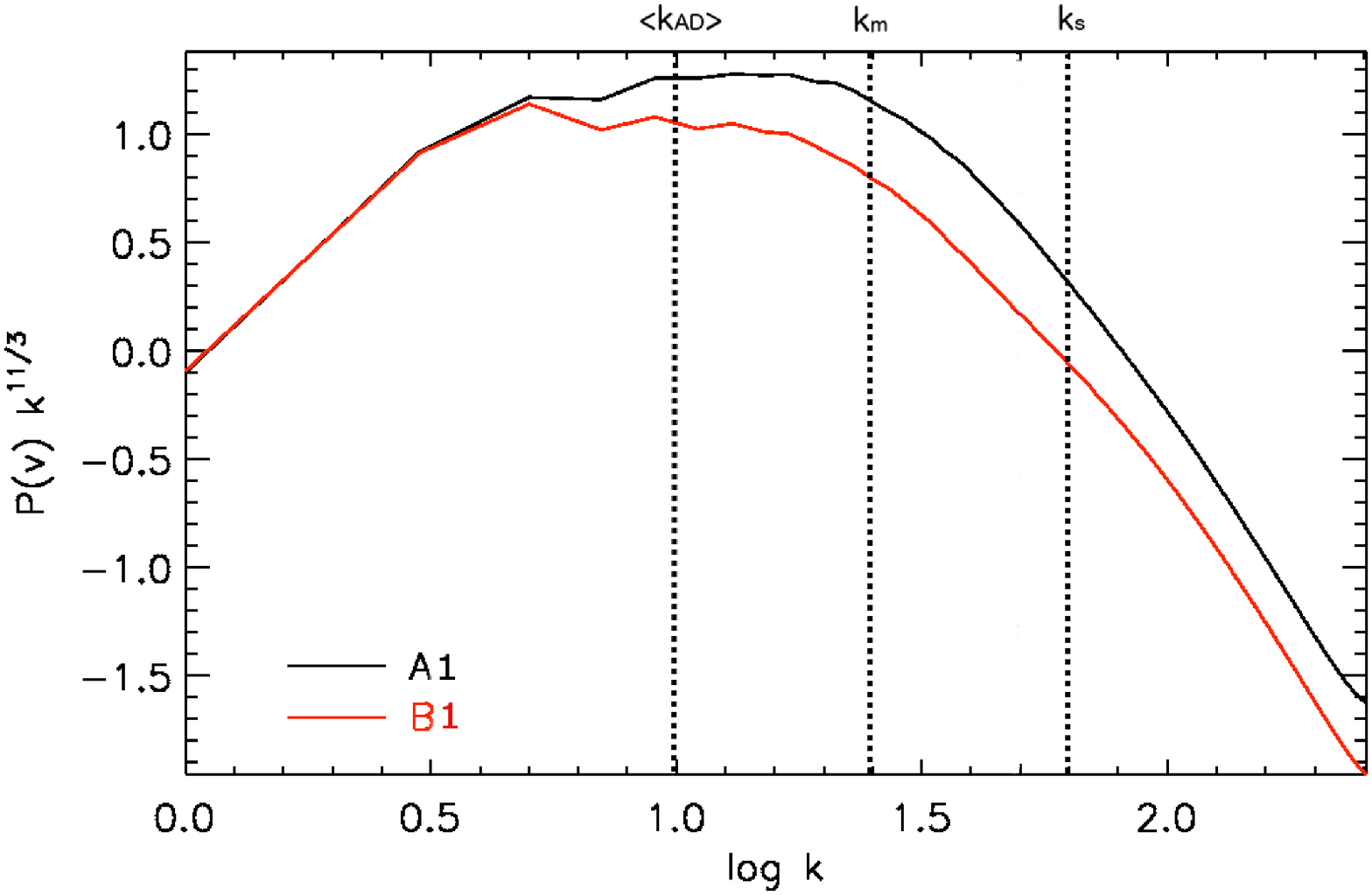}
   \includegraphics[width=\linewidth]{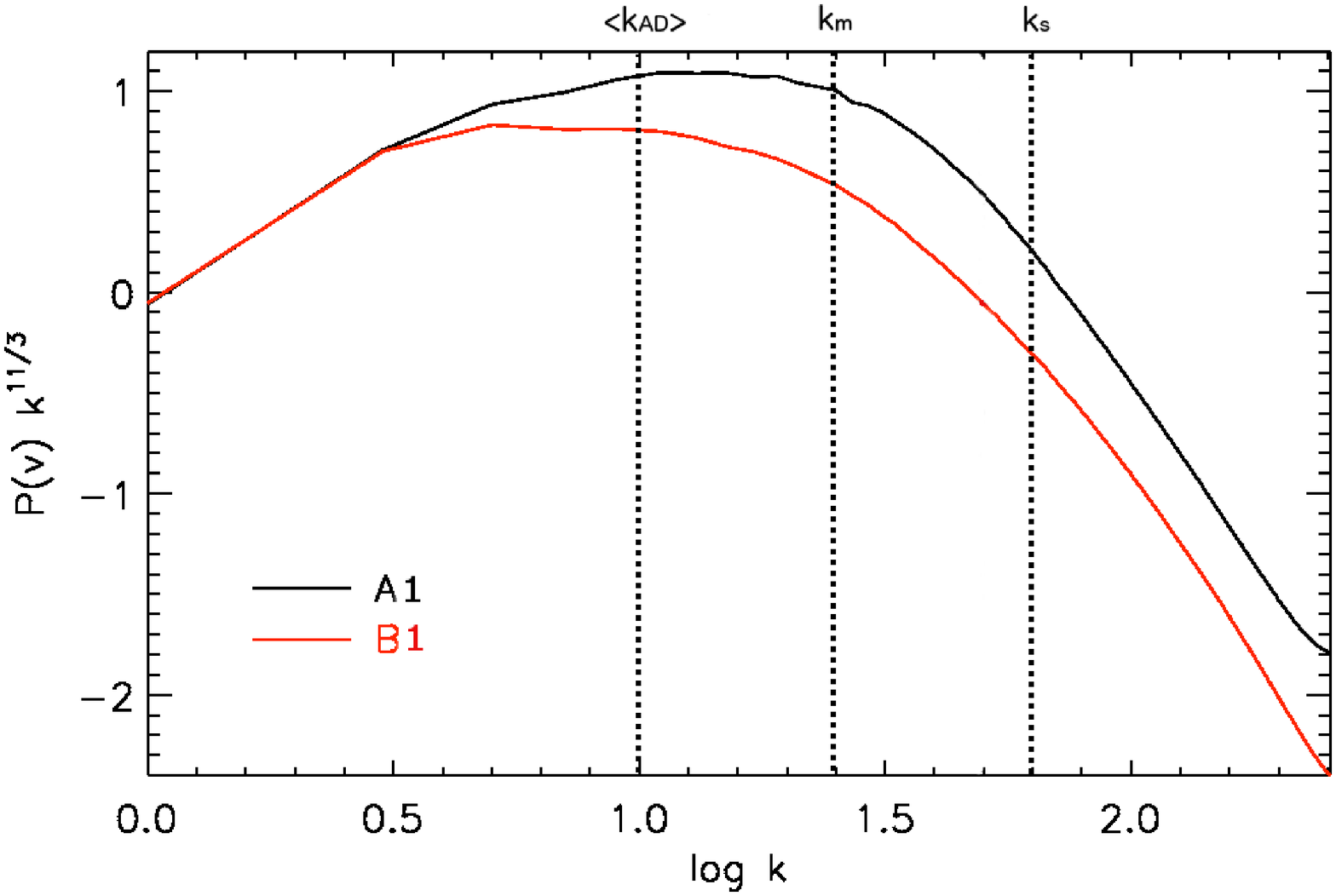}
 \caption{Compensated velocity power spectra for runs A1 and B1.  
Red lines correspond to run A0 and blue lines to run B0. The spectra are shown for two simulation times:  At the top, t=$5\times 10^{-3}$ and at the bottom, t=$10^{-2}$.
The vertical dotted lines show the mean ambipolar diffusion scale,
$<k_{AD}>$, the numerical dissipation scales, k$_m$ and the sonic scale, k$_s$.}
   \label{vel_spec_a1_b1}
\end{figure}
\begin{figure}[!htb]
   \includegraphics[width=\linewidth]{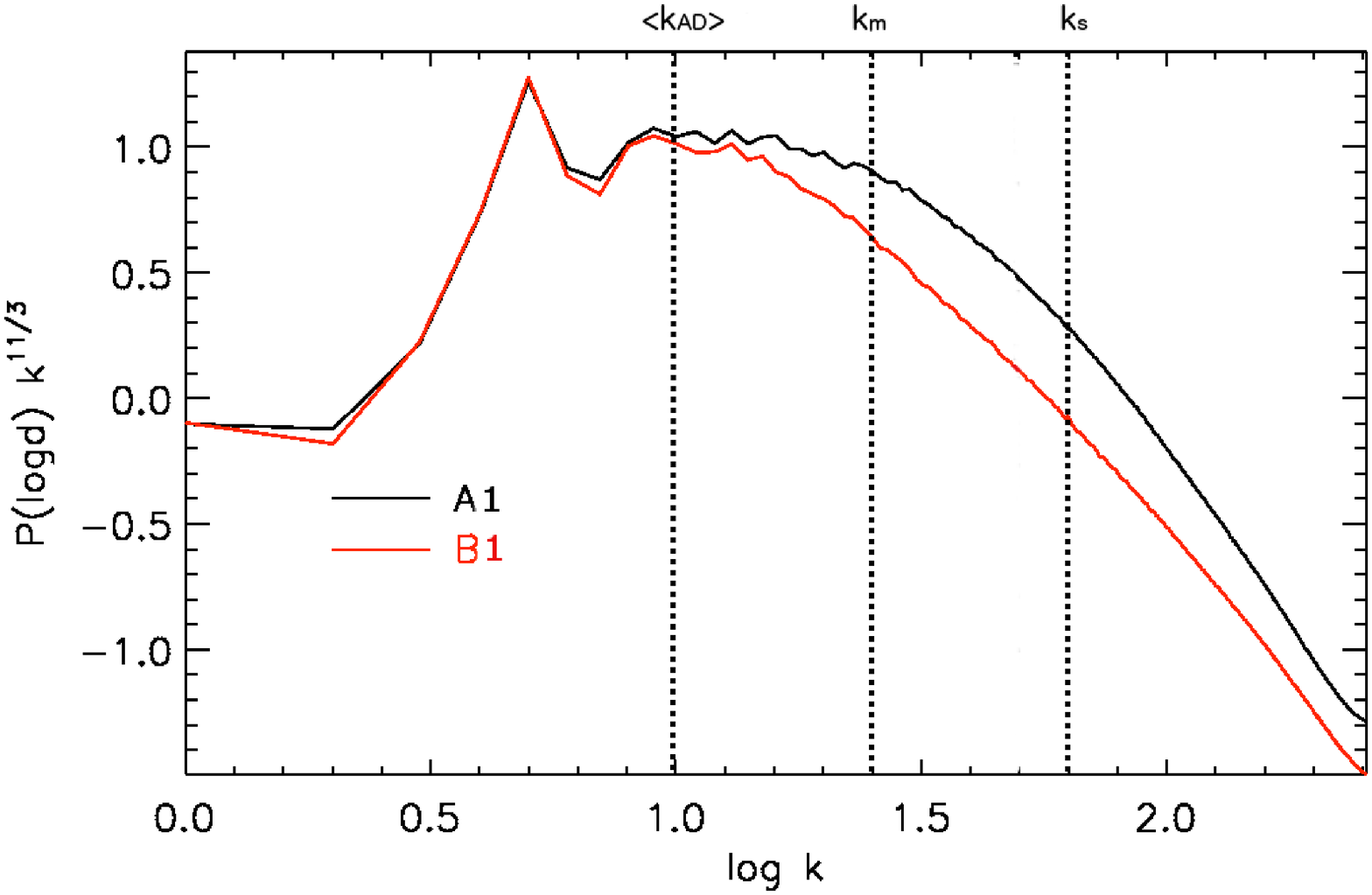}
   \includegraphics[width=\linewidth]{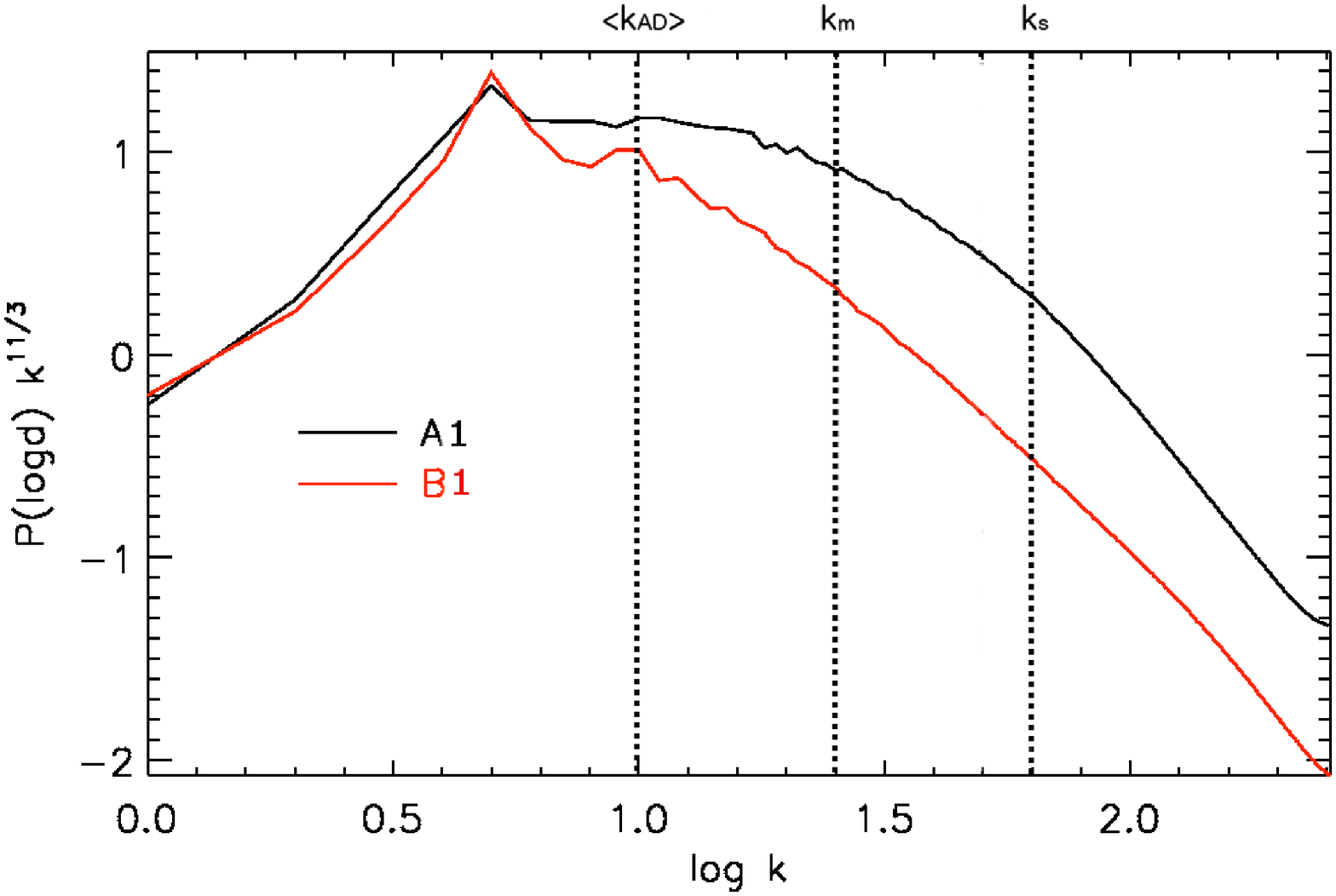}
   \caption{Compensated power spectra of the logarithm of the density at t=$5\times 10^{-3}$ (left) and t=$10^{-2}$ (right). Red lines correspond to run A1 and blue lines to run B1.}
   \label{den_spec_a1_b1}
\end{figure}

\begin{figure*}[!htb]
   \includegraphics[width=0.49\linewidth]{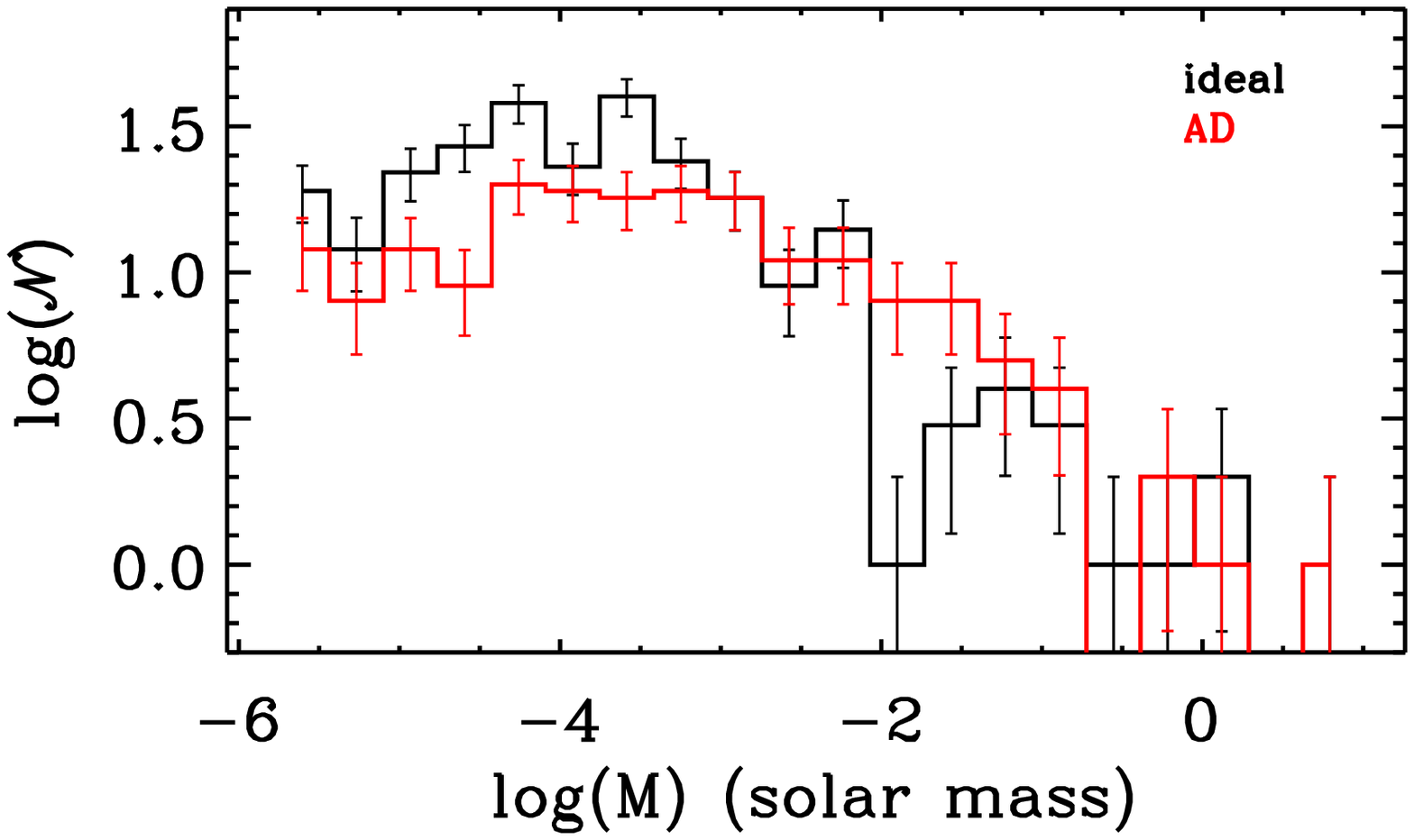}
   \includegraphics[width=0.49\linewidth]{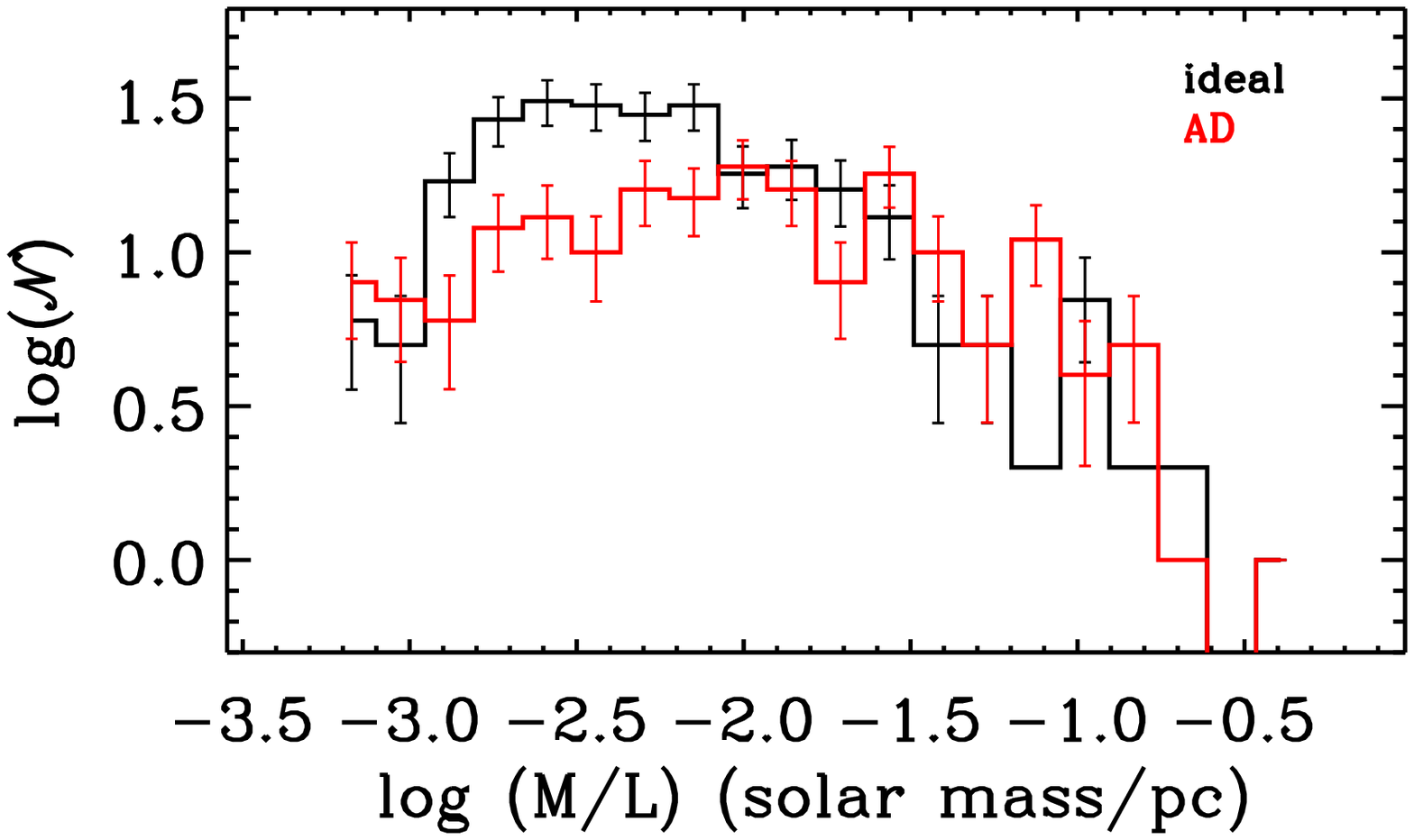}
   \includegraphics[width=0.49\linewidth]{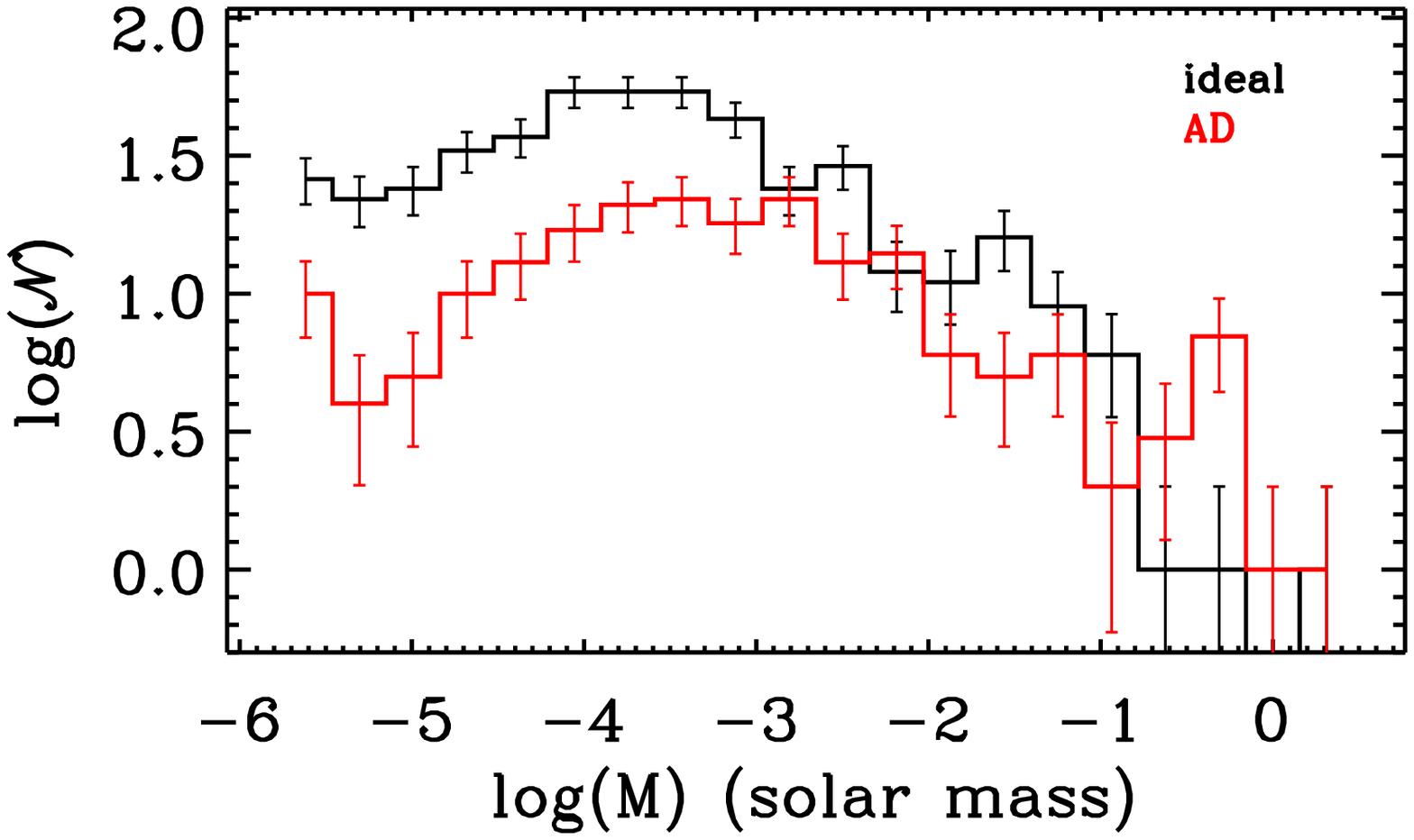}
   \includegraphics[width=0.49\linewidth]{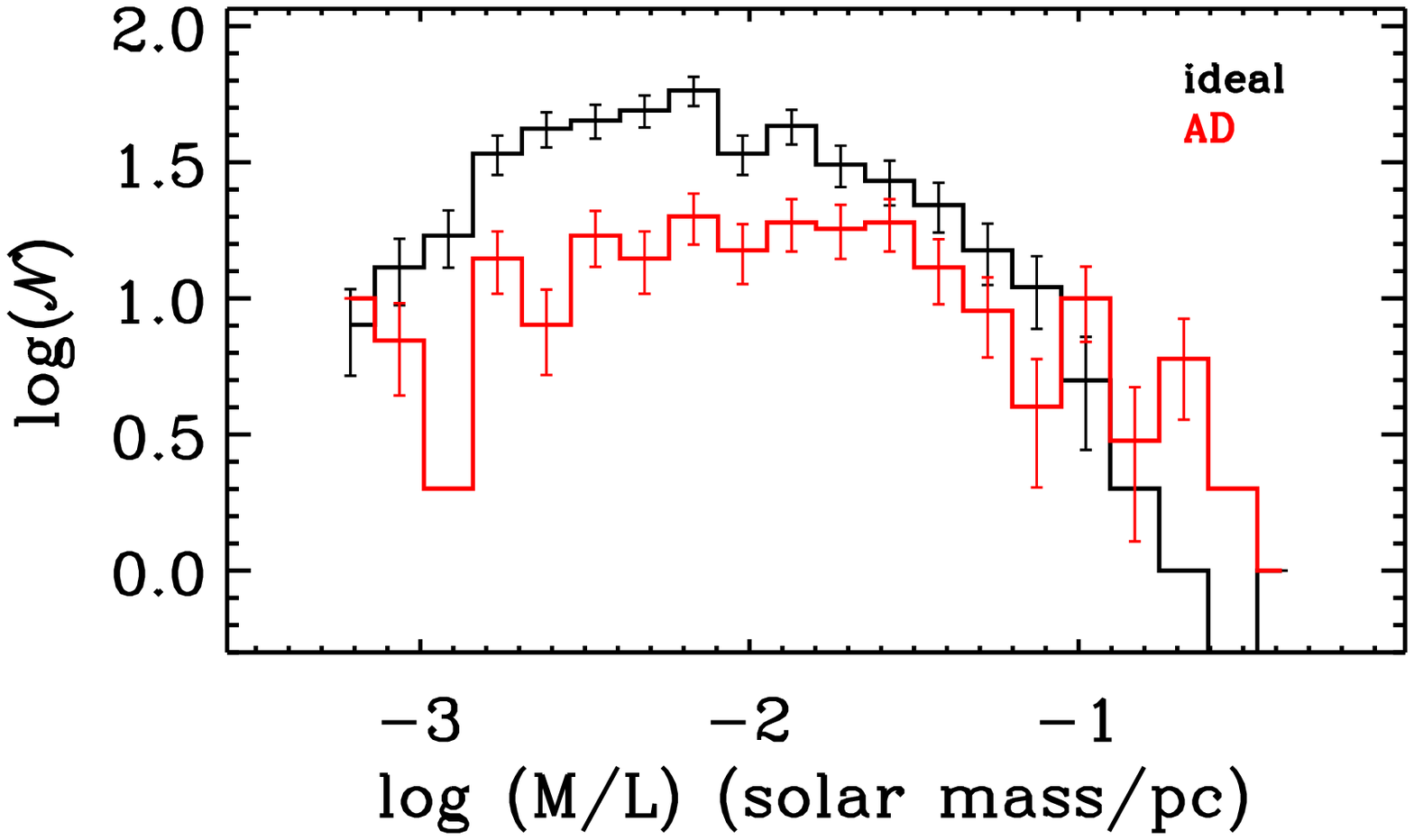}
    \caption{Filament mass (left) 
and mass per unit length (right) distributions. Black lines show run A1 and red lines run B1.  
  The top panel corresponds to $t=5\times 10^{-3}$ and the bottom to $t=10^{-2}$. 
  The density threshold for defining a filament is 2000 cm$^{-3}$}.
 \label{mass_histograms_2_2000}
\end{figure*}
\begin{figure*}[!htb]
   \includegraphics[width=0.49\linewidth]{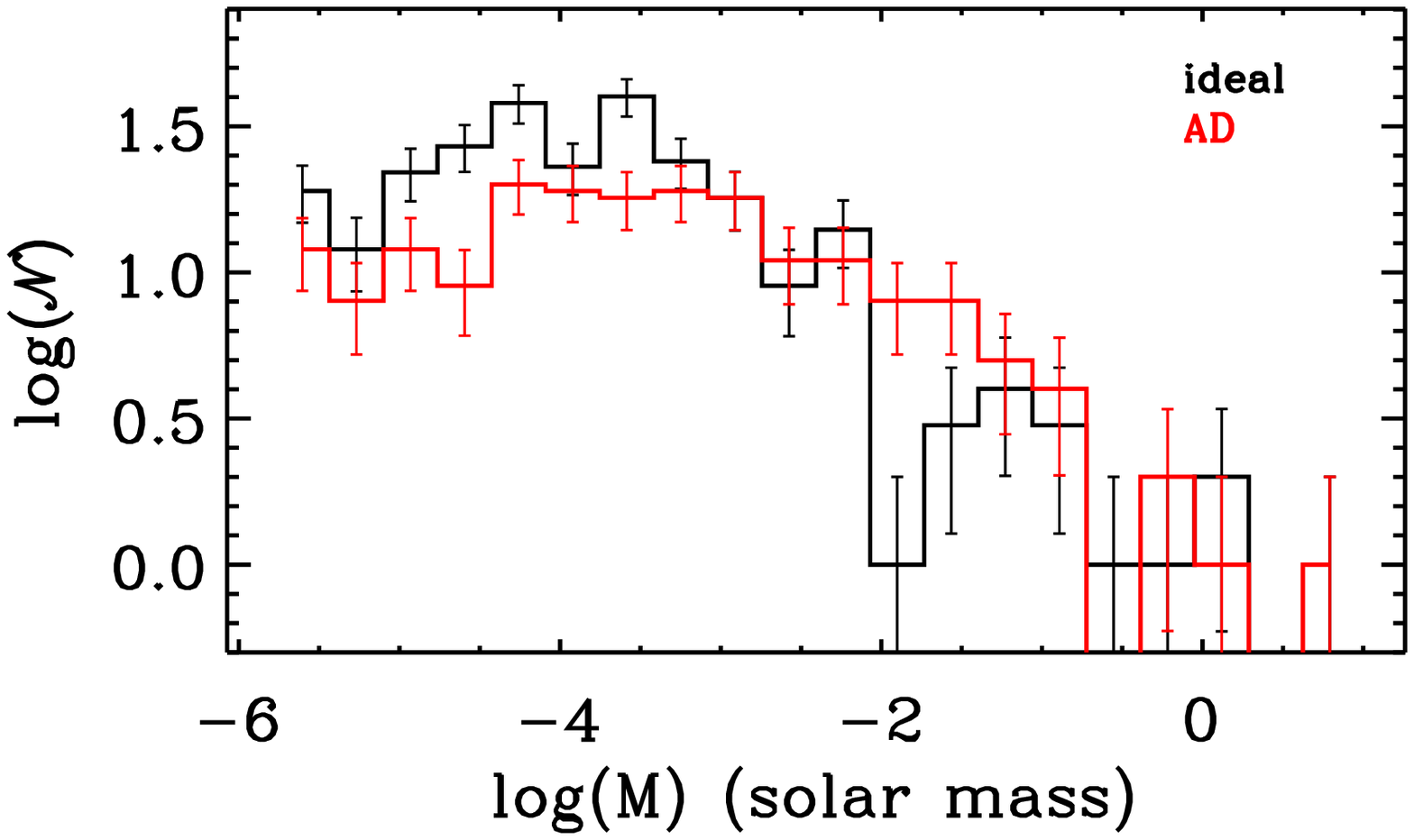}
   \includegraphics[width=0.49\linewidth]{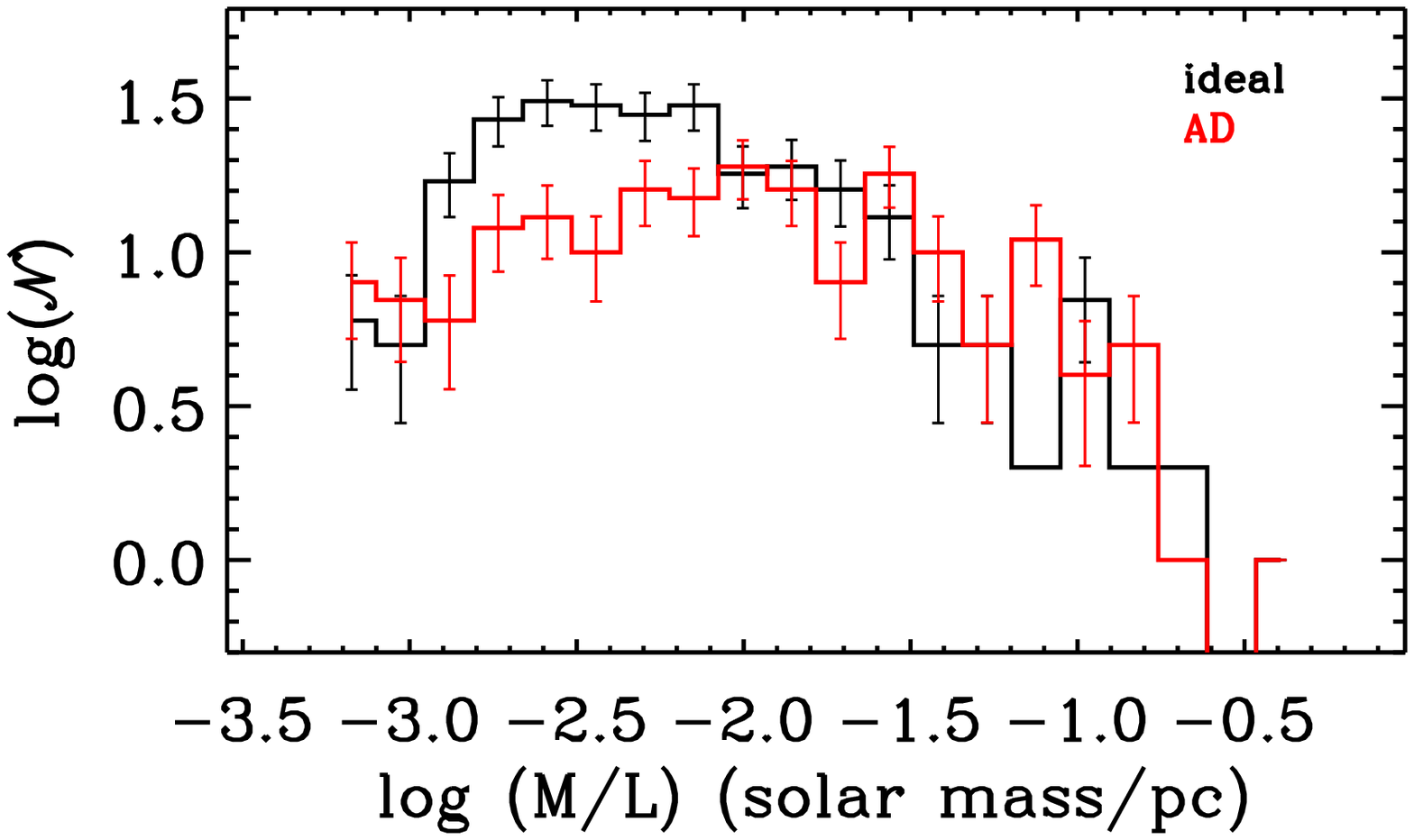}
   \includegraphics[width=0.49\linewidth]{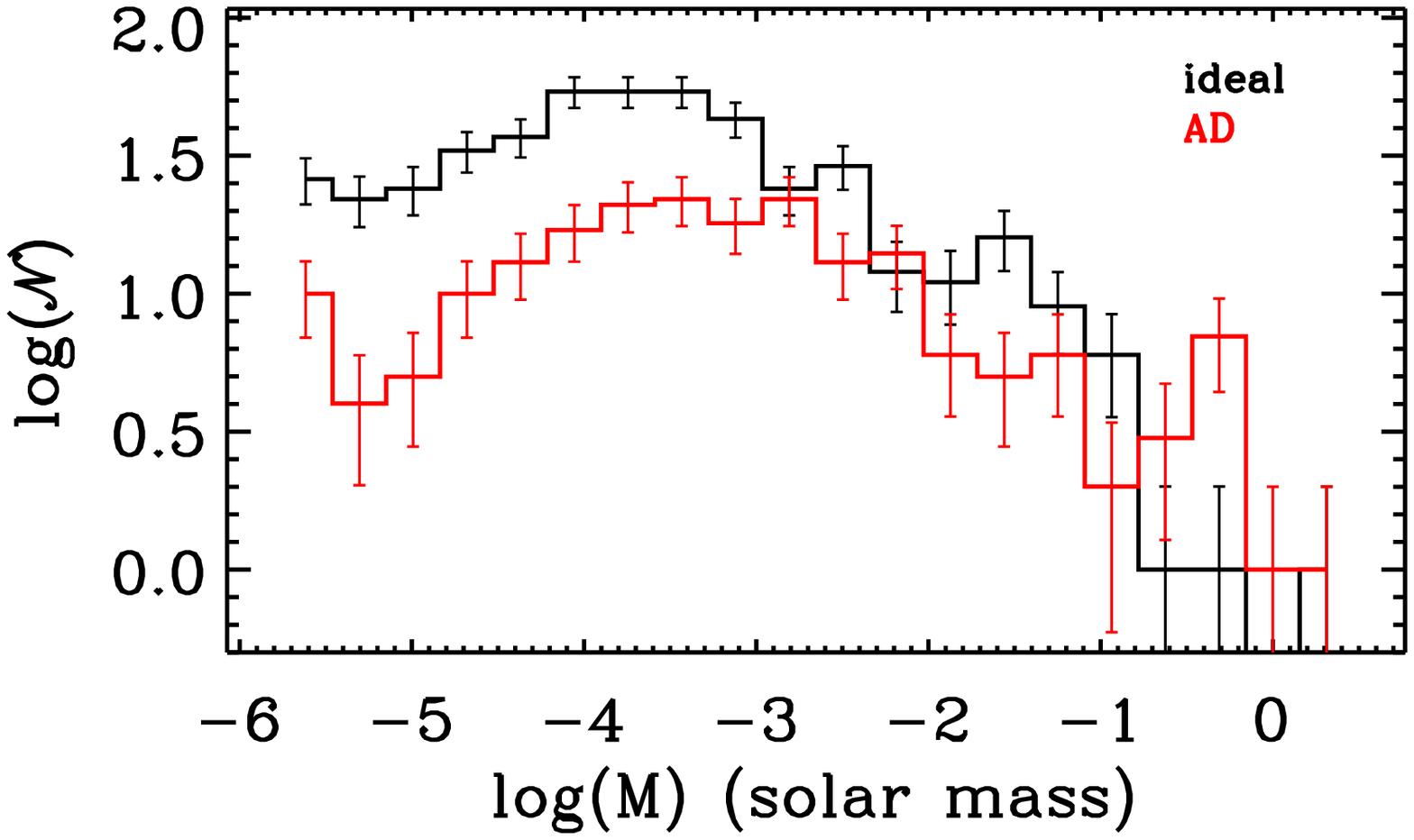}
   \includegraphics[width=0.49\linewidth]{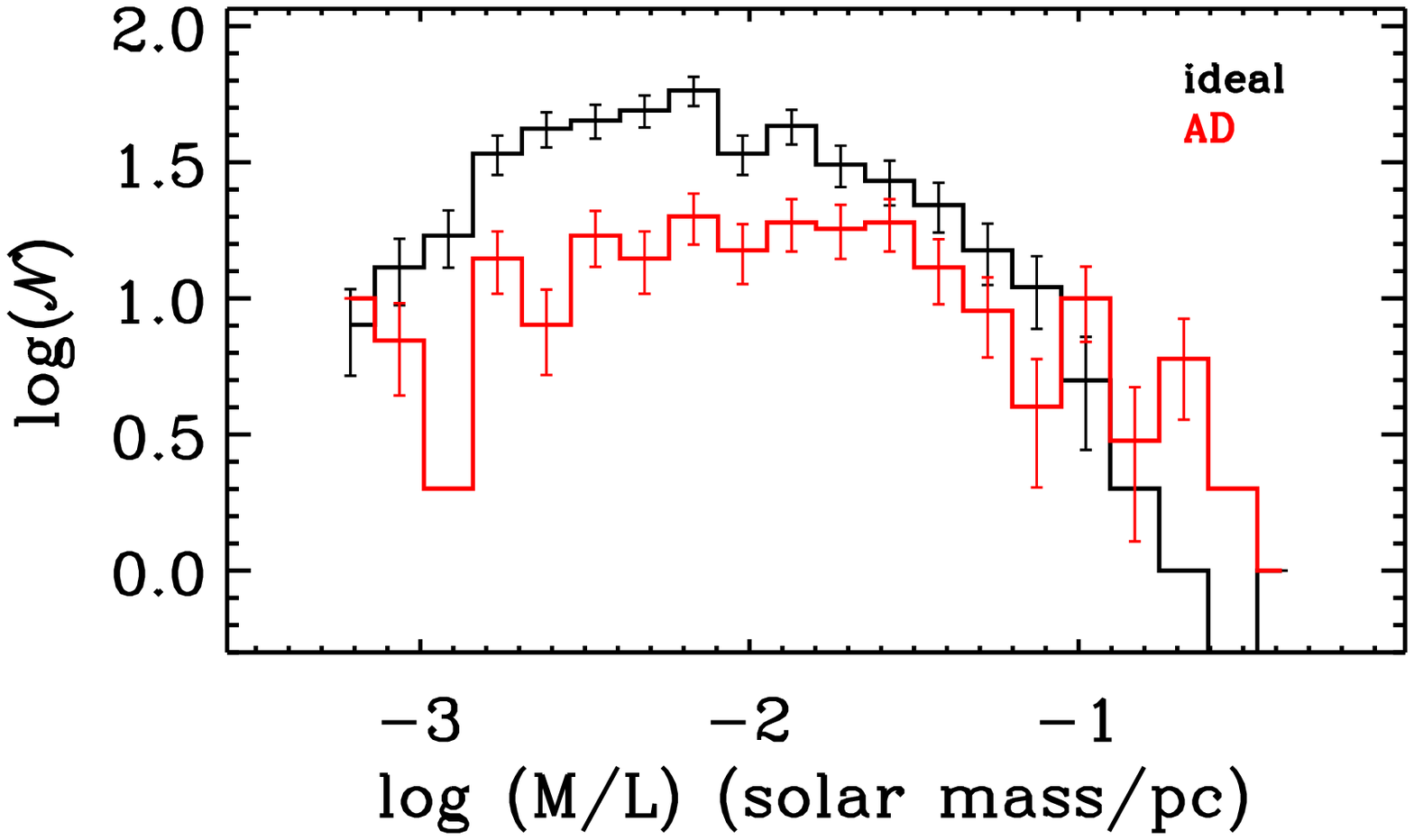}
    \caption{Same as Figure \ref{mass_histograms_2_2000}, for
  a density threshold equal to 5000~cm$^{-3}$}.
 \label{mass_histograms_2_5000}
\end{figure*}

Figures \ref{vel_spec_a1_b1} and \ref{den_spec_a1_b1} show the power spectra of the velocity and of the logarithm of the density.
The inertial range is clearly smaller in comparison to runs A0 and B0.
Distinctive peaks appear in both the density and the velocity power spectra, corresponding to the driving scales.
In addition, the ambipolar diffusion scale is larger here compared to the decaying runs, which also contributes to the loss of scales
in the inertial range.

Nonetheless, the non-ideal run shows less power in small scales both in density and in velocity, which
is the same trend we observed in the case of decaying turbulence.  
This is clear evidence that ambipolar diffusion dominates over numerical dissipation in our simulations.

\subsection{Filament properties}

\begin{figure}[!htb]
   \includegraphics[width=\linewidth]{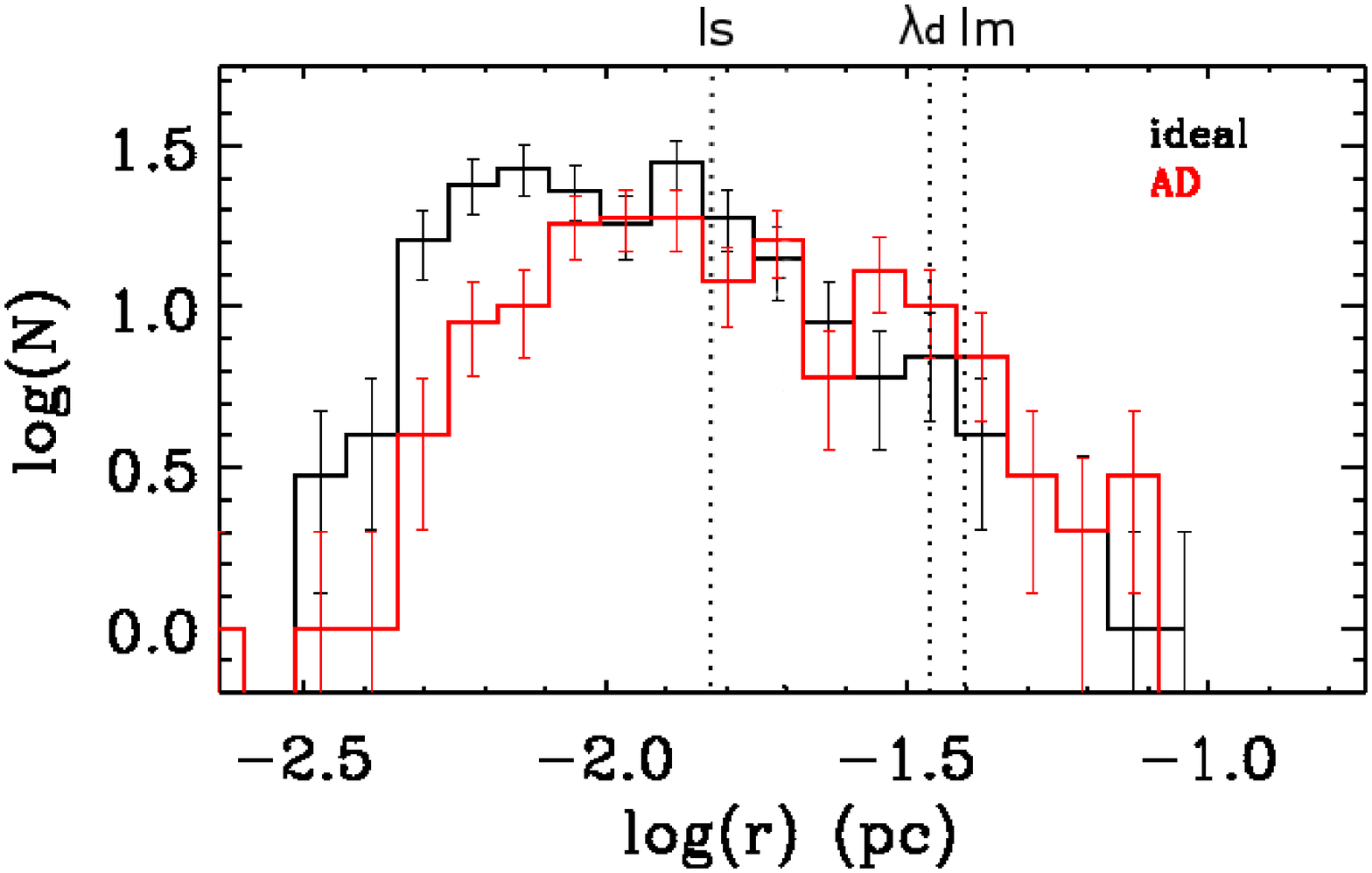}
   \includegraphics[width=\linewidth]{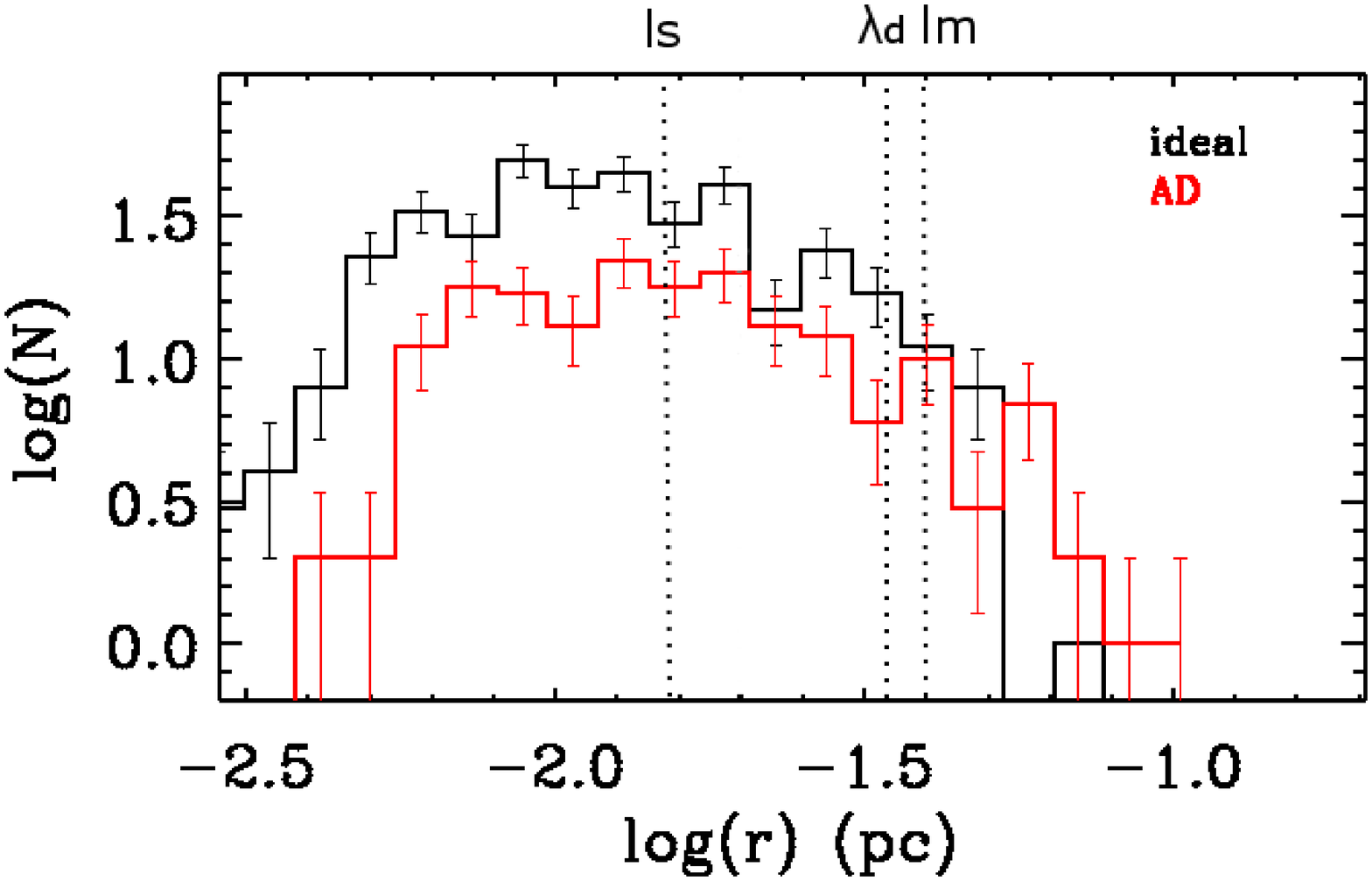}
   \caption{Filament thickness distributions in runs A1 (black lines) and B1 (red lines) at times $t=5\times10^{-3}$ (top) and  $t=10^{-2}$ (bottom). The density threshold for defining a filament is 2000 cm$^{-3}$. The dotted lines indicate the mean ambipolar diffusion dissipation length $\lambda_d$, calculated from Eq. (\ref{ldiss}) for this density threshold, the resolution length $l_{m}$, equal to the length of 20 cells, and the sonic length $l_s$, defined by Eq. (\ref{ls}), for an rms Mach number of 4.}
 \label{logthick_histograms_2_2000}
\end{figure}
\begin{figure}[!htb]
   \includegraphics[width=\linewidth]{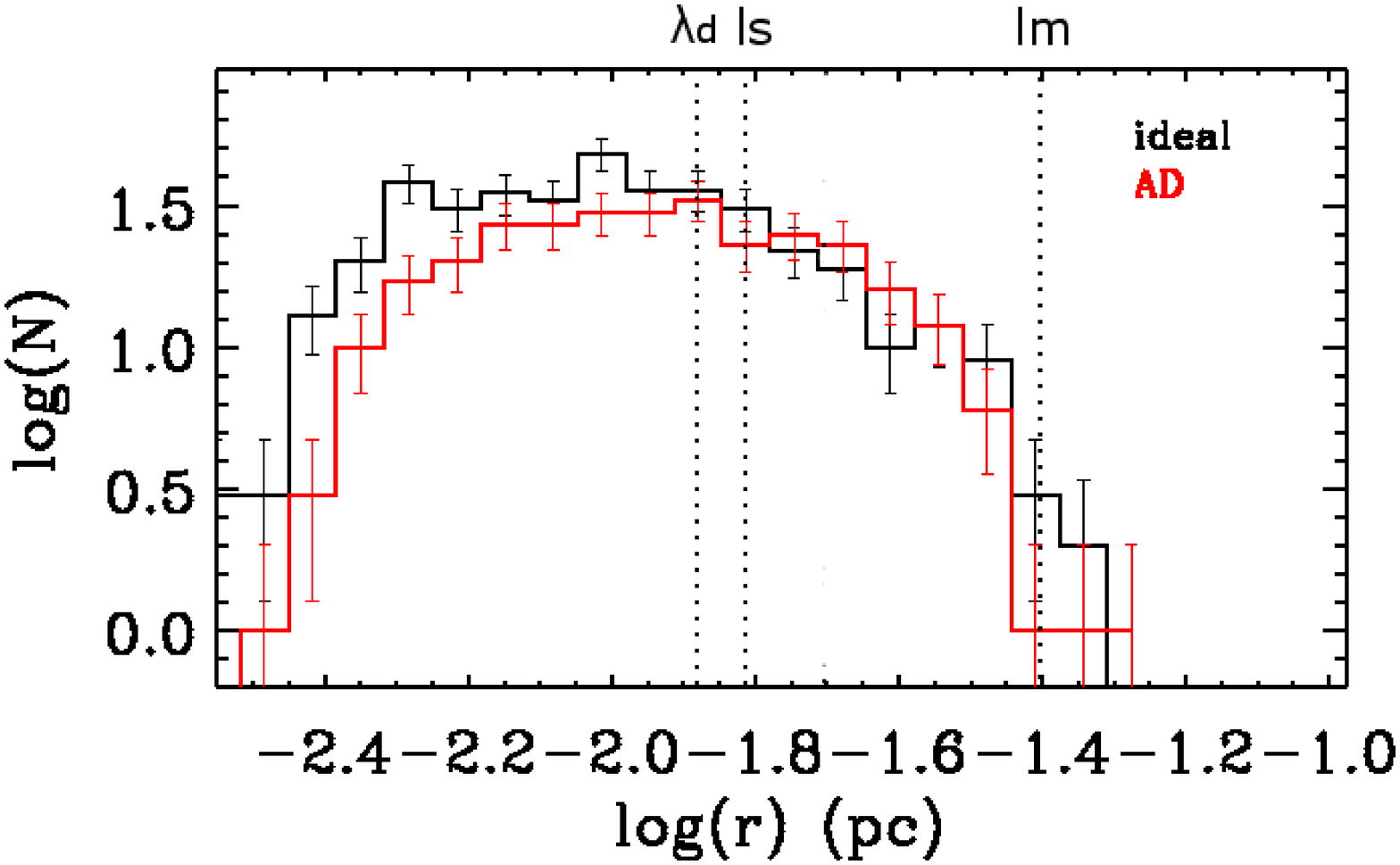}
   \includegraphics[width=\linewidth]{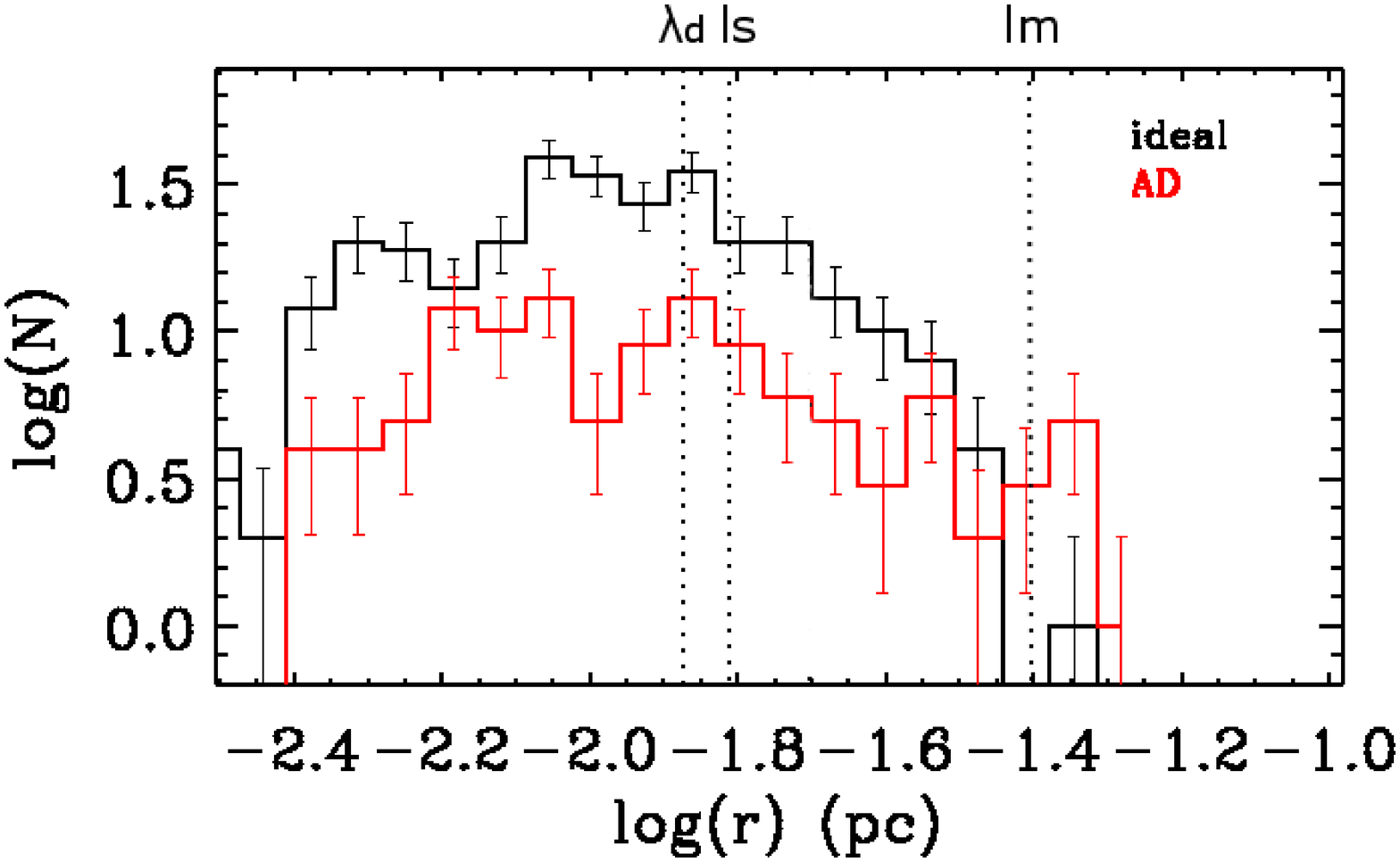}
  \caption{Same as Figure \ref{logthick_histograms_2_2000}, with a density threshold of 5000 cm$^{-3}$.}
 \label{logthick_histograms_2_5000}
\end{figure}

\begin{figure}[!htb]
  \includegraphics[width=\linewidth]{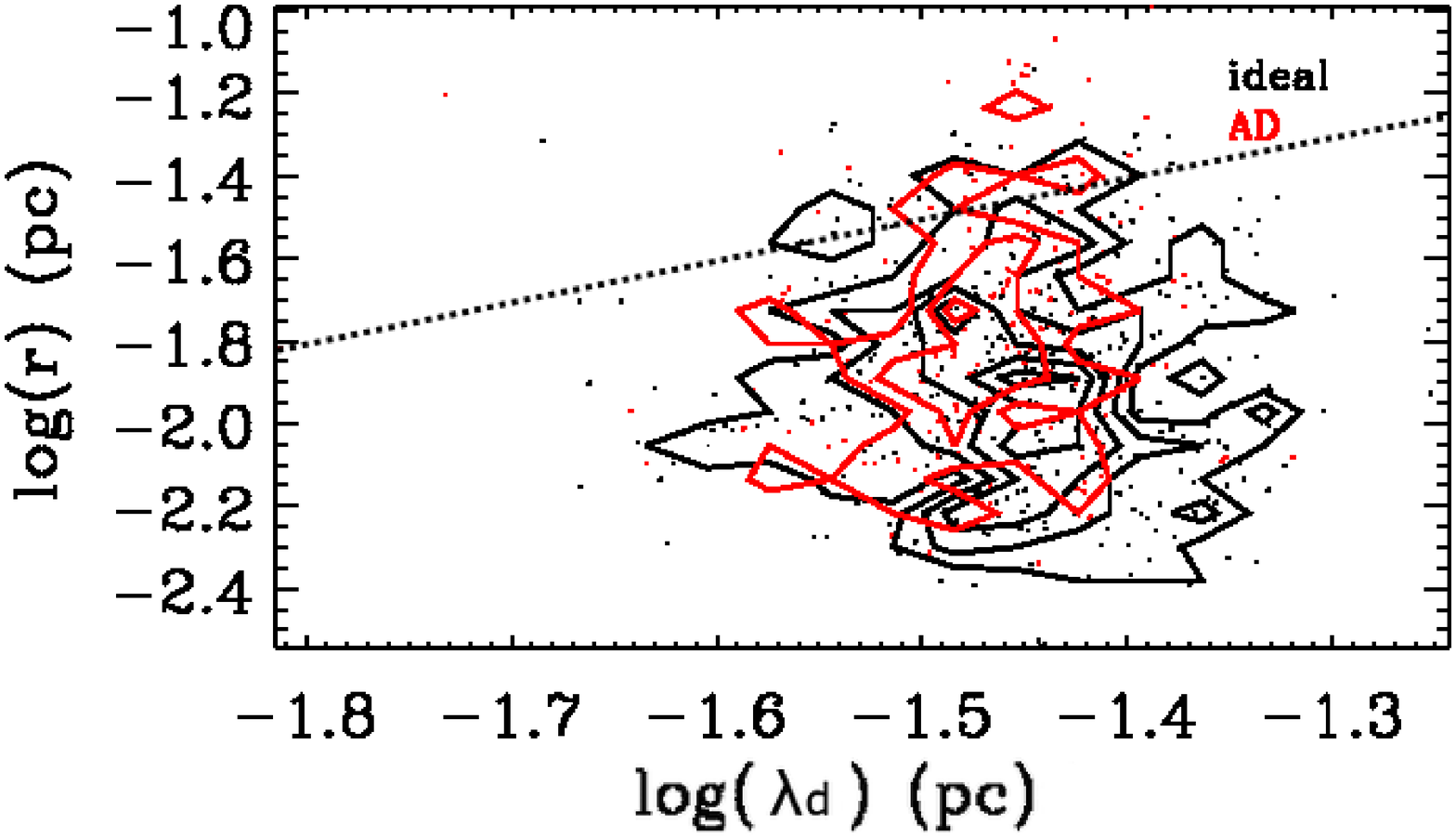}
   \includegraphics[width=\linewidth]{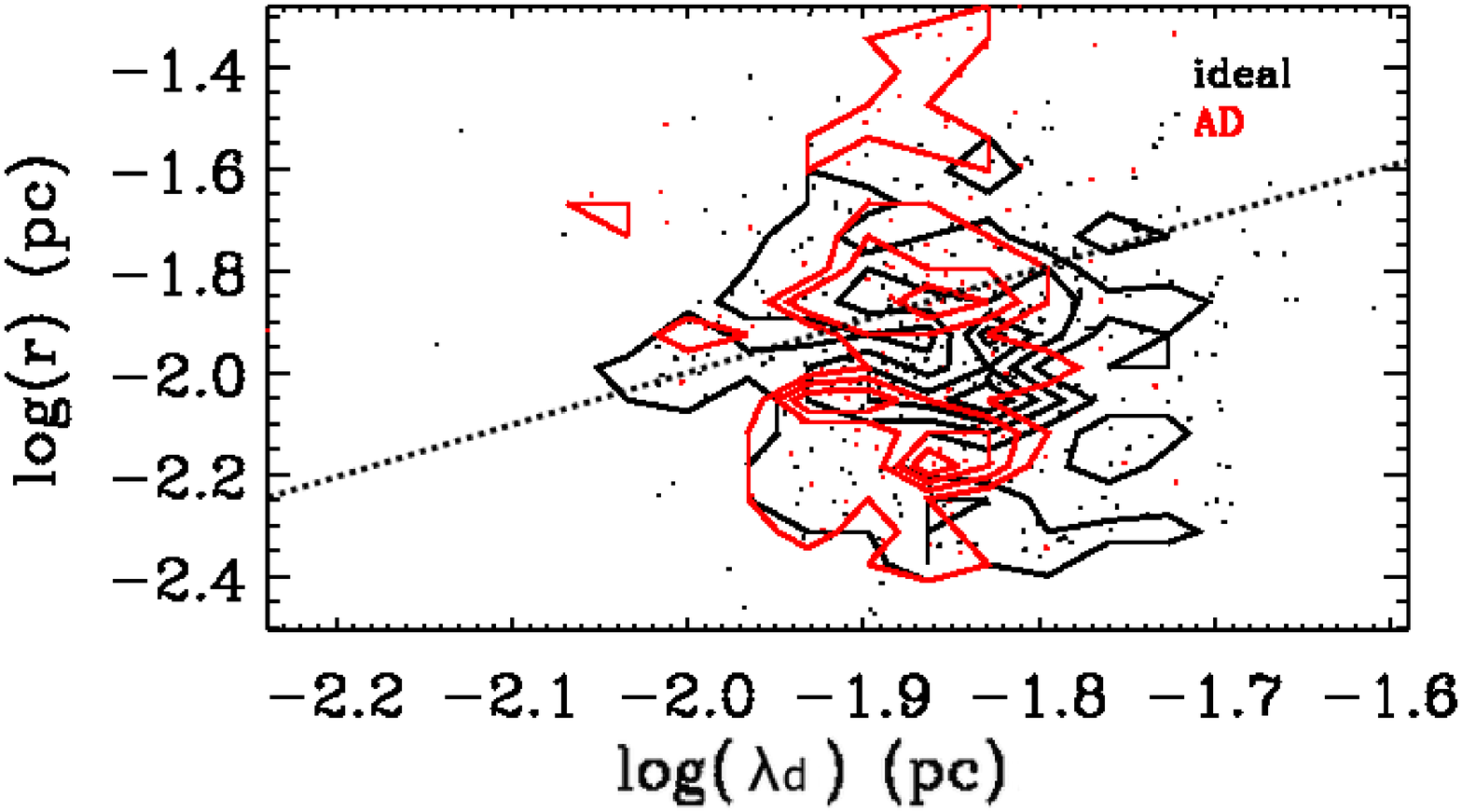}
   \caption{Same as Figure \ref{thick_ldiss_decay}, for runs A1 (black) and B1 (red).} 
 \label{thick_ldiss_driven}
\end{figure}

Figures \ref{mass_histograms_2_2000} and \ref{mass_histograms_2_5000} 
show the mass and mass per unit length distributions of the filaments in runs A1 and B1.
The shift of the peak mass to higher values with ambipolar diffusion is also observed here,
but it is much clearer at later times.  

The same trend appears in the distributions of the mean filament thickness, shown in
Figures \ref{logthick_histograms_2_2000} and \ref{logthick_histograms_2_5000}. 
While at early times the distributions
appear similar, as time advances they separate by a factor of 1.8.   

Figure \ref{thick_ldiss_driven} shows the relative distributions of critical length and filament thickness,
like Figure \ref{thick_ldiss_decay} in Section \ref{thick}. The picture is the same as the one
we described for the decaying turbulence situation: The thickness of the filaments we
measure agrees very well with the estimate of the local critical length for ambipolar diffusion,
with the critical length being much better defined in the non-ideal simulations. 

Overall, runs A1 and B1, although at a different Mach number
and with turbulence constantly driven, point to the same result as the decaying turbulence
runs A0 and B0, that ambipolar diffusion cuts the turbulent cascade off at a larger scale, creating
on average dense structures of larger thickness.

\end{document}